\documentclass[useAMS,usenatbib]{mn2e}

\usepackage{graphicx}
\usepackage{rotating}

\newif\ifAMStwofonts                        
\newcommand{\lsimeq}{{_<\atop^{\sim}}}
\newcommand{\gsimeq}{{_>\atop^{\sim}}}

\title[The PEP HerMES Luminosity Function]{The {\textit{Herschel}}\thanks{{\it Herschel} is an ESA space observatory with science instruments provided by European-led Principal Investigator consortia and with important participation from NASA} PEP/HerMES Luminosity Function -- I: Probing the Evolution of PACS selected Galaxies to {\textit z}$\simeq$4}

\author[C.~Gruppioni, F.~Pozzi, G.~Rodighiero et al.]
{\parbox{\textwidth}{\raggedright C.~Gruppioni$^{(1)}$\thanks{E-mail: \texttt{carlotta.gruppioni@oabo.inaf.it}},
F.~Pozzi$^{(2)}$,
G.~Rodighiero$^{(3)}$,
I.~Delvecchio$^{(2)}$,
S.~Berta$^{(4)}$,
L.~Pozzetti$^{(1)}$, 
G.~Zamorani$^{(1)}$,
P.~Andreani$^{(5)}$,
A.~Cimatti$^{(2)}$,
O.~Ilbert$^{(6)}$
% ...
}\vspace{0.4cm}\\
\parbox{\textwidth}{\raggedright $^{(1)}$INAF - Osservatorio Astronomico di Bologna, via Ranzani 1, I--40127 Bologna, Italy.\\
$^{(2)}$Dipartimento di Astronomia, Universit\`a di Bologna, via Ranzani 1, I--40127 Bologna, Italy.\\
$^{(3)}$Dipartimento di Astronomia, Universit\`a di Padova, vicolo dell'Osservatorio 3, I--35122 Padova, Italy.\\
$^{(4)}$Max-Planck-Institut f\"{u}r Extraterrestrische Physik (MPE), Postfach 1312, D-85741 Garching, Germany. \\
$^{(5)}$ESO, Karl-Schwarzschild-Strasse 2, D-85748, Garching, Germany.\\
$^{(6)}$Laboratoire d'Astrophysique de Marseille, CNRS-Universit\`e de Provence, rue Fr\`ed\`eric Joliot-Curie 38, 13388 Marseille Cedex 13, France.\\
% ...
}}

\begin{document}

\date{Accepted 2013 February 18.  Received 2013 February 13; in original form 2012 October 23}

\pagerange{\pageref{firstpage}--\pageref{lastpage}} \pubyear{2012}

\maketitle

\label{firstpage}
\begin{abstract}
We exploit the deep and extended far-infrared data-sets (at 70, 100 and 160\,$\mu$m) of the {\em Herschel} GTO PACS Evolutionary Probe (PEP) Survey, in combination with the HERschel Multi-tiered Extragalactic Survey (HerMES) data at 250, 350 and 500\,$\mu$m, to derive the evolution of the rest-frame 35-$\mu$m, 60-$\mu$m, 90-$\mu$m, and total infrared (IR) luminosity functions (LFs) up to $z$$\sim$4. 
We detect very strong luminosity evolution for the total IR LF (L$_{\rm{IR}}\propto$(1$+$$z$)$^{3.55\pm0.10}$ up to $z$$\sim$2, and $\propto$(1$+$$z$)$^{1.62\pm0.51}$ at 2$<$$z$$\lsimeq$4) combined with a density evolution ($\propto$(1$+$$z$)$^{-0.57\pm0.22}$ up to $z$$\sim$1 and $\propto$(1$+$$z$)$^{-3.92\pm0.34}$ at 1$<$$z$$\lsimeq$4). 
In agreement with previous findings, the IR luminosity density ($\rho_{\rm{IR}}$) increases steeply to $z$$\sim$1, then flattens between $z$$\sim$1 and $z$$\sim$3 to decrease at $z$$\gsimeq$3.
Galaxies with different SEDs, masses and sSFRs evolve in very different ways and this large and deep statistical sample is the first one allowing us to separately study the different evolutionary behaviours of the individual IR populations contributing to $\rho_{\rm{IR}}$. 
Galaxies occupying the well established SFR--stellar mass main sequence (MS) are found to dominate both the total IR LF and $\rho_{\rm{IR}}$ at all redshifts, with 
the contribution from off-MS sources ($\geq$0.6~{\rm dex} above MS) being nearly constant ($\sim$20\% of the total $\rho_{\rm{IR}}$) and showing no significant signs of increase 
with increasing $z$ over the whole 0.8$<$$z$$<$2.2 range.
Sources with mass in the range 10$\leq$log($M$/M$_{\odot}$)$\leq$11 are found to dominate the total IR LF, with more massive galaxies prevailing at the
bright end of the high-$z$ ($\gsimeq$2) LF.  A two-fold evolutionary scheme for IR galaxies is envisaged: on the one hand, a starburst-dominated phase in which the SMBH grows and is obscured by dust
, is followed by an AGN-dominated phase, then evolving toward a local elliptical. On the other hand, 
moderately star-forming galaxies containing a low-luminosity AGN have various properties suggesting they are good candidates for systems in a transition phase preceding the formation of steady spiral galaxies.
\end{abstract}

\begin{keywords}
cosmology: observations --  galaxies: active -- galaxies: evolution -- galaxies: luminosity function -- galaxies: starburst -- infrared: galaxies.
\end{keywords}

\section{Introduction}

Understanding the origin and growth of the galaxies we observe today is one of the main problems of current cosmology. 
The luminosity function (LF) provides one of the fundamental tools to probe the distribution of galaxies over cosmological time, since it allows us to assess the statistical nature of galaxy formation and evolution. When computed at different redshifts, the LF constitutes the most direct method for exploring the evolution of a galaxy population, describing the relative number of sources of different luminosities counted in representative volumes of the Universe.
The LF computed for different samples of galaxies can provide a crucial comparison between the distribution of different galaxy types, i.e. galaxies at different redshifts, in different enviroments or selected at different wavelengths.

It has now become clear that we cannot understand galaxy evolution without accounting for the energy absorbed by dust and re-emitted at longer wavelengths (e.g, \citealt{genz00}), in the infrared (IR) or sub-millimetre (sub-mm). 
Dust is responsible for obscuring the ultraviolet (UV) and optical light from galaxies: since star-formation occurs within dusty molecular clouds, far-IR and sub-mm data, where the absorbed radiation is re-emitted, are essential for providing a complete picture of the history of star-formation through cosmic time, which is one of the fundamental instruments we have to reconstruct how galaxies have evolved since their formation epoch. For these reasons, extragalactic surveys in the rest-frame IR represent a key tool for understanding galaxy formation and evolution. 

Surveys of dust emission performed with the former satellites exploring the Universe in the mid- and far-IR domain, i.e. the {\em Infrared Astronomical Satellite} ({\em IRAS}; \citealt{neu84}) and the {\em Infrared Space Observatory} ({\em ISO}; \citealt{kes96}), allowed the first studies of the IR-galaxy LF 
at $z$$\lsimeq$0.3 (\citealt{saun90}) and $z$$\lsimeq$1 (\citealt{poz04}), respectively. 
With {\em Spitzer} 24-$\mu$m data, it was possible to study the evolution of the mid-IR LF up to $z$$\sim$2 (e.g. \citealt{lef05}, \citealt{cap07}, \citealt{rod10a}), while, even with the deepest {\em Spitzer Space Telescope} (\citealt{wer04}) 70-$\mu$m data, only $z$$\sim$1--1.2 could be reached in the far-IR (\citealt{mag09}; \citealt{pat12}) ) -- though \citet{mag11} reached $z$$\sim$2 through stacking. 
Since the rest-frame IR spectral energy distributions (SEDs) of star-forming galaxies and AGN peak at 60--200\,$\mu$m, to measure their bolometric luminosity and evolution with $z$ we need to observe in the far-IR/sub-mm regime. However, the detection of large numbers of high-$z$ sources at the peak of their IR SED was not achievable before the {\em Herschel Space Observatory} (\citealt{pil10}), due to source confusion and/or low detector sensitivity, and our knowledge of the far-IR luminosity function in the distant Universe is still affected by substantial uncertainties.
Ground-based and balloon-borne observations in the mm/sub-mm range, probing the evolution of the most distant ($z$$\gsimeq$2) and luminous dusty galaxies, have so far been limited to the identification of sources at the very bright end of the luminosity function (e.g., \citealt{chap05}).
All of these works detected strong evolution in both luminosity and/or density, indicating that IR galaxies were more luminous and/or more numerous in the past. 
Strong observational evidence of high rates of evolution for IR galaxies has been obtained also through the detection of a large amount of energy contained in the Cosmic Infrared Background (CIRB; \citealt{hau01}), and the source counts from several deep cosmological surveys (from 15\,$\mu$m to 850\,$\mu$m)
largely exceeding the no-evolution expectations (e.g. \citealt{smail97}; \citealt{elb99}; \citealt{pap04}; \citealt{beth10}; \citealt{marsd11}). Both the CIRB and the source counts require a strong increase in the IR energy density between the present time and $z$$\sim$1--2. At higher redshifts the total emissivity of IR galaxies is poorly constrained, due to the scarcity of {\em Spitzer} galaxies at $z$$>$2, the large spectral extrapolations to derive the total IR luminosity from the mid-IR (see e.g. \citealt{elb10}, \citealt{nor10} and \citealt{nor12} for descriptions of the failure, at least at $z$$>$1.5, of previous total IR luminosity extrapolations from the mid-IR, although we must note that this failure mainly affects luminosity-dependent methods like, e.g., that of \citealt{ce01}) and the incomplete information on the $z$-distribution of sub-mm sources (Chapman et al. 2005).

{\em Herschel}, with its 3.5-m mirror, is the first telescope which allows us to detect the far-IR population to high redshifts ($z$$\sim$4--5) and to derive its rate of evolution through a detailed LF analysis. The new extragalactic surveys provided by {\em Herschel} in the far-IR/sub-mm domain, like the wide and shallow {\it Herschel}-ATLAS (\citealt{eal10a}; \citealt{dun11}), the complementary {\it Herschel} Multi-tiered Extragalactic Survey (HerMES; \citealt{oliv12}) and PACS Evolutionary Probe (PEP; \citealt{lutz11}) covering the most popular cosmological fields, and the deep, pencil beam, {\it Herschel}-GOODS project (\citealt{elb11}), will be crucial to assess galaxy and AGN evolution in the IR at $z$$>$2. They will give us the opportunity to study in detail the population of IR galaxies and their evolution with cosmic time since the Universe was about a billion years old. In particular, the {\em Photodetector Array Camera \& Spectrometer} ({\em PACS}; \citealt{pog10}), with its high sensitivity and resolution at 70-$\mu$m, 100-$\mu$m and 160-$\mu$m, is the best suited instrument to detect faint IR sources by overcoming the source confusion and blending problems that affected the previous far-IR missions. 

This is the first of two papers aiming at deriving the far- and total IR LFs from the {\em Herschel} PACS$+${\em Spectral and Photometric Imaging Receiver} ({\em SPIRE}; \citealt{grif10}) data obtained within the PEP and HerMES extragalactic survey projects. 
In the present paper, we derive the rest-frame 35-$\mu$m, 60-$\mu$m, 90-$\mu$m and total IR (8--1000\,$\mu$m) LFs from a sample selected at PACS 70, 100 and 160\,$\mu$m wavelengths in the GOODS (GOODS-S and GOODS-N), Extended Chandra Deep Field South (ECDFS) and COSMOS areas. We use the full 70--500\,$\mu$m PACS$+$SPIRE data to determine $L_{\rm {IR}}$ and SED properties of
the PACS selected sources. In a related paper, Vaccari et al. (in prep.) derive rest-frame 100-, 160- and 250-$\mu$m and total IR LFs for a SPIRE selected sample. 
In addition, a third work aimed at studying the total IR LF based on the 24-$\mu$m selected sample, using all the PEP$+$HerMES data in the COSMOS field,
is ongoing (Le Floc'h et al., in preparation).

PEP is one of the major {\em Herschel} Guaranteed Time extragalactic key-projects, designed specifically to determine the cosmic evolution of dusty star-formation and of the IR luminosity function. It is structured as a ``wedding cake'', based on four different layers covering different areas to different depths at 100 and 160\,$\mu$m (in the GOODS-S field also at 70\,$\mu$m), from the large and shallow COSMOS field to the deep, pencil beam GOODS-S field. PEP includes the most popular and widely studied extragalactic fields with extensive multi-wavelength coverage available, in particular deep optical, near-IR and {\em Spitzer} imaging and spectroscopic and photometric redshifts: COSMOS; Lockman Hole; Extended Groth Streep (EGS); ECDFS; GOODS-N; and GOODS-S (see \citealt{ber10}, \citealt{ber11} and \citealt{lutz11} for a detailed description of the fields and observations). Coordinated observations 
of the PEP fields at 250, 350 and 500\,$\mu$m with SPIRE have been obtained by the HerMES Survey (\citealt{oliv12}). HerMES, analogously to PEP but extending to a much wider area, is a legacy programme designed to map a set of nested fields ($\sim$380 deg$^2$ in total) of different sizes and depths, using SPIRE (at 250, 350 and 500\,$\mu$m), and PACS (at 100 and 160\,$\mu$m, shallower than PEP), with the widest component of 270 deg$^2$ with SPIRE alone. In the fields covered by PEP, the two surveys are closely coordinated to provide an optimized sampling over wavelength.

In \citet{grup10} we started to determine the evolution with redshift of the galaxy and AGN LF in the far-IR domain by
exploiting the PEP data obtained in GOODS-N by the PEP Science Demonstration Programme (SDP).  
Here we extend the analysis to the wider and shallower fields -- COSMOS and ECDFS -- and to the deepest field -- GOODS-S -- observed by PEP, and we also take advantage of the HerMES sub-mm data in the same fields to derive improved SED classifications and accurate total IR luminosities for our sources. This allows us to have statistically significant samples of IR galaxies at different redshifts and over a broad range of luminosities, to make a detailed study of the LF at several $z$ intervals, all the way from $z$$=$0 to $z$$\simeq$4.
The measure of the total IR luminosity obtained by integrating the SEDs, well constrained over the entire mid- and far-IR domain (and also in the sub-mm thanks to the available SPIRE data), allows us to derive the total IR LF and its evolution directly from far-IR data for unbiased samples selected at wavelengths close to the peak of dust emission.
Moreover, the availability of deep multi-wavelength catalogues in the PEP fields is crucial for analysing the SEDs, obtaining k-corrections and total IR luminosities, and classifying the PEP sources into different IR populations, in order to separately study their LFs and evolutionary behaviour. This is the first study ever based on such a statistically wide and deep far-IR sample, to be able to provide LFs for different IR populations of galaxies and AGN. Here the evolution of the far- and total IR LFs (and luminosity density, hereafter $\rho_{\rm IR}$) are derived up to unprecedented high redshifts ($\sim$4) both globally (e.g. for all the populations together) and separately, for each SED class.

Despite the abundance of information available in the literature about the stellar mass function (MF; \citealt{font04}; \citealt{pozz10}; \citealt{ilb10}; \citealt{dom11}), very little
is known about the corresponding total IR LFs and star-formation rate (SFR) densities at different masses (an attempt based on {\em Spitzer} data was made by \citealt{perez05}).
From stellar MF studies one finds a clear increase with $z$ of the relative fraction of massive 
(log($M$/M$_{\odot}$$>$11) star-forming objects, starting to contribute significantly to the massive-end of the MFs at $z$$>$1 (\citealt{font04}; \citealt{ilb10}).  
Their evolution and contribution to the total SFR history is however still uncertain, since only few studies have tried to reconstruct the evolution of the SF history of massive objects 
from optical/near-IR or mid-IR surveys (\citealt{jun05}; \citealt{perez05}; \citealt{sant09}; \citealt{font12}) but none from far-IR selected surveys (providing a more direct indicator of the galaxy SF activity). 
In this work, we have derived the IR luminosity function and density in three different mass ranges (from log($M$/M$_{\odot}$)$=$8.5 to log($M$/M$_{\odot}$)$=$12),
extending previous studies (limited to $z$$=$1.8--2 for the most massive galaxies) to $z$$\sim$4.\\
Finally, our PEP data-sets have allowed us to quantify the relative contribution of the two main modes of star formation (a relatively steady one in disk-like galaxies, defining a tight SFR-stellar mass sequence, and a starburst mode in outliers) to the total IR LF and $\rho_{\rm IR}$ in three redshift intervals (0.8$<$$z$$<$1.25, 1.25$<$$z$$<$1.8 and 1.8$<$$z$$<$2.2) and to test the
SED-classes belonging to each mode. 

The paper is structured as follows. The PEP Survey with the far-IR and multi-wavelength data, together with the SED characterisation and redshift distribution of the PEP sources, are described in Sect.\,2. 
The LFs (rest-frame 35-$\mu$m, 60-$\mu$m, 90-$\mu$m and total IR), their evolution (derived for different SED-classes, mass and specific star-formation rate intervals) are discussed in Sect.\,3. In Sect.\,4 we present the number and IR luminosity densities of the different galaxy types, while in Sect.\,5 we discuss our results. In Sect.\,6 we present our conclusions.

\noindent Throughout this paper, we use a Chabrier initial mass function (IMF) and we assume a $\Lambda$CDM cosmology with $H_{\rm 0}$\,=\,71~km~s$^{-1}$\,Mpc$^{-1}$, $\Omega_{\rm m}$\,=\,0.27, and $\Omega_{\rm \Lambda}\,=\,0.73$. 

\section{The Data}
\label{sec_data}
The PEP fields where we computed the LFs are: COSMOS, 2 deg$^2$ observed down to 3$\sigma$ depths of $\sim$5 mJy and 10.2 mJy at 100\,$\mu$m and 160\,$\mu$m, respectively; ECDFS, $\sim$700 arcmin$^2$ down to 3$\sigma$ depths of $\sim$4.5 mJy and 8.5 mJy at 100\,$\mu$m and 160\,$\mu$m, respectively; GOODS-N, $\sim$300 arcmin$^2$ to 3 and 5.7 mJy at 100\,$\mu$m and 160\,$\mu$m, respectively; and GOODS-S, $\sim$300 arcmin$^2$ to 1.2 mJy, 1.2 mJy and 2.4 mJy at 70\,$\mu$m, 100\,$\mu$m and 160\,$\mu$m, respectively.
Our reference samples are the blind catalogues at 70 (in GOODS-S only), 100 and 160\,$\mu$m to the $3\sigma$ level, which contain 373 (all in GOODS-S), 7176 (GOODS-S: 717, GOODS-N: 291, ECDFS: 813, COSMOS: 5355) and 7376 (GOODS-S: 867, GOODS-N: 316, ECDFS: 688, COSMOS: 5105) sources at 70, 100 and 160\,$\mu$m, respectively. 
We refer to \citet{ber10} and \citet{ber11} for a detailed description of the data catalogues and source counts. 

\subsection{Multi-wavelength Identification}
\label{sec_id}
The PEP fields benefit from an extensive multi-wavelength coverage. We have therefore associated our sources to the ancillary catalogues by means of a multi-band likelihood ratio technique (\citealt{suth92}; \citealt{cil01}), starting from the longest available wavelength (160\,$\mu$m, PACS) and progressively matching 100\,$\mu$m (PACS), 70\,$\mu$m (PACS, GOODS-S only) and 24\,$\mu$m ({\em Spitzer}/MIPS). In the GOODS-S field, we have associated to our PEP sources the 24-$\mu$m catalogue by \citet{mag09}, that we have matched with the optical$+$near-IR$+$IRAC MUSIC catalogue of \citet{gra06}, revised by \citet{sant09}, which includes spectroscopic and photometric redshifts. To maximise the fraction of identifications, we limited our study to the area covered by the MUSIC catalogue ($\sim$196 arcmin$^2$), obtaining 233, 468 and 492 sources at 70, 100 and 160\,$\mu$m, respectively, with flux density greater than the flux limits reported above (all with either spectroscopic or photometric redshifts).
In the GOODS-N field, as described in \citet{ber10}, \citet{ber11} and \citet{grup10}, a PSF-matched multi-wavelength catalogue\footnote{publicly available at \\
\tt{http://www.mpe.mpg.de/ir/Research/PEP/public$\_$data$\_$releases.php}} was created, including photometry from the far-UV ({\em GALEX}) to the mid-IR ({\em Spitzer}). As in GOODS-S, to maximise the identifications, we limited our study in GOODS-N to the area covered by the ACS ($\sim$150 arcmin$^2$), obtaining 176 and 186 sources with flux density greater than the flux limit at 100 and 160\,$\mu$m, respectively (all with redshifts). We have matched our sources in the ECDFS with the multi-wavelength Survey by Yale-Chile (MUSYC) by \citet{card10}, obtaining 687 sources at 100\,$\mu$m and 625 sources at 160\,$\mu$m (578 and 547 with redshifts, $\sim$45\% spectroscopic).
Finally, in COSMOS, we have matched our catalogue with the deep 24-$\mu$m sample of \citet{lef09} and with the IRAC-based catalogue of \citet{ilb10}, including optical and near-IR photometry and photometric redshifts. After the removal of PEP sources within flagged areas of the optical and/or IRAC COSMOS catalogues, we ended up with two catalogues consisting of 4110 and 4118 sources, with flux densities $\geq$5.0 and $\geq$10.2 mJy at 100 and 160\,$\mu$m respectively (3817 and 3849 with either spectroscopic or photometric redshifts).
Throughout this paper and specifically for the SED fits described in Section~\ref{sec_class}, we adopt these spectroscopic or rest-frame UV to near-IR photometric redshifts for the various fields.
\begin{figure*}
\includegraphics[width=17cm]{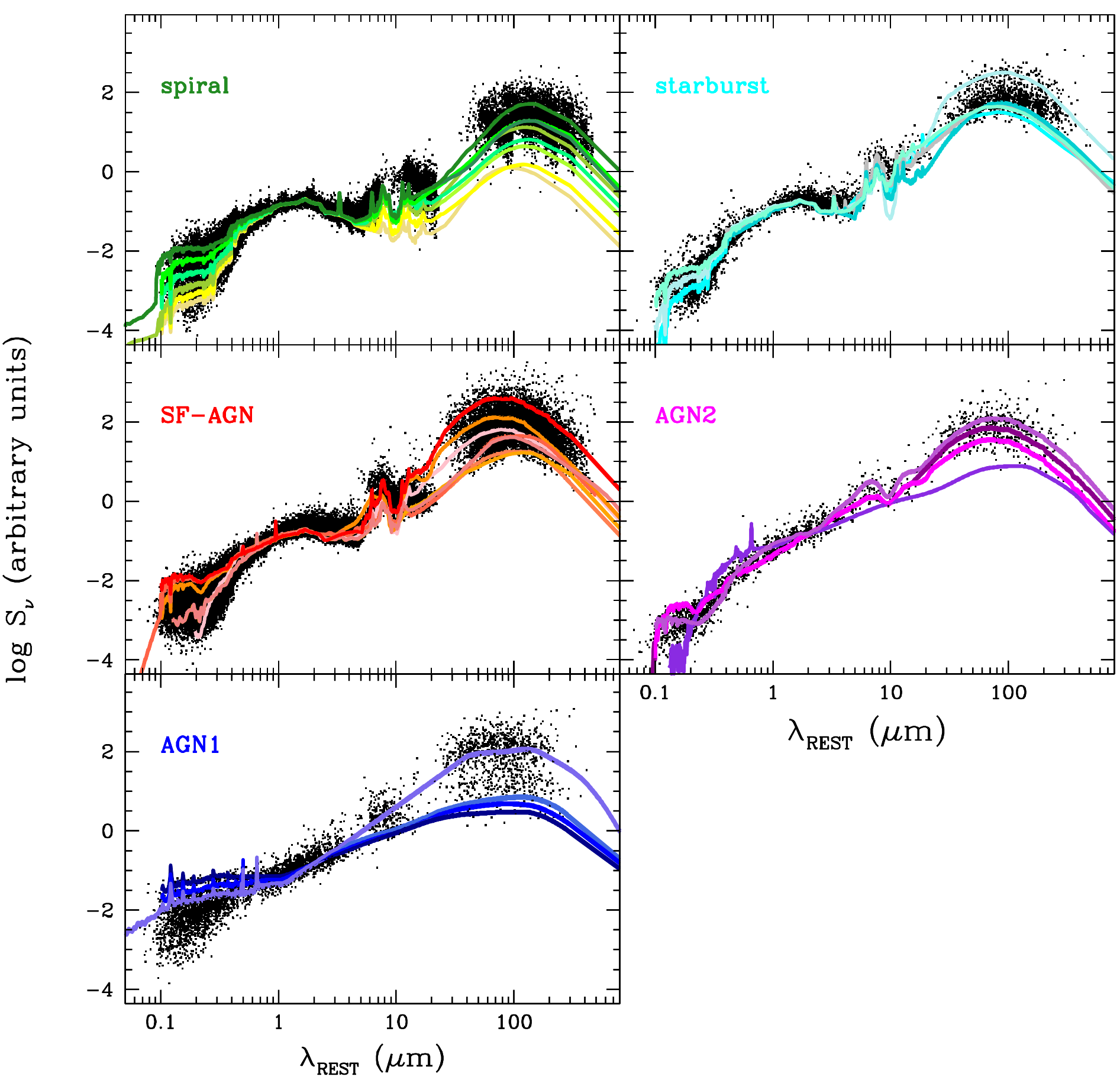}
\caption{Observed rest-frame SEDs of the PEP sources (black dots) divided by population (as shown in the plot) and normalized to the $K_{\rm s}$-band. The more representative templates for each SED-class have been overplotted in different colours.}
\label{fig_medsed}
\end{figure*}

The HerMES extragalactic survey (\citealt{oliv12}) performed coordinated observations with SPIRE at 250, 350 and 500\,$\mu$m in the same fields covered by PEP. In particular, in HerMES a prior source extraction was performed using the method presented in \citet{rose10}, based on MIPS-24\,$\mu$m positions. The 24-$\mu$m sources used as priors for SPIRE source extraction are the same as those associated with our PEP sources through the likelihood ratio technique.
We have therefore associated the HerMES sources with the PEP sources by means of the 24-$\mu$m sources matched to both samples. For most of our PEP sources ($\sim$87 per cent) we found a $>$3$\sigma$ SPIRE counterpart in the HerMES catalogues.  

\subsection{Galaxy Classification}
\label{sec_class}
We made use of all the available multi-wavelength data to derive the SEDs of our PEP sources, which we interpreted and classified by performing a $\chi^2$ fit (using the {\tt Le Phare} code\footnote{available at \\
\tt{http://www.cfht.hawaii.edu/$\sim$arnouts/LEPHARE/lephare.html}}; \citealt{arn02} and \citealt{ilb06}) with the semi-empirical template library of \citet{pol07}, representative of different classes of IR galaxies and AGN. To this library we added some templates modified in their far-IR part to better reproduce the observed {\em Herschel} data (see \citealt{grup10}), and three starburst templates from \citet{riek09}.
If required to improve the fit, different extinction values ($E_{\rm (B-V)}$ from 0.0 to 0.5) have been applied to the templates, by letting the code free to choose the most suitable extinction curve. 
The considered set of templates included SEDs of elliptical galaxies of different ages, lenticular, spirals (from Sa to Sdm), starburst galaxies (SB), type 1 QSOs, type 2 QSOs, Seyferts, LINERs and composite ULIRGs (containing both starburst and obscured AGN component), in the wavelength range between 0.1 and 1000\,$\mu$m. The latter templates, are empirical ones created to reproduce the SEDs of the heavily obscured AGN.
Two of these SEDs (the broad absorption-line QSO Markarian~231 \citep{ber05} and the Seyfert 2 galaxy IRAS~19254$-$7245 South \citep{ber03}) are similar in shape, containing a
powerful starburst component, mainly responsible for their far-IR emission, and an AGN component that contributes to -- and dominates -- the mid-IR (Farrah et al. 2003), reproducing the SEDs 
of ``obscured'' AGN regardless of their optical spectra (i.e. broad or narrow lines in the optical; \citealt{grup08}). 
Hereafter, we will refer to this class of templates and to the sources reproduced by them as to type 2 AGN ({\tt AGN2}).
Three other empirical templates, reproducing the observed SEDs of nearby ULIRGs containing an obscured AGN (i.e. IRAS 20551-4250; IRAS 22491-1808; NGC 6240) have been associated to the 
Seyfert 1.8/2, LINER ones, since they all contain an AGN, but this AGN does not dominate the observed energetic output at any wavelength (from UV to far-IR/sub-mm), showing up just in the range where the host galaxy SED has a minimum (i.e. the mid-IR). The AGN in these objects is either obscured or of low luminosity. We refer to this class as to star-forming galaxies containing an AGN ({\tt SF-AGN}), since their IR luminosity is largely dominated by star-formation.\\
\begin{table*}
 \caption{SED Classification of the PEP Sources}
\begin{tabular}{|l|c|c|c|c|c|c|c|}
\hline \hline
 field & {\tt spiral} & {\tt starburst} & {\tt SF-AGN} & {\tt AGN2} & {\tt AGN1} & {\tt SF-AGN(SB)} & {\tt SF-AGN(Spiral)} \\ 
\hline \hline 
 GOODS-S           &  N  & N & N & N  & N & N & N \\   \hline
  70\,$\mu$m &  53 & 22 & 142 & 5 & 12  & 26 & 116 \\
                       & (23\%) & (9\%) & (61\%) & (2\%) & (5\%) & &  \\
 100\,$\mu$m & 117   & 60 & 250 & 10 & 31 & 54 & 96 \\
                        & (25\%) & (13\%) & (53\%) & (2\%) & (7\%) & &  \\
 160\,$\mu$m &  123 & 55 & 277 & 11 & 26  & 73 & 204 \\ 
                         & (25\%) & (11\%) & (56\%) & (2\%) & (6\%) & &  \\ \hline
 GOODS-N &   N  & N & N & N  & N & N & N\\    \hline
 100\,$\mu$m &  68  & 20 & 78 & 7 &  3  & 21 & 57 \\
                        & (39\%) & (11\%) & (44\%) & (4\%) & (2\%) &  & \\
 160\,$\mu$m &  67 & 21 & 85 & 10 & 3  & 21 & 64 \\ 
                         & (36\%) & (11\%) & (46\%) & (5\%) & (2\%) &  &  \\ \hline
 ECDFS      &   N  & N & N & N  & N  & N & N\\   \hline
 100\,$\mu$m &  253  & 49 & 245  & 8 &  23 & 83 & 162 \\
                        & (44\%) & (9\%) & (42\%) &  (1\%) & (4\%) & &  \\
 160\,$\mu$m &  233 &  49&  231& 12 &  22 & 99 & 132 \\ 
                        & (43\%) & (9\%) & (42\%) &  (2\%) & (4\%) &  & \\ \hline
 COSMOS   &   N  & N & N & N  & N & N & N\\   \hline
 100\,$\mu$m &  1637  & 232 &  1689 & 76 &  183 & 580 & 1109 \\
                        & (43\%) & (6\%) & (44\%) &  (2\%) & (5\%) &  & \\ 
 160\,$\mu$m &  1483 &  243 & 1847 &  103&  173 & 777 & 1070 \\
                        & (39\%) & (6\%) & (48\%) &  (3\%) & (4\%) &  & \\ 
\hline
 TOTAL  &    N  & N & N & N  & N & N & N \\   \hline  
100\,$\mu$m & 2075   & 361 & 2262 & 101 & 240 & 738 & 1424 \\
                     & (41\%)&(7\%)&(45\%)&(2\%)&(5\%)&&\\
 160\,$\mu$m &  1906 &  368 & 2440 & 136 & 224 & 970 & 1470 \\   
                     & (38\%)&(7\%)&(48\%)&(3\%)&(4\%)&&\\
 \hline \hline
\end{tabular}
\label{tab_class}
\end{table*}
In our analysis, we make the basic assumption that the SED shapes seen at low redshifts are also able to represent the higher redshift objects. In any case, to further
increase the range of SEDs in the fit, we have applied additional extinction with different extinction curves to our templates. All SED fits adopt fixed spectroscopic
or photometric redshifts described in Section~\ref{sec_id}.\\
The template library used to fit our data contains a finite number of SEDs (38), representative of given classes of local infrared
objects, which do not vary with continuity from one class to another (there are large gaps in the parameter space). Therefore,
the quality of the fit depends not only on the photometric errors, but also on the template SED uncertainties. For this reason,
in our fitting procedure, in addition to the photometric errors on data, we need to take into account also the uncertainties due to
the template SEDs discretisation and additional extinction. To do this, we have proceeded as described in detail by \citet{grup08}
and summarised as follows. 
First, we have run {\tt Le Phare} on our PEP SEDs considering the nominal errors from catalogues, computing the distributions of the $(S_{\rm object}-S_{\rm template})_{\rm band}/(\sigma_{\rm object})_{\rm band}$ values in each of the considered photometric band (where $S_{\rm object}$ and $\sigma_{\rm object}$ are the flux density and the
relative error of the source, and $S_{\rm template}$ the flux density of the template in
the considered band), iteratively increasing the photometric errors
until we have obtained a Gaussian distribution with $\sigma$$\sim$1. This
corresponds to reduced $\chi^2$ distributions peaked around 1 (as expected
in the case of good fit). With the new photometric uncertainties (on average, significantly increased mainly in the optical/near-IR and SPIRE bands), 
we have run {\tt Le Phare} on our sources for the second time, obtaining what we have taken as the final SED-fitting results.

The majority of our PEP sources are reproduced by templates of normal spiral galaxies ({\tt spiral}), SB galaxies ({\tt starburst}), and Seyfert2/1.8/LINERS/ULIRGs$+$AGN ({\tt SF-AGN}), although different classes prevail at different redshifts and luminosities. The {\tt spiral} SEDs show no clear signs of enhanced SF or nuclear activity (see Fig.\,\ref{fig_medsed}), the far-IR bump being characterised by relatively cold dust ($T_{\rm dust}$$\sim$20 K). On the other hand, SB templates are characterised by warmer ($T_{\rm dust}$$\sim$40--45 K), more pronounced far-IR bumps and significant UV extinction, indicative of intense star-formation activity. Templates of star-forming galaxies containing either a low-luminosity or obscured AGN ({\tt SF-AGN}) are characterised by a ``flattening'' in the 3--10\,$\mu$m spectrum (suggesting detection of an AGN in the wavelength range where the host galaxy SED has a minimum) and a far-IR bump dominated by star-formation, which is intermediate (in terms of both energy and $T_{\rm dust}$) between spirals and SBs. Although they can be considered as star-forming galaxies at the wavelengths relevant to this work, we prefer to refer to them as {\tt SF-AGN} throughout the paper, to keep in mind that they probably contain an AGN, whose presence, though not dominant in the far-IR, might be very important for analysis in other bands (e.g., in the X-rays or the mid-IR). \\
We note that the well-studied high-z ($\simeq$2.3) SED of the strongly lensed sub-mm galaxy SMM J2135-0102, known as ``the Cosmic Eyelash'' (\citealt{ivi10}; \citealt{swin10}), best-fitted with
our procedure by an extincted (E(B-V)$\sim$0.2) IRAS 22491-1808 template (though rather poorly in the near-IR), does not represent the bulk of our population at high-$z$ ($>$1.5), whose SEDs are indeed well reproduced by our library of templates. 

\begin{figure*}
\includegraphics[width=14.5cm,height=11.cm]{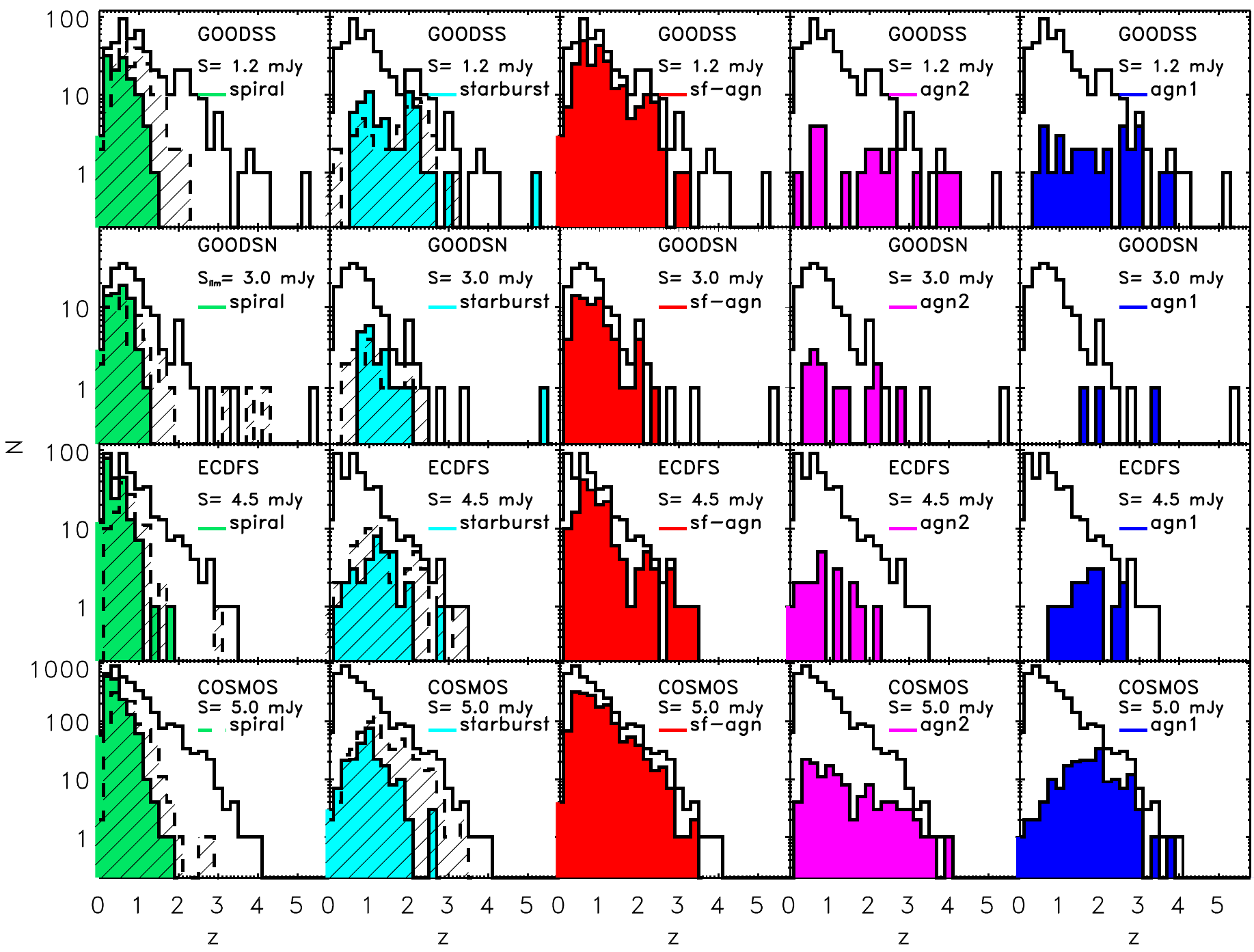}
\includegraphics[width=14.5cm,height=11.cm]{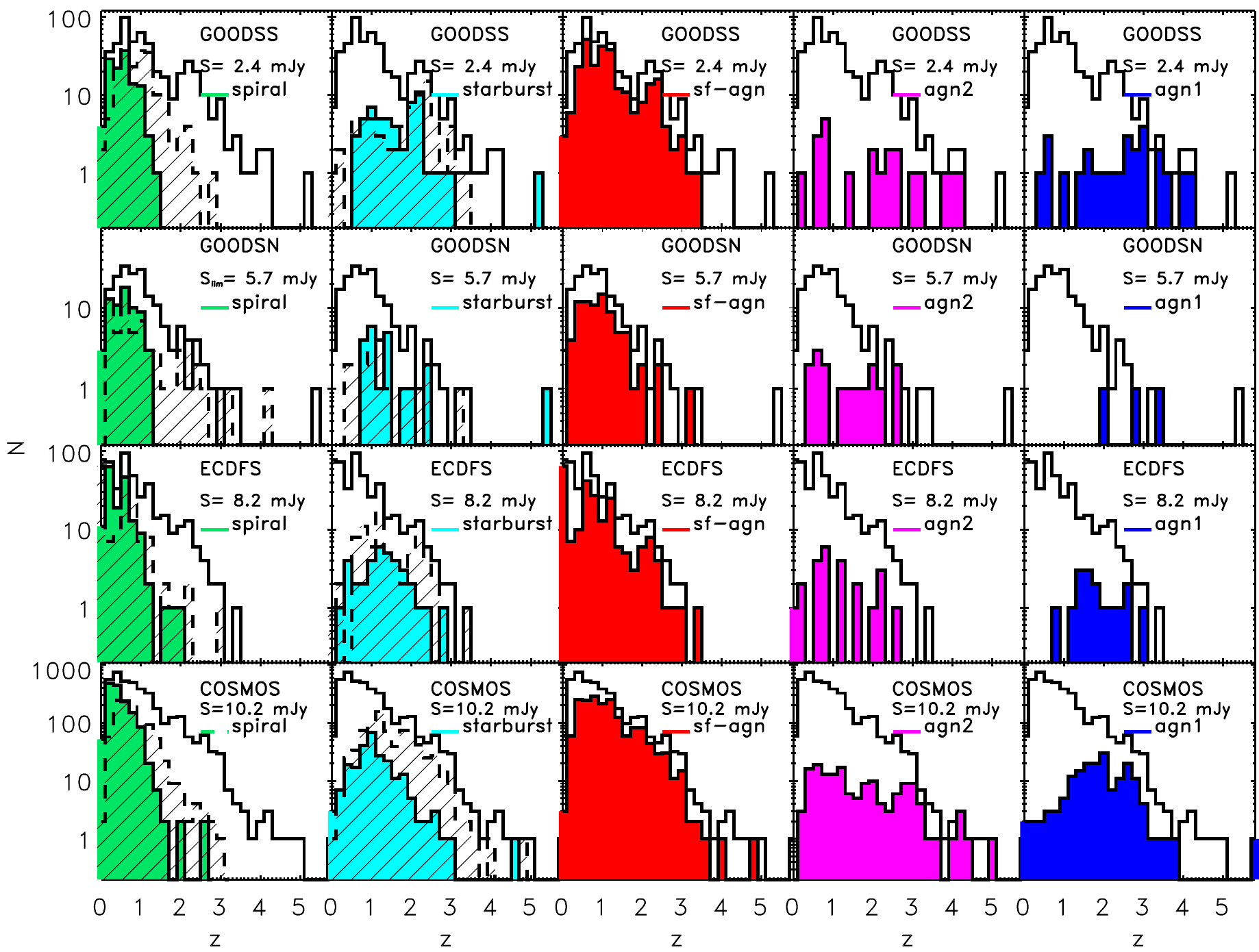}
\caption{Redshift distributions of 100-$\mu$m ($top$) and 160-$\mu$m ($bottom$) sources in the four PEP fields (first row from top, GOODS-S; second row, GOODS-N; third row, ECDFS; bottom row, COSMOS) to different limiting fluxes (as shown in the plot). The redshift distributions of the five different populations have been plotted in different colours (green, {\tt spiral}; cyan, {\tt starburst}; red, {\tt SF-AGN}; magenta, {\tt AGN2}; blue, {\tt AGN1}) and compared to the total distribution (black solid histogram). The line-filled dashed histograms shown in the {\tt spiral} and {\tt starburst} panels represent the redshift distributions of the 
{\tt SF-AGN$\_$Spiral} and {\tt SF-AGN$\_$SB} sub-classes, respectively.}
\label{fig_zdist}
\end{figure*}

In fact, the considered template set provides very good fits to the SEDs of our PEP sources. In Fig. \ref{fig_medsed} we show the rest-frame SEDs (black dots) of the PEP sources belonging to the different ``broad'' SED classes ({\tt spiral}, {\tt starburst}, {\tt SF-AGN}, {\tt AGN1} and {\tt AGN2}), compared to the template SEDs of those classes  normalised to the $K_{\rm s}$-band flux density. In Table~\ref{tab_class} we report the fraction of sources belonging to each SED class: we find that in all the fields $\sim$41(38) per cent of the 100(160)-$\mu$m sources are reproduced by a {\tt spiral} template SED, 7(7) per cent with a {\tt starburst} template SED, 45(48) per cent with a {\tt SF-AGN} template SED, 2(3) per cent with an {\tt AGN2} SED and 5(4) per cent with an {\tt AGN1} SED. We note that the fraction of {\tt SF-AGN}
derived in this work is in agreement with results from mid-IR spectroscopy (with {\em Spitzer-IRS}) of local star-forming
galaxies from the SINGS sample by \citet{smith07}, who found that $\sim$50 per cent of local galaxies (though of lower luminosities than ours) do harbour
low-luminosity AGN (of LINER or Seyfert types). Recently, \citet{saji12} found an even higher fraction ($\sim$70 per cent) of objects hosting an AGN in the mid-IR ({\em Spitzer} 24-$\mu$m)
selected samples ($\sim$23 per cent AGN-dominated and $\sim$47 per cent showing both AGN and starburst activity).
However, since the far-IR SED of the {\tt SF-AGN} is dominated by star-formation and at these wavelengths resembles either {\tt starburst} or {\tt spiral} galaxy templates, 
we have also divided the {\tt SF-AGN} class into {\tt SF-AGN(SB)} and {\tt SF-AGN(Spiral)} sub-classes, based of their far-IR/near-IR colours (e.g. $S_{\rm 100}$/$S_{\rm 1.6}$) and SED resemblance (apart from the rest-frame mid-IR flattening, which is detected in all of the {\tt SF-AGN} SEDs). 
Specifically, galaxies best-fitted by the Seyfert2/1.8 templates (either
the original ones from \citealt{pol07} or those modified by \citealt{grup10}) have been classified as {\tt SF-AGN(Spiral)}, while 
galaxies best-fitted by the NGC 6240, IRAS 20551-4250 or IRAS 22491-1808 templates have been classified as {\tt SF-AGN(SB)}.
The number of sources belonging to the former and the latter sub-classes 
are also reported in Table~\ref{tab_class} as additional information.  

\subsection{Redshift Distribution}
\label{sec_zdistr}

A large number of spectroscopic redshifts have been measured in the GOODS, ECDFS and COSMOS regions. In the GOODS-S and ECDFS area a collection of more than 5000 spectroscopic redshifts are available (\citealt{cris00}; \citealt{croo01}; \citealt{bun03}; \citealt{dick04}; \citealt{stan04}; \citealt{strol04}; \citealt{szok04}; \citealt{vanderw04}; \citealt{doh05}; \citealt{lefe05}; \citealt{mig05}; \citealt{van08}; \citealt{pop09}; \citealt{sant09}; \citealt{bal10}; \citealt{coop12}). In the GOODS-N area more than 2000 spectroscopic redshifts come from various observations (\citealt{coh00}; \citealt{wirth04}; \citealt{cow04}; \citealt{barg08}). Finally, in COSMOS we could use a collection of $\sim$3000 spectroscopic redshifts from either the public zCOSMOS bright database or the non-public zCOSMOS deep database (\citealt{lil07}; \citealt{lil09}).
For the PEP sources without spectroscopic redshift available, we have adopted the photometric redshifts derived from multi-wavelength (UV to near-IR) photometry by different authors in the different fields, as mentioned in Section~\ref{sec_id}.
In the GOODS-S field the MUSIC photometric redshift catalogue (\citealt{gra06}; \citealt{sant09}) provided photo-$z$s for most of our PEP sources without spectroscopic data, while in the GOODS-N field, photo-$z$s were obtained by \citet{ber10} for almost all the PEP sources within the ACS area. The \citet{card10} and \citet{ilb09} catalogues provided photometric redshifts for a large fraction of the PEP sources in the ECDFS and COSMOS areas, respectively.
When considering both the spectroscopic and photometric redshifts, in our PEP fields the redshift incompleteness is very low.
In particular, in the GOODS-S field we have either a spec-$z$ or a photo-$z$ for $\sim$100 per cent of the PEP sample within the MUSIC areas ($\sim$80 per cent spectroscopic, though most of them lie at $z$$<$2.5; see \citealt{ber11}).
In the GOODS-N field we have a redshift completeness of $\sim$100 per cent of sources (70 per cent spectroscopic) within the ACS area. In the ECDFS and COSMOS fields we have a redshift completeness of 88 per cent and 93 per cent respectively (45 per cent and 40 per cent spectroscopic). 

The uncertainty in the photometric redshifts has been evaluated by means of a comparison with the available spec-$z$s by the different authors providing photo-$z$ catalogues in the PEP fields. 
In particular, \citet{ber11} have compared the photometric and the available spectroscopic redshifts in GOODS-S, GOODS-N and COSMOS, finding a fraction of outliers, defined as objects having $\Delta z$/(1$+$$z_{\rm spec}$)$>$0.2, of $\sim$2 per cent for sources with a PACS detection. 
Most of these outliers are sources with few photometric points available, or SEDs not well reproduced by the available templates. 
The median absolute deviation of the $\Delta z$/(1$+$$z_{\rm spec}$) distribution in the three fields analised by \citet{ber11} is 0.04 for the whole catalogue, and 0.038 for PACS-detected objects.
In GOODS-S, \citet{gra06} found an excellent agreement between photometric and spectroscopic redshifts over the fully accessible redshift range 0$<$$z$$<$6 ($\sigma[\Delta z/(1+z)]$$\simeq$0.045), with a very limited number of catastrophic errors.  
In COSMOS, \citet{ilb10} estimated the photometric redshift uncertainties of their 3.6-$\mu$m catalogue matched with the COSMOS photo-$z$ multi-wavelength catalogue of \citet{ilb09},
finding $\sigma[\Delta z/(1+z)]$$=$0.008 (and $<$1 per cent of catastrophic failures) at $i^+_{\rm AB}$$<$22.5, $\sigma[\Delta z/(1+z)]$$=$0.011 
at 22.5$<$$i^+_{\rm AB}$$<$24	and $\sigma[\Delta z/(1+z)]$$=$0.053 at 24$<$$i^+_{\rm AB}$$<$25.
In the ECDFS, by comparing non-X-ray sources with high-quality spectroscopic redshifts, \citealt{card10} found $\sigma[\Delta z/(1+z)]$$=$0.008 to $z$$\sim$1.2 $=$0.027
at 1.2$\leq$$z$$\leq$3.7 and $=$0.016 at $z$$>$3.7.
Note that we have checked all the $z$$>$2.5 photometric redshifts through {\tt Le Phare}, assigning the {\tt Le Phare} derived value in case of significant disagreement with that from the catalogue (though most resulted in very good agreement).
The fractions of spectroscopic redshifts in the 2.5$<$$z$$<$3.0 interval amount to just $\sim$6\% and $\sim$25\% in COSMOS and GOODSS respectively, dropping to $\sim$4\% and  $\sim$6\% at 3.0$<$$z$$<$4.2.
We note that from our comparison between photo- and spec-$z$ (when available) we find a general good agreement in all fields.

Photometric redshift errors may, in principle, affect the shape of the luminosity function at the bright end: by scattering objects to higher redshifts they make the steep fall-off at high luminosities appear shallower (e.g. \citealt{drory03}). 
To study the impact of redshift uncertainties on the inferred infrared LF, we have performed Monte Carlo simulations, as discussed in detail in Section~\ref{sec_totLF}.
It is indeed very difficult to estimate the effect of catastrophic failures at $z$$>$2.5, where mainly photometric redshifts are available and very little reliable spectroscopic data can be used to validate them.
Moreover, for the limited high-$z$ samples with spectroscopic information, different results are found in the different fields: i.e., in GOODS-S \citet{ber11} found about 25 per cent of catastrophic failures for PACS detected sources above $z$$\sim$2, with the tendency to have a higher than real photometric redshift, while
in COSMOS the catastrophic failures (20 per cent) found by \citet{ilb09} for MIPS selected sources at 2$<$$z$$<$3 were mostly for photo-$z$'s smaller than spectroscopic ones. 
In addition to that, sometimes high-$z$ spectroscopic redshifts can be even more uncertain than photometric ones and great care must be taken when selecting spec-$z$s for comparison (i.e. we need to choose those with  high quality flags). 
For all these reasons, we limited our analysis of photo-$z$ uncertainties to the Monte Carlo simulations described in Section~\ref{sec_totLF}, without trying to derive uncertainties also due to catastrophic failures.
In Section~\ref{sec_rflf} we also note that the different (far-IR) photo-z approach of \citet{lapi11} produces consistent LF results in the common part of parameter space.

The median redshift of the 70-$\mu$m sample in GOODS-S is $z$$_{\rm med}$(70)$=$0.67 (the mean is $\langle$$z$$\rangle_{70}$$=$0.86), while those of the
100- and 160-$\mu$m samples are different in each of the fields, given the different flux density depths reached by PEP in each area. In Table~\ref{tabzmed} we report the median and the mean redshifts found for the different fields (and for the combined sample) at the different selection wavelengths. As expected, GOODS-S reaches the highest redshifts, while the surveys in the COSMOS and ECDFS fields are shallower and sample lower redshifts.
On average, the 160-$\mu$m selection favours higher redshifts than the 100-$\mu$m one (see also \citealt{ber11}). 
\begin{table}
 \caption{Average redshifts in the PEP Survey fields}
\begin{tabular}{|l|c|c|c|c|c|c|}
\hline \hline
 field &  \multicolumn{2}{c}{70\,$\mu$m} & \multicolumn{2}{c}{100\,$\mu$m} & \multicolumn{2}{c}{160$\mu$m}   \\ 
        &   $z$$_{\rm med}$  &  $\langle$$z$$\rangle$  &  $z$$_{\rm med}$  &  $\langle$$z$$\rangle$  & $z$$_{\rm med}$  &  $\langle$$z$$\rangle$   \\   \hline \hline 
 GOODS-S &  0.67 & 0.86 & 0.85 & 1.07 & 0.98 & 1.16 \\
 GOODS-N &          &          & 0.73 & 0.84 & 0.84 & 0.94 \\
 ECDFS      &          &          & 0.66 & 0.76 & 0.69 & 0.85 \\
 COSMOS   &          &          & 0.59 & 0.74 & 0.70 & 0.88 \\ 
 \hline
 ALL FIELDS &  0.67 & 0.86 & 0.64 & 0.78 & 0.73 & 0.91 \\ 
 \hline \hline
\end{tabular}
\label{tabzmed}
\end{table}

In Fig.~\ref{fig_zdist} we show the redshift distribution of the PEP sources selected at 100\,$\mu$m and 160\,$\mu$m in the four different fields. The black solid histogram is the total redshift distribution in the field (one for each row of the plot), while the filled histograms in different colours represent the redshift distributions of the different populations (green, {\tt spiral}; cyan, {\tt starburst}; red, {\tt SF-AGN}; magenta, {\tt AGN2}; blue, {\tt AGN1}). The line-filled dashed histograms shown in the {\tt spiral} and {\tt starburst} panels represent the redshift distributions of the {\tt SF-AGN(Spiral)} and {\tt SF-AGN(SB)} sub-classes, respectively.
In addition to the principal redshift peak, in GOODS-S a secondary peak centred at $z$$\sim$2 is clearly visible at both 100 and 160\,$\mu$m. A similar result has been shown and discussed also by \citet{ber11}, while an extensive analysis of PACS GOODS-S large-scale structure at $z$$=$2--3 and of a $z$$\sim$2.2 filamentary overdensity have been presented by \citet{maglio11}.

\section{The Luminosity Function}
\label{sec_lf}
The sizes and depths of the PEP samples are such as to allow a direct and accurate determination of the far-IR LF in several redshift bins, from $z$$\simeq$0 up to $z$$\sim$4. PEP$+$HerMES is the unique {\em Herschel} survey to allow such analysis over such a wide redshift and luminosity range, sampling both the faint and bright ends of the far- and total IR LFs with sufficient statistics.
Because of the redshift range covered by PEP, we would need to make significant extrapolations in wavelength when computing the rest-frame LFs at any chosen wavelengths. 
In order to apply the smallest extrapolations for the majority of our sources, we choose to derive the far-IR LFs at the rest-frame wavelengths corresponding to the median redshift of each sample. 
Given the median redshift of the 70-$\mu$m sample in GOODS-S ($\sim$0.67, see Table~\ref{tabzmed}), we use that sample to derive the rest-frame luminosity function at 35\,$\mu$m. 
With the 100- and 160-$\mu$m PEP samples (whose median redshifts are $\sim$0.64 and 0.73 respectively), we derived the rest-frame LFs at 60 and 90\,$\mu$m.
Note that, given the excellent multi-wavelength coverage available for most of our sources (thanks also to the HerMES data available in all the PEP fields and providing reliable counterparts for most of our PEP sources), their SEDs are very well determined from the UV to the sub-mm. 
The extrapolations are therefore well constrained by accurately defined SEDs, even at high redshifts (i.e. at $z$$\sim$3.5 the rest-frame 90-$\mu$m luminosity corresponds to $\lambda_{\rm observed}$$\sim$400\,$\mu$m, which is still in the range covered by HerMES). 
\begin{figure*}
\includegraphics[width=14cm]{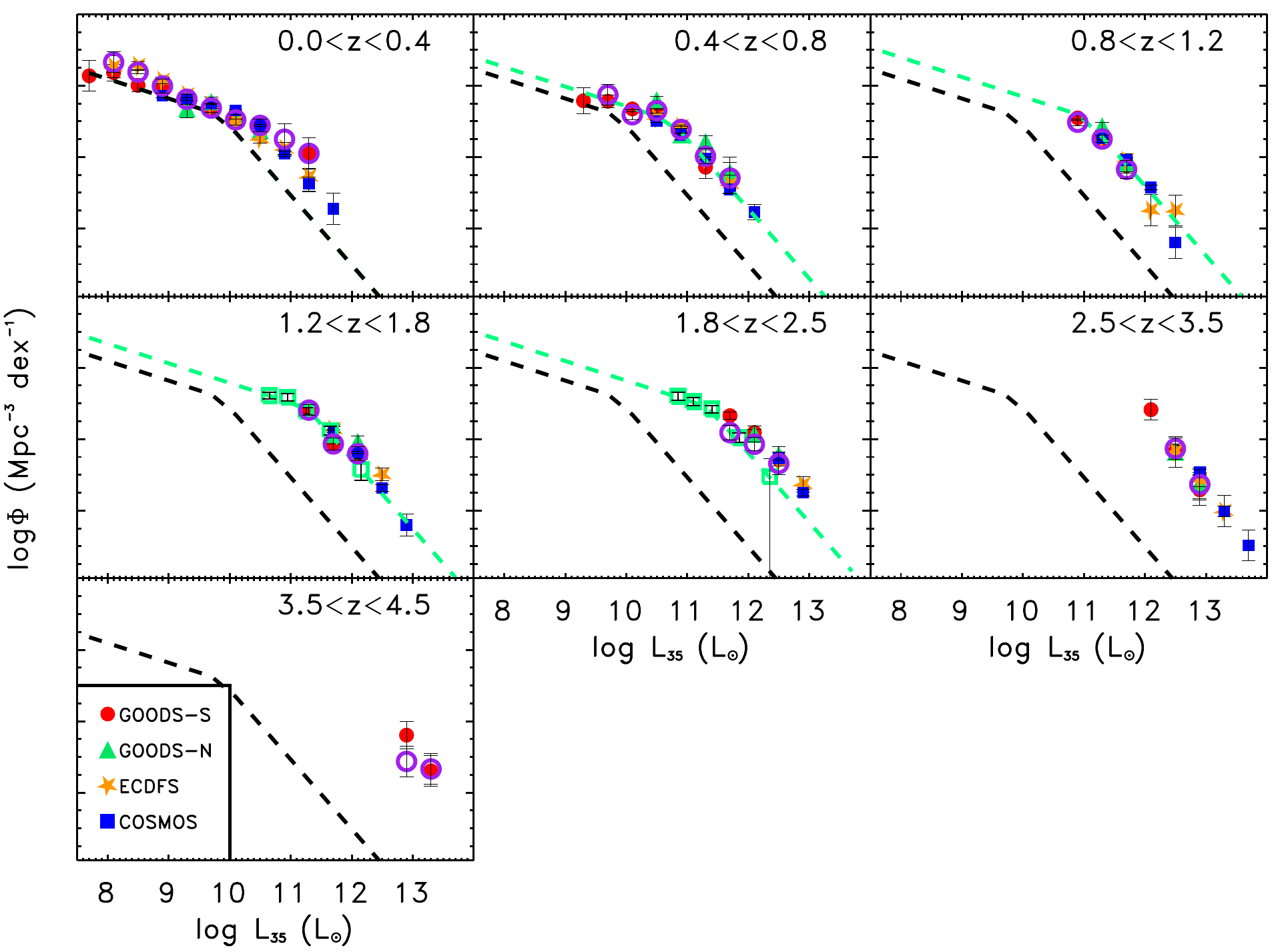}
\caption{Rest-frame 35\,$\mu$m Luminosity Function estimated with the 1/V$_{\rm max}$ method from the PEP 70-$\mu$m sample in the GOODS-S field (purple open circles) and independently in the four PEP fields from the 100-$\mu$m selected samples (red filled circles, GOODS-S; green filled triangles, GOODS-N; orange filled stars, ECDFS; blue filled squares, COSMOS). 
The error-bars in the data points represent the Poissonian uncertainties. For comparison, we also plot the determination of \citet{mag11} at 1.2$<$$z$$<$2.5, shown as green open squares, and the double power-law fit of \citet{mag09} and \citet{mag11}, shown as a green dashed line. The black dashed line is the $z$$=$0 determination of \citet{mag09}.}
\label{figLF35um}
\end{figure*}
\begin{figure*}
\includegraphics[width=15cm]{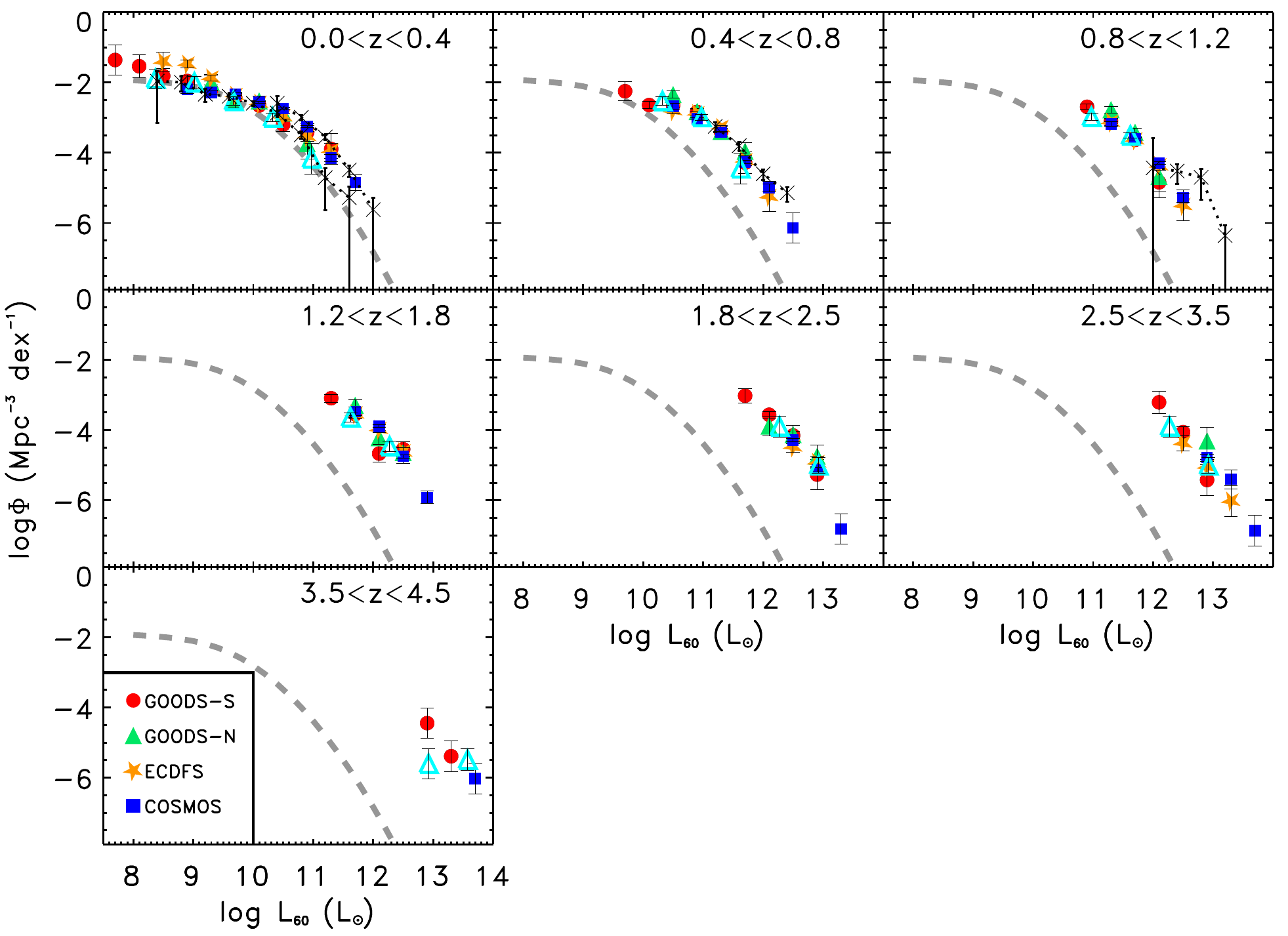}
\caption{Rest-frame 60\,$\mu$m Luminosity Function estimated independently from the 100-$\mu$m selected samples in the four PEP fields (red filled circles, GOODS-S; green filled triangles, GOODS-N; orange filled stars, ECDFS; blue filled squares, COSMOS). 
The diagonal crosses connected by the dotted line are the {\em Spitzer} 70-$\mu$m derivations of \citet{pat12} (in the lower $z$-bin we report their LFs derived both at 0.0$<$$z$$<$0.2 and at 0.2$<$$z$$<$0.4). 
The cyan open triangles show the PEP 60-$\mu$m LF from the SDP data in GOODS-N by \citet{grup10} (the redshift bins are not exactly the same). The grey dashed line is the \citet{saun90} local LF.}
\label{figLF60um}
\end{figure*}

\begin{figure*}
\includegraphics[width=15cm]{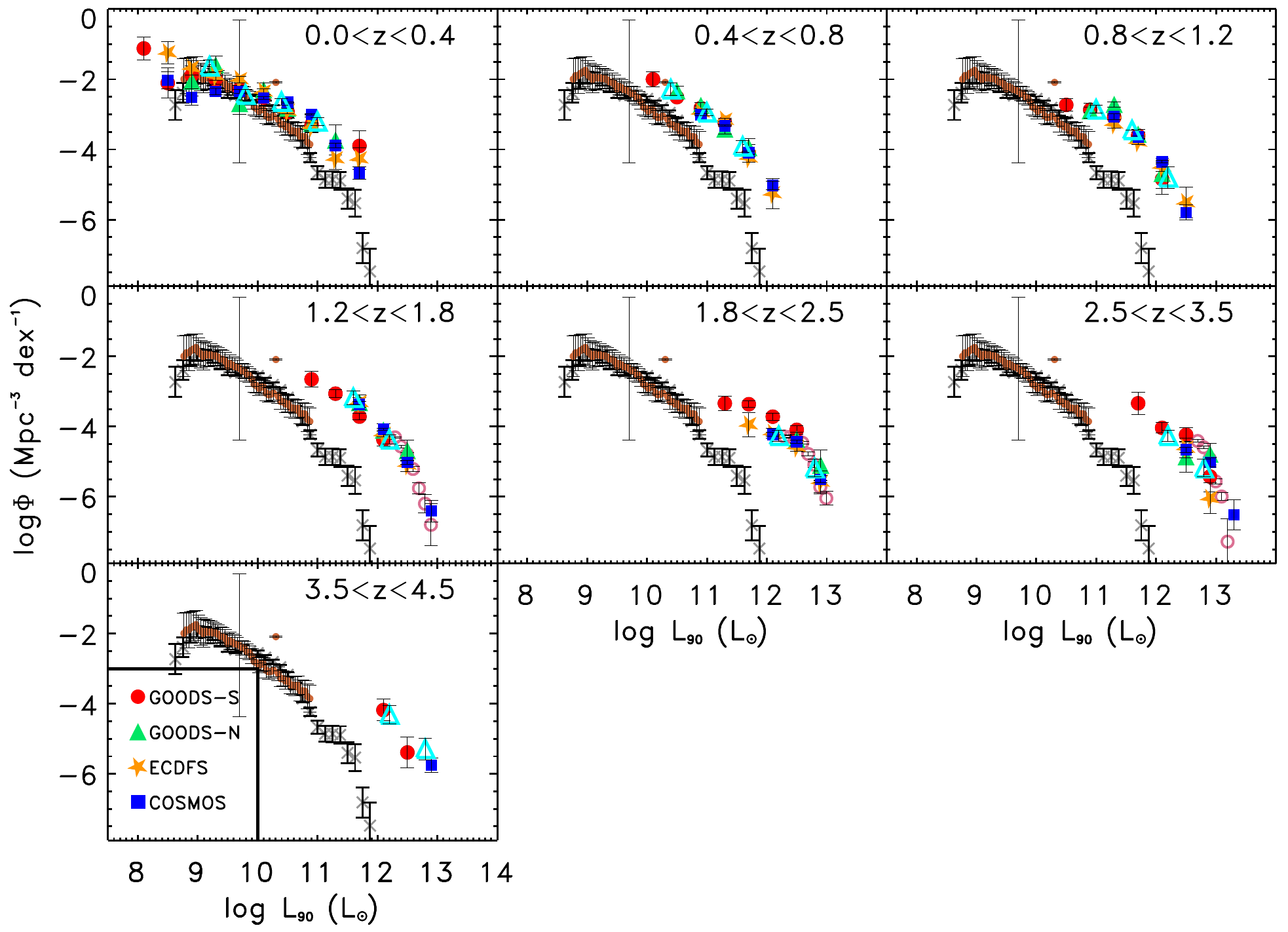}
\caption{Rest-frame 90\,$\mu$m Luminosity Function estimated independently from the 160-$\mu$m selected samples in the four PEP fields with the 1/V$_{\rm max}$ method (with the same symbols as in Fig.~\ref{figLF60um}). The diagonal crosses represent the local LF of \citet{serj04}, the brown dots with error bars are the local LF of \citet{serj11} in the {\it AKARI} Deep Field, while the pink open circles are the 100-$\mu$m LF derivation of \citet{lapi11} in the H-ATLAS fields. The cyan open triangles show the PEP 90-$\mu$m LF computed by \citet{grup10} from the SDP data in the GOODSN (not exactly the same redshift bins).}
\label{figLF90um}
\end{figure*}

\subsection{Method}
\label{sec_method}
The LFs are derived using the $1/V_{\rm max}$ method (Schmidt 1968). This method is non-parametric and does not require any assumptions on the LF shape, but derives the LF directly from the data. 
We have first derived the LFs in each field separately, in order to check for consistency and to test the role of cosmic variance. 
Successively, we have made use of the whole data-sets to derive the monochromatic and total IR LFs, by means of the \citet{avni80} method for coherent analysis of independent data-sets.
We have divided the samples into different redshift bins, over the range 0$\lsimeq$$z$$\lsimeq$4, selected to be almost equally populated, at least up to $z$$\sim$2.5. 
In each redshift bin we have computed the comoving volume available to each source belonging to that bin, defined as
$V_{\rm max}$$=$$V_{\rm z_{max}}$$-$$V_{\rm z_{min}}$, where $z_{\rm max}$ is the minimum between the maximum redshift at which a source would still be included in the
sample given the limiting flux of the survey (different for each field) and the upper boundary of the considered redshift bin, while $z_{min}$ is just the lower boundary of the considered redshift bin.\\ 
When combining the four samples, we have constructed a complete sample over the whole GOODS-S$+$GOODS-N$+$ECDFS($-$GOODS-S)$+$COSMOS region, including all the observed objects (see details in Section~\ref{sec_rflf}).
The depth of the sample is not constant throughout the region, but an object with a given flux density (included in the list of observed objects irrespective of the identity of its parent sample)
can {\em a priori} be found in one (or more) region if its redshift is $\leq$$z_{\rm max}^{\rm fld}(S_{\rm limit})$ of that region (e.g. sources detected in the COSMOS area are detectable over the whole joint area, while the fainter
sources detected in GOODS-S are detectable in GOODS-S only). The maximum volume of space which is available to such an object to be included in the joint sample is then defined by
\begin{eqnarray}
\label{eq_vmax}
V_{\rm zmax,i}&=&\frac{\Omega_{\rm GS}}{4\pi} V_{\rm z_{max}}^{\rm GS}+\frac{\Omega_{\rm GN}}{4\pi} V_{\rm z_{max}}^{\rm GN}+\frac{\Omega_{\rm E}}{4\pi} V_{\rm z_{max}}^{\rm E}+\\
&&+\frac{\Omega_{\rm C}}{4\pi} V_{\rm z_{max}}^{\rm C} \nonumber \\
& & (if ~z_{\rm max,i}\leq z_{\rm max}^{\rm C}) \nonumber \\
 &=&\frac{\Omega_{\rm GS}}{4\pi} V_{\rm z_{max}}^{\rm GS}+\frac{\Omega_{\rm GN}}{4\pi} V_{\rm z_{max}}^{\rm GN}+\frac{\Omega_{\rm E}}{4\pi} V_{\rm z_{max}}^{\rm E} \nonumber \\
& & (if ~z_{\rm max}^{\rm C} < z_{\rm max,i} \leq z_{\rm max}^{\rm E})  \nonumber \\
 &=&\frac{\Omega_{\rm GS}}{4\pi} V_{\rm z_{max}}^{\rm GS}+\frac{\Omega_{\rm GN}}{4\pi} V_{\rm z_{max}}^{GN} \nonumber  \\
& & (if ~z_{\rm max}^{\rm E} < z_{\rm max,i} \leq z_{\rm max}^{\rm GN})  \nonumber  \\
 &=&\frac{\Omega_{\rm GS}}{4\pi} V_{\rm z_{max}}^{\rm GS}  \nonumber  \\
& & (if ~z_{\rm max}^{\rm GN} < z_{\rm max,i}) \nonumber 
\end{eqnarray}
where V$_{\rm z_{max}}^{\rm fld}$ (with $\rm fld$=$\rm GS$, $\rm GN$, $\rm E$, $\rm C$ corresponding to GOODS-S, GOODS-N, ECDFS and COSMOS, respectively) is the comoving volume available to each source
in that field, in a given redshift bin, while $\Omega_{\rm fld}$ is the solid angle subtended by that field sample on the sky. \\
For each luminosity and redshift bin, the LF is given by:
\begin{equation}
\centering
\Phi(L,z)=\frac{1}{\Delta L}\sum_{\rm i}{\frac{1}{w_{\rm i} \times V_{\rm max,i}}}
\end{equation}
where $V_{\rm max,i}$ is the comoving volume over which the $i$-${\rm th}$ galaxy could be observed, $\Delta L$ is the size of the luminosity bin, and $w_{\rm i}$ is the completeness correction factor of the $i$-${\rm th}$ galaxy. These completeness correction factors are a combination of the completeness corrections given by \citet{ber10} and \citet{ber11}, 
derived as described in \citet{lutz11}, to be applied to each 
source as function of its flux density, together with a correction for redshift incompleteness (for the ECDFS and COSMOS only, see Section~\ref{sec_zdistr}). 
Since, as mentioned in \ref{sec_zdistr}, the redshift incompleteness in the COSMOS and ECDFS areas is independent on PACS flux density, in these fields we have applied the corrections regardless of the source luminosity and redshift (i.e. by multiplying $\Phi(L,z)$ by 1.07 and 1.14 in COSMOS and ECDFS, respectively).  
However, the redshift incompleteness does not affect our conclusions, since $\gsimeq$95 per cent of all our sources do have a redshift.

Uncertainties in the infrared LF values depend on photometric redshift uncertainties. 
To quantify the effects of the uncertainties in photometric redshifts on our luminosity functions, we
performed a set of Monte Carlo simulations, as described in Section~\ref{sec_totLF}. 

\subsection{The Rest-Frame 35-, 60- and 90-$\mu$m Luminosity Function}
\label{sec_rflf}
By following the method described above, we have derived the 35-$\mu$m, 60-$\mu$m and 90-$\mu$m rest-frame LFs from the 70-$\mu$m (in GOODS-S only), 
100-$\mu$m and 160-$\mu$m samples, respectively.
In order to check the consistency between the catalogues and the effects of cosmic variance, we have first derived the monochromatic LFs in each field separately. 
Note that, since the 70-$\mu$m data are available in the GOODS-S field only, to have a better sampling of the LF especially at the bright-end, we have also computed the rest-frame 35-$\mu$m LF from the 100-$\mu$m samples in the four fields and compared them with that obtained from the 70-$\mu$m sample (see Fig~\ref{figLF35um}; Table~3). The agreement between the two derivations is very good, implying correct extrapolations in wavelength due to the good and complete SED coverage.
 
We have divided the samples into seven redshift bins: 
0.0$<$$z$$\leq$0.4; 0.4$<$$z$$\leq$0.8; 0.8$<$$z$$\leq$1.2; 1.2$<$$z$$\leq$1.8; 1.8$<$$z$$\leq$2.5; 2.5$<$$z$$\leq$3.5; and 3.5$<$$z\leq$4.5. 
The results of the computation of our 35-$\mu$m (reported in Table 3), 60-$\mu$m and 90-$\mu$m LFs are shown in Figs.~\ref{figLF35um}, \ref{figLF60um} and \ref{figLF90um},
respectively.
The LFs in the four different fields appear to be consistent with each other within the error bars ($\pm$1$\sigma$) in most of the common luminosity bins. The COSMOS and GOODS-S Surveys are
complementary, with the faint end of the LFs being mostly determined by data in GOODS-S, and the bright end by COSMOS data. 
After having checked the field-to-field consistency, we have combined the 100- and 160-$\mu$m samples in all fields, obtaining the global rest-frame 60- and 90-$\mu$m LFs (reported in Tables 4 and 5, respectively).

The data from each field in each $z$-bin have been plotted (and considered in the combination) only in the luminosity bins where we expect our sample to be complete, given that at fainter luminosities not all galaxy types are observable (depending on their SEDs; Ilbert et al. 2004).  For comparison, we overplot the LFs at 35\,$\mu$m from \citet{mag09} and \citet{mag11}, the local LFs at 60\,$\mu$m from \citet{saun90} and those at 90\,$\mu$m from \citet{serj04} and \citet{serj11} and at 100\,$\mu$m from \citet{lapi11}, respectively. 
The comparison between the 35-$\mu$m LF, derived from the 70-$\mu$m PEP sample in GOODS-S, and the results of \citet{mag09} and \citet{mag11}, based on a 24-$\mu$m prior extraction and stacking analysis on {\em Spitzer} maps, shows very good agreement, both with the data and with the double power-law fit. The 1.8$<$$z$$<$2.5 PEP LF is consistent  within $\pm$1$\sigma$ 
with the \citet{mag11} data points, though the power-law fit at bright $L_{\rm 35}$ ($>$10$^{12}$ L$_{\odot}$) seems to be slightly lower than our data. 
At $z$$>$2.5 no comparison data from {\em Spitzer} are available, while our LF derivation can provide hints of evolution at the bright end of the LF. 
In the common redshift intervals (between $z$$\sim$1 and 3.5), our 90-$\mu$m LF is in very good agreement with the 100-$\mu$m \citet{lapi11} derivation from the H-ATLAS survey (although their
redshift bins are somewhat different than ours: 1.2$<$$z$$<$1.6; 1.6$<$$z$$<$2.0; 2.0$<$$z$$<$2.4; and 2.4$<$$z$$<$4.0) and with the previous PEP-SDP derivation (\citealt{grup10}).
The consistency between our 90-$\mu$m LFs and the \citet{lapi11} ones (derived from a different sample, using a different template SED to fit the data and a rest-frame far-IR based method to derive photometric redshifts)
gives us confidence that, at least up to $z$$\sim$3.5, our computation is not significantly affected by incompleteness or by photo-$z$ uncertainties.

\begin{figure*}
\includegraphics[width=15cm,height=13cm]{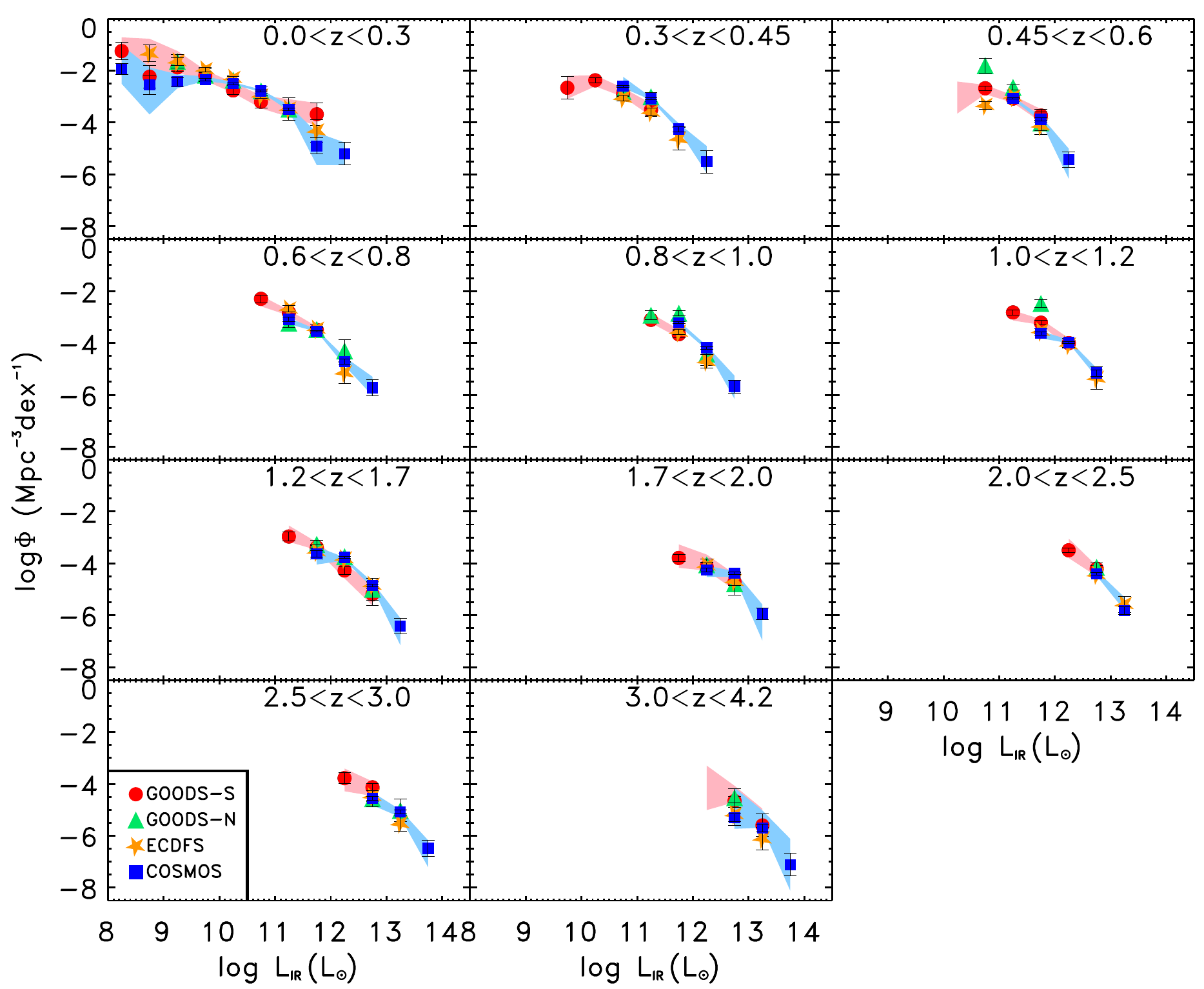}
\caption{Total IR Luminosity Function estimated independently from the 160-$\mu$m samples in the four PEP fields (with the same symbols as in Fig.~\ref{figLF60um}). The pink and sky-blue shaded areas represent the range of values derived with 20 iterations by allowing a change in photo-$z$ in the GOODS-S and COSMOS fields, respectively.}
\label{figLFbolf}
\end{figure*}

\begin{figure*}
\includegraphics[width=16cm,height=15cm]{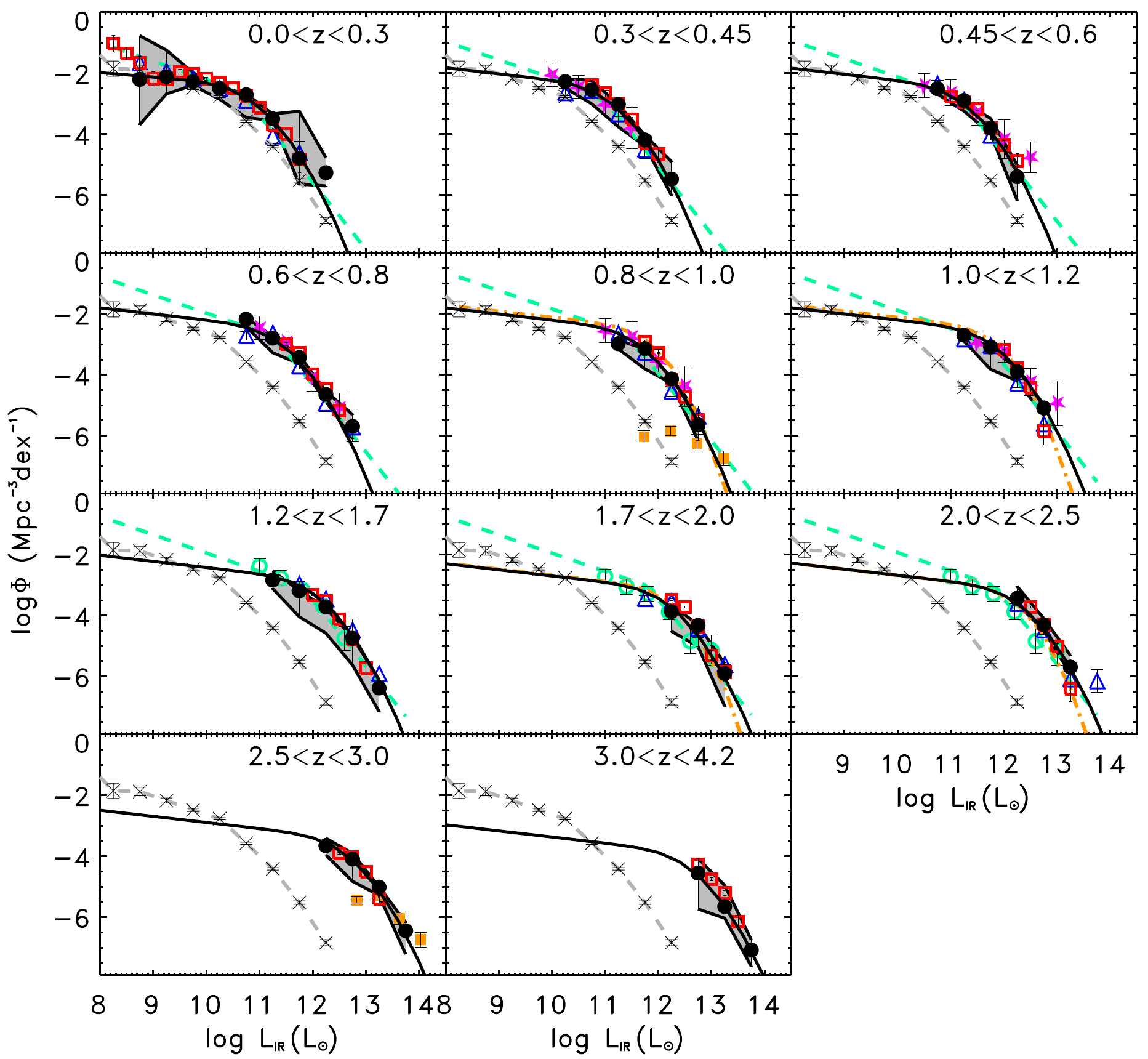}
\caption{Total IR Luminosity Function estimated by combining the data from the four PEP fields using the \citet{avni80} method (black filled circles). The grey filled area shows the uncertainty locus obtained by combining the Poissonian error with the photometric redshift uncertainty derived through Monte Carlo simulations. The black solid line represents our best fit to the PEP data in the different redshift bins, corresponding to the parameters reported in Table~\ref{tabpar}. Other results from the literature are plotted for comparison (diagonal crosses connected by a grey dashed line, LLF of \citealt{sand03}; magenta filled stars, \citealt{lef05}; orange filled squares, \citealt{chap05}; orange dot-dashed line, \citealt{cap07}; blue open triangles, \citealt{rod10a}; green dashed line, \citealt{mag09}, 2011; green open circles, \citealt{mag11}). Note that the \citet{mag11} data correspond to slightly different redshift bins: 1.3$<$$z$$<$1.8 and 1.8$<$$z$$<$2.3, respectively. We have therefore plotted the data points corresponding to the first redshift bin in our 1.2$<$$z$$<$1.7 panel and the data point corresponding to the second redshift bin in both our 1.7$<$$z$$<$2.0 and 2.0$<$$z$$<$2.5 panels. The red open squares are the total IR LFs derived by Vaccari et al. (in preparation) from the HerMES 250-$\mu$m selected COSMOS sample.}
\label{figLFbol}
\end{figure*}

\begin{figure*}
\includegraphics[width=15cm]{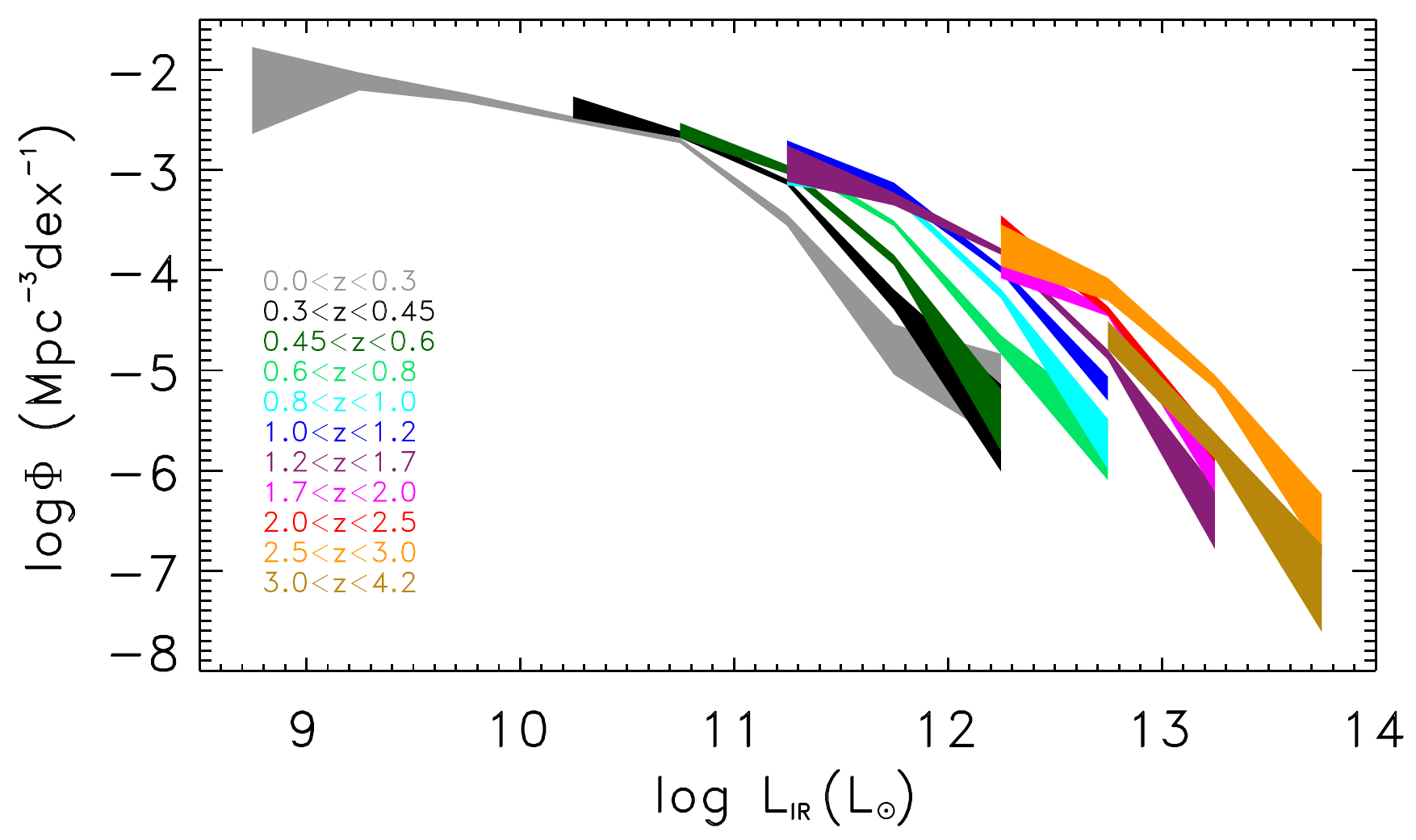}
\caption{Total IR Luminosity Function estimated by combining the data from the four PEP fields using the \citet{avni80} method plotted in all the different redshift intervals considered in this study, from $z$$\sim$0 to $z$$\sim$4. The different colour-filled areas represent the $\pm$1$\sigma$ (Poissonian) uncertainty regions at different redshifts. }
\label{figLFallz}
\end{figure*}

\subsection{The Total Infrared Luminosity Function}
\label{sec_totLF}

We integrate the best-fit SED of each source over 8$\leq$$\lambda_{\rm rest}$$\leq$1000\,$\mu$m to derive the total IR luminosities ($L_{\rm IR}$$=L$[8--1000\,$\mu$m])
in 11 redshift bins (0.0--0.3; 0.3--0.45; 0.45--0.6; 0.6--0.8; 0.8--1.0; 1.0--1.2; 1.2--1.7; 1.7--2.0; 2.0--2.5; 2.5--3.0; and 3.0--4.2), selected to be almost equally populated, at least up to $z$$\sim$2.5. 
Our approach is similar to that of other studies based on mid-IR selected galaxy samples (e.g. \citealt{lef05}; \citealt{rod10a}; \citealt{mag11}), but this is the first time that the SEDs have been accurately constrained by sufficiently deep data in the far-IR/sub-mm domain and not simply extrapolated from the mid-IR to the longer wavelengths or from average flux density ratios. 

As mentioned in Section~\ref{sec_zdistr}, we have studied the impact of redshift uncertainties on the total IR LF by performing Monte Carlo simulations.
As test cases, we used the COSMOS and GOODS-S samples (which are basically defining the bright and faint ends of
the LF in an almost complementary way), and we checked the effect on the total IR luminosity function. We iterated the computation of
the total IR LF by each time varying the photometric redshift of each source (assigning a randomly selected value, according to a Gaussian distribution, within the 68 per cent
confidence interval). Each time, we then recomputed the monochromatic and total IR luminosities, as well as the $V_{\rm max}$ value, but we
did not perform the SED-fitting again, keeping the previously found best-fit template for each object (the effect on the k-correction is not relevant in
the far-IR wavelength range). The results of this Monte Carlo simulation are reported in Fig.~\ref{figLFbolf}, where we show the total IR LFs derived in each of the PEP fields independently:
the red and blue filled circles represent the estimates of the GOODS-S and COSMOS LFs, with their range of values derived with 20 iterations by allowing a change in photo-$z$
represented by the pink and sky-blue shaded areas, respectively. The comparison shows that the effect of the uncertainty of the photometric redshifts 
on the error bars is slightly larger than the simple Poissonian value (1/$\sqrt{N}$), and affects mainly the lower and the higher redshift bins (especially at low and high luminosities).  
Using these Monte Carlo simulations, we find no evident
systematic offsets caused by the photometric redshift uncertainties (see Fig.~\ref{figLFbolf}). 
This is due to the very accurate photometric redshifts available in these fields.
For the total uncertainty in each luminosity bin in GOODS-S and COSMOS, we have therefore assumed the dispersion given by the Monte Carlo simulations (as shown in Fig.~\ref{figLFbol}).
We note that at the higher $z$ ($>$2.5), where we must rely on a majority of photometric redshifts, the true uncertainties (taking into account also catastrophic 
failures or incompleteness effects) might be larger than derived with simulations. The unavailability of a ``true'' comparison sample (i.e. a large comparison sample
with accurate spectroscopic redshifts and fully representative for the PACS flux selection) at high z does not allow to properly quantify this statement. 

In Fig.~\ref{figLFbol} the total IR LFs obtained by combining the 160-$\mu$m selected samples with the \citet{avni80} technique is plotted and compared with other derivations available in the literature. The total IR LF of \citet{sand03} is plotted as a local reference, in addition to the LFs
of  \citet{lef05}, \citet{rod10a}, \citet{cap07}, \citet{mag09} and \citet{mag11} in various redshift intervals. 
Globally, data from surveys at different wavelengths agree relatively well over the common $z$-range. 
No data for comparison are available at $z$$>$2.5, apart from the IR LF of sub-mm galaxies from \citet{chap05} and \citet{wall08} at $z$$\sim$2.5, which represent reasonably well just the very bright end of the total IR LF. Our derivation is the first at such high redshifts, especially in the 3$<$$z$$\leq$4.2 range. Note the good agreement between our PEP-based total IR LF and the HerMES-based one derived by Vaccari et al. (in preparation), shown by the red open squares in Fig.~\ref{figLFbol}. Though consistent within the error-bars, in the highest redshift bin the HerMES LF is slightly higher than ours. Since the 250-$\mu$m HerMES selection favours the detection of higher redshift sources than the PEP one, it is more likely that the PEP LF in the higher-z bin is affected by some flux/redshift incompleteness rather than by the presence of low-$z$ sources 
erroneously placed at high-$z$ by incorrect photometric redshift assignment (catastrophic failures).  
The values of our total IR LF for each redshift and luminosity bin are reported in Table~6.

\subsection{Evolution}
\label{sec_evol}
In order to study the evolution of the total IR LF, we derive a parametric estimate of the
luminosity function at different redshifts. For the shape of the LF we assume a modified-Schechter function
(i.e. Saunders et al. 1990), where $\Phi(L)$ is given by
\begin{equation}
\Phi(L){\rm dlog}L=\Phi^{\star}\left(\frac{L}{L^{\star}}\right)^{1-\alpha} \exp\left[-\frac{1}{2\sigma^2}\log_{10}^2\left(1+\frac{L}{L^{\star}}\right)\right]{\rm dlog}L, 
\label{eq:schechter}
\end{equation}
behaving as a power law for $L \ll L^{\star}$ and as a Gaussian in $\log L$ for $L\gg L^{\star}$.
The adopted LF parametric shape depends on 4 parameters ($\alpha$, $\sigma$, $L^{\star}$ and $\Phi^{\star}$), whose best fitting values and uncertainties
have been found using a non-linear least squares fitting procedure. In detail, while in the first $z$-bin all the parameters
have been estimated, starting from the second $z$-bin, the values of $\alpha$ and $\sigma$ have been frozen
at the values found at lower redshift, leaving only $L^{\star}$ and $\Phi^{\star}$ free to vary (see Table~\ref{tabpar}). 
Note that, in the highest redshift bin (3.0$<$$z$$<$4.2) we are not able to constrain the LF break, while we are up to $\sim$3. Therefore, the results 
found at this redshift are affected by larger uncertainties than the $z$$\lsimeq$3 ones.
However, although there is some degeneracy in the values of  $L^{\star}$ and $\Phi^{\star}$ at 3.0$<$$z$$<$4.2, the range of allowed value combinations 
still giving a reasonable fit to the three observed data points 
(constraining the bright-end of the LF) is limited and do not significantly affect our results.
\begin{figure}
\includegraphics[width=8cm,height=7cm]{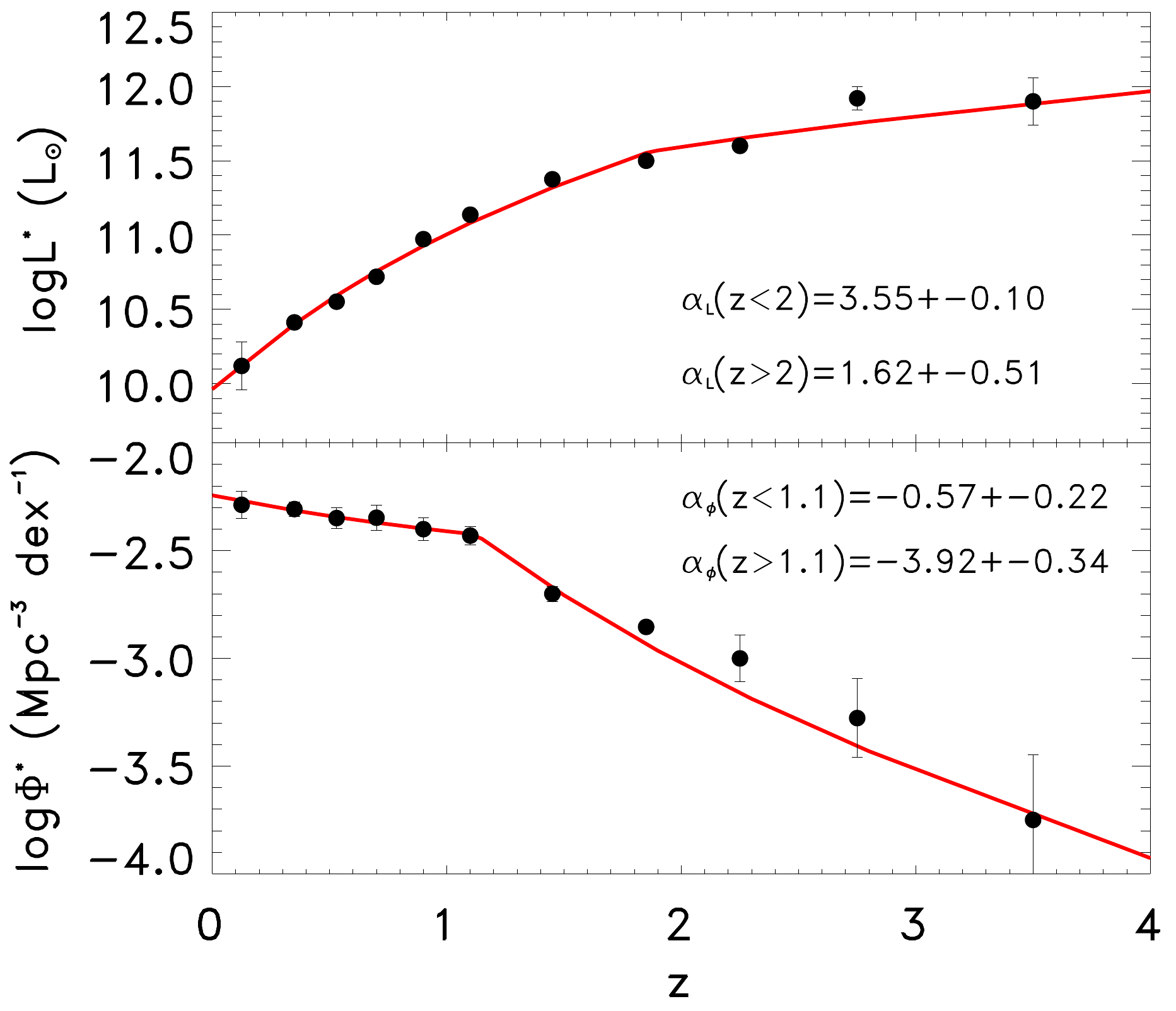}
\caption{Evolution of L$^{\star}$ and $\Phi^{\star}$ as function of $z$ (L$^{\star}$$\propto$$(1+z)^{3.55 \pm 0.10}$ at $z$$\lsimeq$1.85, $\propto$$(1+z)^{1.62 \pm 0.51}$ at $z$$\gsimeq$1.85; $\Phi^{\star}$$\propto$$(1+z)^{-0.57 \pm 0.22}$ at $z$$\lsimeq$1.1, $\propto$$(1+z)^{-3.92 \pm 0.34}$ at $z$$\gsimeq$1.1.}
\label{figlpstarz}
\end{figure}

In Fig.~\ref{figLFallz} we plot the total IR LFs at all redshifts with the $\pm$1$\sigma$ Poissonian uncertainty regions (different colours for different $z$-intervals). There is a clear luminosity evolution with
redshift, at least up to $z$$\sim$3. The apparent ``fall'' of the 
$z$$>$3 LF with respect to those at lower redshifts, if real, might imply a global negative evolution of the IR galaxies and/or AGN starting at $z$$\gsimeq$3. 
However, as mentioned above, we must point out that in the highest redshift bin PACS data could be affected by incompleteness, since with increasing redshift (and for intrinsically ``cold'' sources) the true PACS fluxes might fall below the detection limit (i.e. faint sources should be missed even if completeness corrections are perfect), while, if luminous enough, they can still be detectable by SPIRE. 
This effect is expected to be more relevant in COSMOS, where SPIRE data are quite deep relatively to PACS, while it should not happen in GOODS-S, where PACS data are very deep compared to the SPIRE
ones (which are limited by confusion). 
The high-$z$ incompleteness of PACS surveys might also be emphasised by the redshift incompleteness of the COSMOS sample, that could affect the highest redshift bins more than the lower ones (although the redshift incompleteness seems to be independent of redshift; see \citealt{ber11}). 
However, we must point out that a decrease at $z$$\geq$2.7--3 similar to that observed in our data, is also observed in the space density of X-ray (see \citealt{bru09}; \citealt{civ11})
and optically selected AGN (\citealt{rich06}), and of sub-mm galaxies (\citealt{wall08}), as well as in the HerMES total IR LF -- though less evident -- (from 250-$\mu$m data; Vaccari et al., in preparation: see Fig.~\ref{figLFbol}). 
Moreover, our result is in agreement with
the recent finding of \citet{smit12}, that the characteristic value of the galaxy SFR exhibits a substantial, linear decrease as a function of redshift from $z$$\sim$2 to $z$$\sim$8.
\\
In Fig.~\ref{figlpstarz} we show the values of $L^{\star}$ and $\Phi^{\star}$ at different redshifts, with the best curve ($\propto(1+z)^k$) fit to the data points. 
The values of the curve slopes and of the redshifts corresponding to evolutionary breaks have been chosen to be those which minimise the $\chi^2$ of the fit with two power-laws.
We find a significant variation of L$^{\star}$ with $z$, which increases as $(1+z)^{3.55\pm0.10}$ up to $z$$\sim$1.85, and as $(1+z)^{1.62\pm0.51}$ between $z$$\sim$1.85 and $z$$\sim$4. 
The variation of $\Phi^{\star}$ with redshift starts with a slow decrease as $(1+z)^{-0.57\pm0.22}$ up to $z$$\sim$1.1, followed by a rapid decrease $\propto$(1$+$$z$)$^{-3.92\pm0.34}$ at $z$$\gsimeq$1.1 and up to $z$$\sim$4. \\
Previous estimations of the evolution of $L^{\star}$ and $\Phi^{\star}$ (i.e. \citealt{cap07}, \citealt{bet11} and \citealt{marsd11}) discussed a decrease in the density of far-IR sources between $z$$=$1 and $z$$=$2. 
In particular, \citet{bet11} and \citet{marsd11}, by evolving a parameterised far-IR LF, explored the evolution required by the source counts in the parameter space. 
The results of these works (especially those of \citealt{bet11}, showing a decrease of $\Phi^{\star}$ at $z$$>$1 and a flatter trend on the evolution of $L^{\star}$ at $z$$>$2), are close 
to ours, though with the source counts only it was not possible to constrain the evolution of IR sources at $z$$>$2.
\setcounter{table}{6}
\begin{table*}
 \caption{Parameter values describing the curve fitted to the total IR LF}
\begin{tabular}{ccccc}
\hline \hline
 redshift range &  $\alpha$ & $\sigma$ & log$_{10}$($L^{\star}$/L$_{\odot}$) & log$_{10}$($\Phi^{\star}$/Mpc$^{-3}$ dex$^{-1}$)   \\ 
        &     &    &   &     \\   \hline 
 0.0$<$$z$$<$0.3 &   1.15$\pm$0.07 &    0.52$\pm$0.05 &     10.12$\pm$0.16   &  $-$2.29$\pm$0.06\\
0.3$<$$z$$<$0.45 &  1.2$^{a}$  &   0.5$^{a}$ &   10.41$\pm$0.03   &   $-$2.31$\pm$0.03\\
 0.45$<$$z$$<$0.6 &  1.2$^{a}$  &  0.5$^{a}$  &    10.55$\pm$0.03  &       $-$2.35$\pm$0.05\\
 0.6$<$$z$$<$0.8 & 1.2$^{a}$   & 0.5$^{a}$ & 10.71$\pm$0.03      &     $-$2.35$\pm$0.06\\
 0.8$<$$z$$<$1.0 &  1.2$^{a}$  &  0.5$^{a}$ &     10.97$\pm$0.04   &     $-$2.40$\pm$0.05\\
 1.0$<$$z$$<$1.2 & 1.2$^{a}$  & 0.5$^{a}$ & 11.13$\pm$0.04      &      $-$2.43$\pm$0.04\\
 1.2$<$$z$$<$1.7 & 1.2$^{a}$   & 0.5$^{a}$ &       11.37$\pm$0.03    &      $-$2.70$\pm$0.04\\
 1.7$<$$z$$<$2.0 &  1.2$^{a}$  &  0.5$^{a}$ &      11.50$\pm$0.03    &      $-$3.00$\pm$0.03\\
  2.0$<$$z$$<$2.5 &  1.2$^{a}$  &  0.5$^{a}$ &   11.60$\pm$0.03     &      $-$3.01$\pm$0.11\\
 2.5$<$$z$$<$3.0 & 1.2$^{a}$  &  0.5$^{a}$ &    11.92$\pm$0.08     &      $-$3.27$\pm$0.18\\
 3.0$<$$z$$<$4.2 & 1.2$^{a}$   & 0.5$^{a}$ &      11.90$\pm$0.16    &       $-$3.74$\pm$0.30\\
   \hline \hline
$^a$ fixed value
\end{tabular}
\label{tabpar}
\end{table*}

\subsection{Evolution of the Different IR Populations}
\label{sec_pop}

\begin{figure*}
\includegraphics[width=16cm]{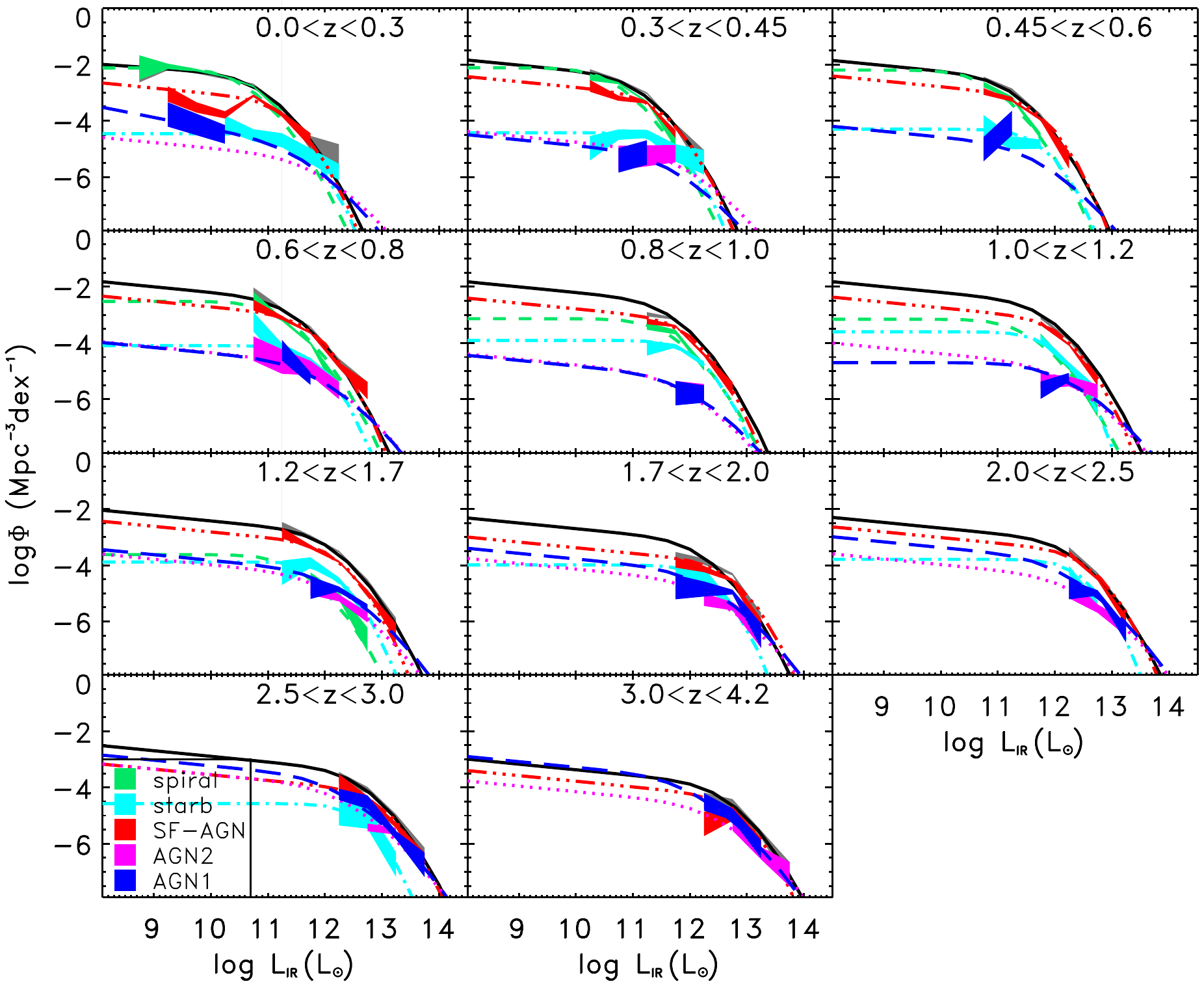}
\caption{Total IR Luminosity Function estimated with the 1/V$_{\rm max}$ method by combining the data from the four PEP fields for the different populations (their $\pm$1$\sigma$ uncertainty regions are shown as coloured filled areas: green for {\tt spirals}; cyan for {\tt starbursts}; red for {\tt SF-AGN}; magenta for {\tt AGN2}; and blue for {\tt AGN1}), compared to the total LF ($\pm$1$\sigma$, grey filled area, same as in Fig.~\ref{figLFbol}). The best-fit modified Schechter functions are also plotted, extrapolated to fainter and brighter luminosities than covered by the data (with the black curve being for the total LF and the same colours used for the filled areas as for the single populations).}
\label{figLFpop}
\end{figure*}

\begin{figure*}
\includegraphics[width=16cm]{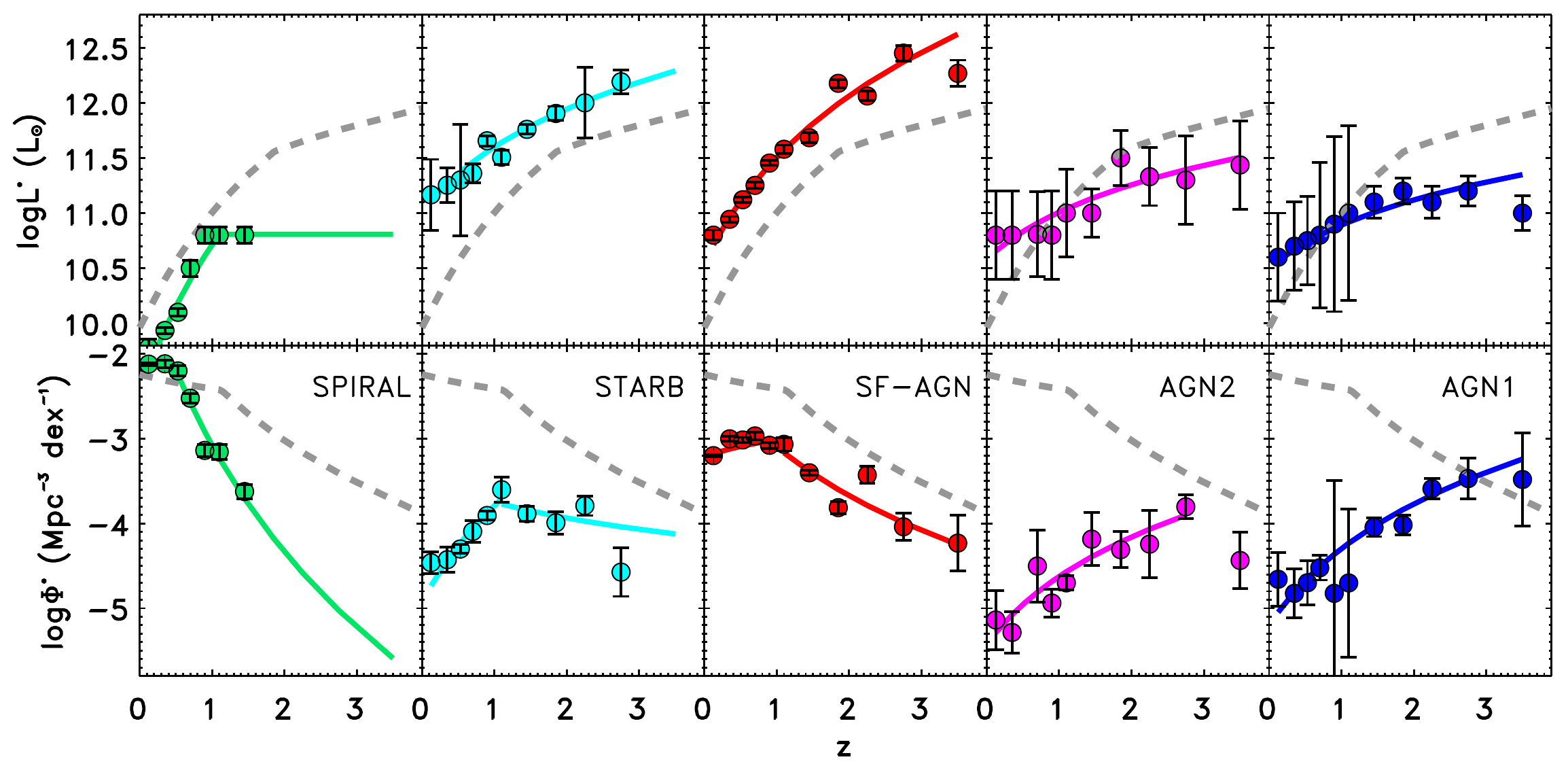} 
\caption{Evolution of $L^{\star}$ ({\em top}) and $\Phi^{\star}$ ({\em bottom}) as a function of redshift (in the form $\propto(1+z)^{\kappa}$) for the different IR populations. For comparison, the $L^{\star}$ and $\Phi^{\star}$ evolution of the ``global'' total IR LF (plotted in Fig.~\ref{figlpstarz}), are shown as grey dashed lines. }
\label{fig:lphievol}
\end{figure*}

The PEP survey, given its size and its coverage in luminosity and redshift, allows us to go further in investigating the evolution of the total IR LF: it gives us the opportunity
to study the evolution of the different galaxy classes that compose the global IR population.
To investigate the different evolutionary paths of the various IR populations, 
we have computed the 1/V$_{\rm max}$ LFs separately for the five galaxy classes defined by the SED-fitting analysis. 
In Fig.~\ref{figLFpop} we show the total IR LFs derived from the combined PEP samples for the 
different SED classes (coloured filled areas). 
The results of the fit to a modified Schechter function (see equation \ref{eq:schechter}) for each population are overplotted on the data.  
In fact, by following the same procedure adopted for the global luminosity function, a parametric fit to the LFs at different redshifts has been performed also for the single populations. The $\alpha$ and $\sigma$  parameters, for each population, have been estimated at the redshift where the corresponding LF is best sampled (not necessarily at the lowest $z$-bin as for the global LF). 
Subsequently, the values of $\alpha$ and $\sigma$ have been frozen at the values found in the ``optimal'' redshift bin, leaving only $L^{\star}$ and $\Phi^{\star}$ free to vary.
In Fig.~\ref{fig:lphievol}, analogously to Fig.~\ref{figlpstarz}, we show the values of $L^{\star}$ and $\Phi^{\star}$ at different redshifts for the different populations, with the best least square fitting curves ($\propto(1+z)^k$) overplotted. The best-fitting values of $\alpha$, $\sigma$, $L^{\star}$ and $\Phi^{\star}$ in the first redshift bin, together with the parameters describing the luminosity ($k_{\rm L,1}$, $k_{\rm L,2}$ and $z_{\rm b,L}$: $\propto$(1$+$$z$)$^{\rm k_{L,1}}$ to $z$$=$$z_{\rm b,L}$, $\propto$(1$+$$z$)$^{k_{\rm L,2}}$ at $z$$>$$z_{\rm b,L}$) and density evolution ($k_{\rm \rho,1}$, $k_{\rm \rho,2}$ and $z_{\rm b,\rho}$: $\propto$(1$+$$z$)$^{k_{\rm \rho,1}}$ out to $z$$=$$z_{\rm b,\rho}$, $\propto$(1$+$$z$)$^{k_{\rm \rho,2}}$ at $z$$>$$z_{\rm b,\rho}$) are reported in Table~\ref{tabpop}.

\begin{table*}
 \caption{Parameter values describing the curve fitted to the total IR LF of the different SED populations}
\small
\begin{tabular}{lcccccccccc}
\hline \hline
 &  $\alpha$ & $\sigma$ & log$_{10}$(L$^{\star}$/ & log$_{10}$($\Phi^{\star}$/  &  $k_{\rm L,1}$ & $k_{\rm L,2}$ & $z_{\rm b,L}$ & $k_{\rm \rho,1}$ & $k_{\rm \rho,2}$ & $z_{\rm b,\rho}$ \\ 
  &          &    &  \multicolumn{1}{r}{L$_{\odot}$)}  & \multicolumn{1}{r}{Mpc$^{-3}$ dex$^{-1}$)}    & & & & & \\   
 & \multicolumn{4}{c}{(0.0$<$$z$$<$0.3)} &    &    &     &   & & \\ \hline
 {\tt spiral} &   1.00$\pm$0.05  &  0.50$\pm$0.01    &  ~9.78$\pm$0.04 & $-$2.12$\pm$0.01 & 4.49$\pm$0.15 & 0.00$\pm$0.46 & 1.1 & $-$0.54$\pm$0.12~~ & $-$7.13$\pm$0.24 & 0.53 \\
 {\tt starburst} &  1.00$\pm$0.20      &  0.35$\pm$0.10        &  11.17$\pm$0.16 & $-$4.46$\pm$0.06 & 1.96$\pm$0.13 &  &   & 3.79$\pm$0.21 & $-$1.06$\pm$0.05 & 1.1 \\
 {\tt SF-AGN} &  1.20$\pm$0.02   &  0.40$\pm$0.10   &  10.80$\pm$0.02 & $-$3.20$\pm$0.01 & 3.17$\pm$0.04 &  &  & 0.67$\pm$0.05 & $-$3.17$\pm$0.15 & 1.1 \\
 {\tt AGN2} &   1.20$\pm$0.20    &   0.70$\pm$0.20    &  10.80$\pm$0.20 & $-$5.14$\pm$0.17 & 1.41$\pm$0.33 & & & 2.65$\pm$0.32 & & \\
 {\tt AGN1} &  1.40$\pm$0.30      &    0.70$\pm$0.20      &  10.50$\pm$0.20 & $-$5.21$\pm$0.11 & 1.31$\pm$0.02 & & & 3.00$\pm$0.25 &  & \\
   \hline \hline
\end{tabular}
\label{tabpop}
\end{table*}
\normalsize

A clear result of our analysis is that the evolution derived for the global IR LF is indeed a combination of different evolutionary paths: the far-IR population does not evolve all together ``as a whole'', as it is often assumed in the literature, but is composed by different galaxy classes evolving differently and independently.
As shown in Fig.~\ref{figLFpop}, the normal {\tt spiral} galaxy population dominates the luminosity function at low-$z$, from the local Universe up to
$z$$\sim$0.5. Moving to higher redshifts, the number density of galaxies with spiral galaxy SEDs sharply decreases, while their luminosity continues
to increase, at least up to $z$$\sim$1 (see the $\Phi^{\star}$ and $L^{\star}$ parameter trends shown in Fig.~\ref{fig:lphievol}). 
We note that what we observe between $z$$\sim$0 and $z$$\sim$1 for the {\tt spiral} SED galaxies is an increase of $L^{\star}$ by a factor of $\sim$5, and a decrease of $\Phi^{\star}$ 
by a factor of $\sim$10. Since the two evolutions are not independent, the ``total'' evolutionary effect results from the combination of the two (as can be observed in the
total IR luminosity density, see Section~\ref{IRlumdens}).
A way to derive the "total" effect of evolution on a LF is to fix at a given volume density value and see how the luminosity corresponding to that value changes: indeed we find an increases by a factor of $\sim$2.5 between $z$$=$0 and $z$$=$1 for the {\tt spiral} LF, in good agreement with previous results, either empirical (for morphologically classified disky galaxies; \citealt{scar07}) or theoretical (from chemical evolution models of Milky Way-like galaxies; \citealt{colav08}).  \\
Over the whole redshift range 0.5$-$3, the ``totall'' luminosity
function is dominated by the {\tt SF-AGN} population. The number density of {\tt SF-AGN} is nearly constant
from the local Universe up to $z$$\sim$1--1.5, showing a slight decrease at higher redshifts, while their luminosities show positive
evolution up to the highest redshifts ($z$$\sim$3.5--4). 
From Fig.~\ref{figLFpop} we note a sort of bimodality in the {\tt SF-AGN} LFs (at $z$$\lsimeq$0.45, where we are able to cover a larger luminosity range).
This bimodality is indeed to be ascribed to the crossing of two contributions: that of the {\tt SF-AGN(Spiral)} population, responsible for the faint-end steepness of the LFs, and that of the {\tt SF-AGN(SB)} population, 
dominating the bright-end of the {\tt SF-AGN} LFs and declining at low $L_{\rm IR}$ (not reported in the figure).\\
The {\tt starburst} galaxy population never dominates. The redshift range where we observe the highest contribution from the {\tt starburst} galaxies
is at $z$$\sim$1--2, while in the local Universe their contribution is almost negligible (i.e. their $\Phi^{\star}$
parameter shows an opposite trend with respect to that of {\tt spiral} galaxies see Fig.~\ref{fig:lphievol}). \\
The {\tt AGN1} and {\tt AGN2} populations show a very similar evolutionary trend as a function of $z$, both in $\Phi^{\star}$ and
$L^{\star}$. These powerful AGN populations dominate only the very bright end of the total IR LF, although their number densities and luminosities keep 
increasing from the local Universe up to the higher redshifts. At $z$$>$2.5 the {\tt AGN1} and {\tt AGN2} populations become as important as the
{\tt SF-AGN} one, with the total IR LF of PACS-selected sources in the redshift range 2.5--4 being totally dominated by objects containing an AGN.

\subsection{Total IR LF in Mass and Specific Star-Formation Rate bins}
\label{sec_mass}
\subsubsection{Stellar masses and SFR from SED fitting}
\label{sec_ssfr}
\begin{figure*}
\includegraphics[width=17cm,height=5cm]{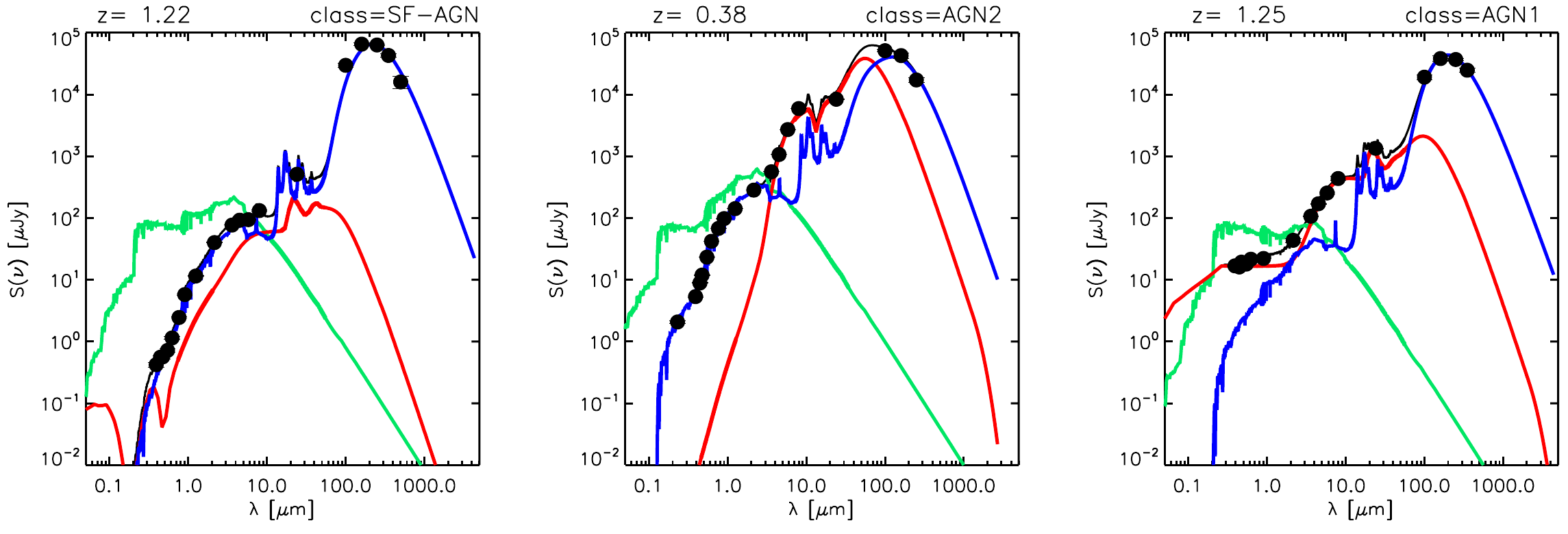}
\caption{Examples of PEP AGN SEDs fitted by the {\tt MAGPHYS} code (\citealt{dacun08}) modified by \citet{berta13} 
to add an AGN component from the \citet{fritz06} library.
From {\it left} to {\it right} the result of the fit to a {\tt SF-AGN}, an {\tt AGN2} and an {\tt AGN1} are shown. The stellar component unattenuated by dust is shown in green, while the dust-attenuated spectrum with 
dust IR re-emission is shown in blue. The AGN contribution is shown in red, while the black curve is the total fitted spectrum, obtained from the sum of the blue and red components. Our data are represented by the black dots.}
\label{fig_magphys}
\end{figure*}

The wealth of multi-wavelength data available in the cosmological
fields included in our work allow us to perform a detailed SED fitting of all sources, in order
to derive their most relevant physical parameters (e.g. stellar masses). 
To derive stellar masses we have fitted the broad-band SEDs of our sources using a modified version of {\tt MAGPHYS} (\citealt{dacun08}),
which is a code describing the SEDs using a combination of stellar light and emission
from dust heated by stellar populations. In particular, the {\tt MAGPHYS} software simultaneously fits the broad-band UV-to-far IR observed SED of each object, 
ensuring an energy balance between the absorbed UV light and that re-emitted in the far-IR regime.
The main assumptions are that the energy re-radiated by dust is equal to that absorbed, and that starlight is
the only significant source of dust heating. We refer to \citet{dacun08} for a thorough formal
description of how galaxy SEDs are build.
At each source's redshift, the code chooses among different combinations of star formation histories, metallicities
and dust contents, associating a wide range of optical models to a wide range of infrared
spectra and comparing to observed photometry, seeking for $\chi^2$ minimization.
Each star formation (SF) history is parameterised in terms of an underlying continuous model with exponentially declining
star formation rate (SFR), on top of which are superimposed random bursts (see \citealt{dacun08}, \citealt{dacun10}). 
We note that, although the {\tt MAGPHYS} assumption of exponentially
declining SFR might not be the best to reproduce the SFR history of $z$$>$1.5 star-forming galaxies (i.e. exponentially increasing or increasing SFR 
would be better choices, as widely discussed by \citealt{maras10} and \citealt{reddy12}), 
in our specific case it does not affect the results. In fact, we do not use the {\tt MAGPHYS} derived SFRs, but we compute them by integrating the best-fitting SED 
(resulting from {\tt Le Phare}).
The models are distributed uniformly in metallicity between 0.2 and 2 times solar.
Since the {\tt MAGPHYS}  code assumes that starlight is the only significant source of dust heating, thus ignoring the presence of a possible AGN component,
\citet{berta13} have developed a modified version of the {\tt MAGPHYS}  code by adding a torus component to the modelled SED emission, 
combining the \citet{dacun08} original code with the \citet{fritz06} AGN torus library (see also \citealt{feltre12}).
The spectral fitting is performed by comparing the observed SED of our galaxies to every model in the 
generated library, at the corresponding redshift. A $\chi^2$ minimisation provides the quality of each fit.
We must point out that the mass derivation for unobscured AGN (i.e. {\tt AGN1}) is a problematic issue, therefore the masses estimated for that class of objects
are the most uncertain ones. One source of uncertainty in the mass measurement for {\tt AGN1} is due to the fact that, in these objects, the UV part of the SED is likely dominated by the AGN
rather than by the host galaxy. This may produce an underestimate of the mass, since, if the AGN contribution is not taken into account, the data can be
fit by a bluer, younger and smaller mass object. 
On the other hand, the mid-IR part of the AGN SED is dominated by dust emission from the dusty torus heated by the central black hole. If a proper decomposition into a stellar and a
torus component is not performed, the use of a pure stellar template for estimating the mass from SED-fitting tends to reproduce the mid-IR emission with an older, redder and larger mass 
object (mass sometimes larger by a factor of 2 than those derived through a decomposition procedure; \citealt{sant12}). These two effects lead to an increase in the uncertainty of the mass derivation, 
although they might somewhat compensate their effects for a large sample of objects.
For this reason, we obtained measurements of the stellar masses of our objects containing an AGN by means of the specific decomposition technique developed 
by \citet{berta13}, to separate stellar and nuclear emission components. Examples of the results of this decomposition applied to {\tt SF-AGN}, {\tt AGN2} and {\tt AGN1} are shown in Fig.~\ref{fig_magphys}.
Masses of AGN estimated with the original {\tt MAGPHYS} and with the \citet{berta13} code have been compared, showing very good agreement and small dispersion around the 1--1 relation.
Similarly, as further check, we have also computed stellar masses with different code ({\tt Hyperz}; \citealt{bolz00}) and stellar library (BC03, \citealt{bc03}, instead of the CB07 used as default by {\tt MAGPHYS}), finding 
good agreement and no sistematics, too.

We have derived the SFRs from the total IR luminosities (estimated from
the SED fitting described in Section~\ref{sec_id}) with the standard \citet{kenn98} relation (converted to Chabrier IMF),
after subtracting the AGN contribution to $L_{\rm IR}$. We note that
the total IR luminosity in PEP sources is usually dominated by star formation, even in objects for which an AGN dominates the optical/near-IR/mid-IR part of the spectrum.

\subsubsection{LFs in different mass bins}
\label{sec_lfmass}
\begin{figure*}
\includegraphics[width=16cm]{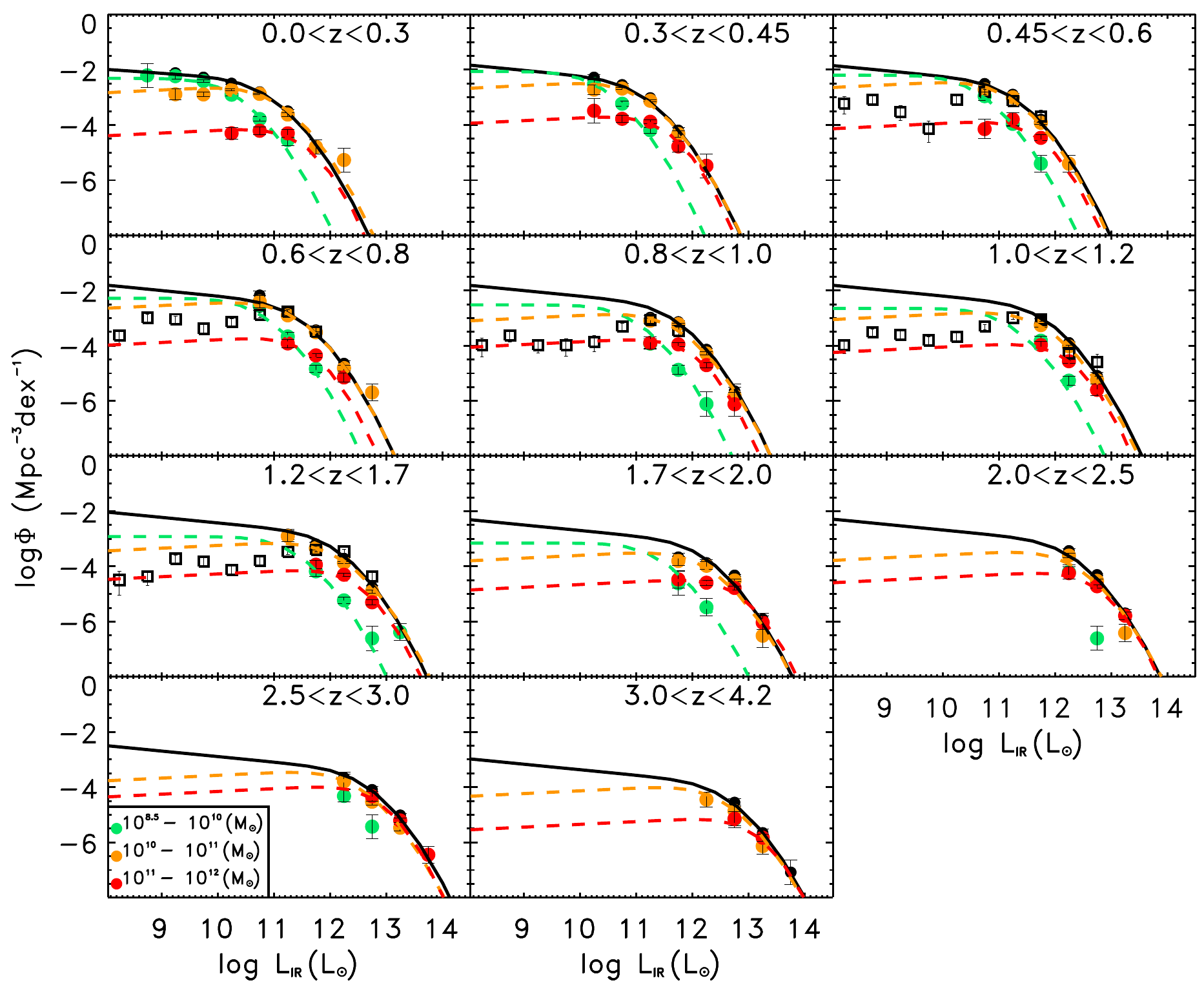} 
\caption{Contribution to the total IR LF from galaxies in three different mass intervals: 8.5$<$log($M$/M$_{\odot}$)$<$10 (green); 10$<$log($M$/M$_{\odot}$)$<$11 (orange); and 11$<$log($M$/M$_{\odot}$)$<$12 (red). For comparison, the SFR function of log($M$/M$_{\odot}$)$>$10 sources in the GOODS-S by \citealt{font12} (converted to total IR LF) is shown as black open squares.}
\label{fig:lfmass}
\end{figure*}

\begin{figure}
\includegraphics[width=8.5cm,height=8cm]{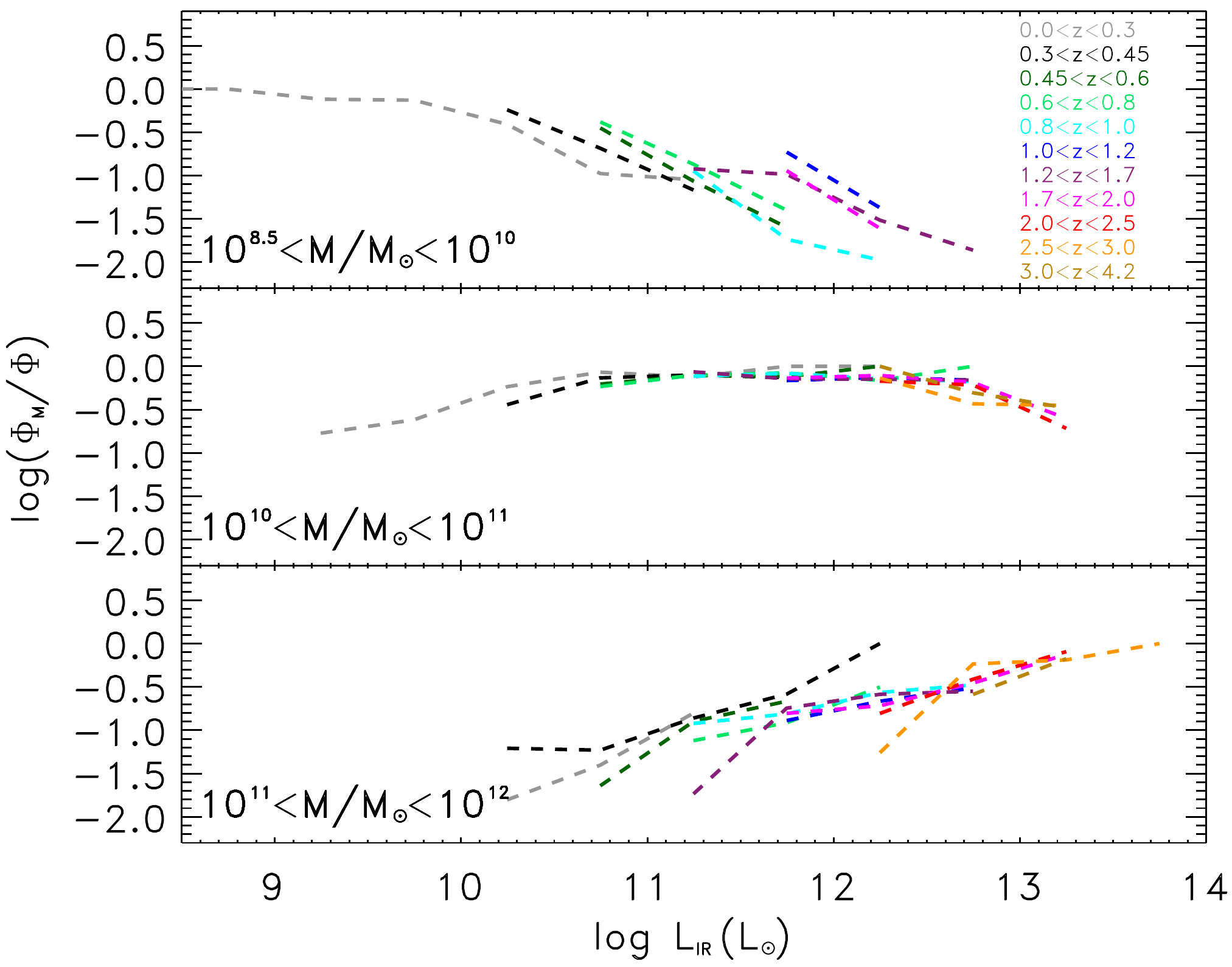} 
\caption{Ratio between the IR LF of galaxies in three different mass intervals: 8.5$<$log($M$/M$_{\odot}$)$<$10 ($top$); 10$<$log($M$/M$_{\odot}$)$<$11 ($middle$); and 11$\leq$log($M$/M$_{\odot}$)$<$12 ($bottom$), as shown in Fig.~\ref{fig:lfmass}, and the total IR LF, plotted in all the different redshift intervals considered in this study, from $z$$\sim$0 to $z$$\sim$4. The different colours represent the different redshifts, as explained in the plot.}
\label{fig:lfmassr}
\end{figure}

We compute the total IR LF for galaxies of different stellar masses: 8.5$\leq$log($M$/M$_{\odot}$)$<$10, 10$\leq$log($M$/M$_{\odot}$)$<$11, and 11$\leq$log($M$/M$_{\odot}$)$<$12, by means of the standard 1/V$_{\rm max}$ formalism, and we show the results in Figs.~\ref{fig:lfmass} (total IR LF in different $z$-bins) and \ref{fig:lfmassr} (ratio between the LF in the mass intervals and the total IR LF).

We have compared our results with the SFR function (SFR converted to IR luminosity using the \citet{kenn98} relation) of 
massive (log($M$/M$_{\odot}$$>$10) galaxies derived by \citet{font12} from the GOODS-MUSIC sample at 
redshift 0.4$<$$z$$<$1.8. In the common redshift and luminosity range we find an excellent agreement 
with our total IR LF, which is dominated by sources with 10$<$log($M$/M$_{\odot}$)$<$11. At lower luminosities, not sampled by our data, the \citet{font12} LFs are characterized 
by a double-peaked structure, interpreted in terms of the well-known bimodality in the 
colour(SFR)-Mass diagram. 
As expected from the SFR-Mass relation, the knee ($L^{\star}$) of our IR LFs in different mass bins moves to 
higher luminosities with increasing masses (i.e. at 0.0$<$$z$$<$0.3, log($L^{\star}$/L$_{\odot}$) changes from $\sim$9.5 for
log($M$/M$_{\odot}$)$=$8.5--10 sources, to $\sim$11.3 for sources with log($M$/M$_{\odot}$)$>$11).\\
The slope of the total IR LF in each mass bin is always similar to (or flatter than) the ``global'' LF (total, including all the masses).
The lower mass galaxies dominate at lower luminosities (log($L$/L$_{\odot}$)$<$9),
while the most massive galaxies (log($M$/M$_{\odot}$)$>$11) contribute only at higher luminosities, even if they never dominate the LF.
At all masses, the LF evolves with redshift, following the evolution of the ``global'' LF.
Fig.~\ref{fig:lfmassr} shows that the main contribution ($>$50 per cent) to the total IR LF is due to 
intermediate-mass objects (log($M$/M$_{\sun}$)$=$10--11) at all redshifts and luminosities, 
with their fraction remaining almost the same from $z=$0 to $z$$\sim$4, simply shifting to higher luminosities. 
Lower mass objects (log($M$/M$_{\odot}$)$=$8.5--10) contribute significantly 
only at log($L$/L$_{\odot}$)$<$10, with their fraction just shifting to higher luminosities with redshift, but 
always being below 30 per cent at $z$$>$0.45 and log($L$/L$_{\odot}$)$>$11. 
The contribution of the most massive objects (log($M$/M$_{\odot}$)$=$11--12) increases with IR luminosity and redshift, becoming 
significant ($>$50 per cent) only at $z$$>$1.7 and log($L$/L$_{\odot}$)$>$12.5. 
Thus the bulk of the IR luminosity is produced by star-forming galaxies of mass around the characteristic mass M$^{\star}$ of the Schecter mass function.

\subsection{Specific-SFR and the main sequence of star-forming galaxies}
\label{sec_ssfr}

\begin{figure}
\includegraphics[width=8.5cm,height=14cm]{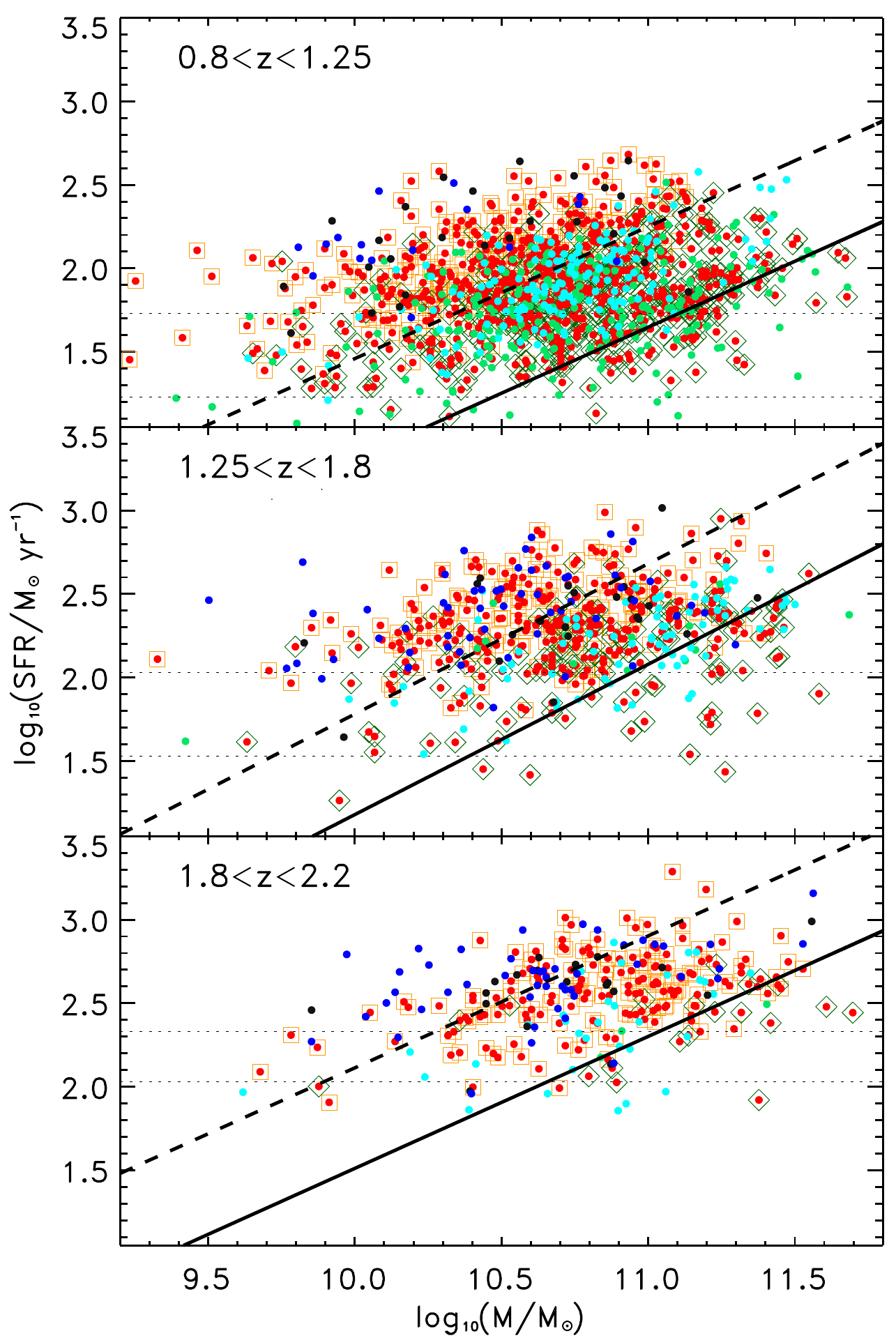}
\caption{
SFR versus stellar mass for our PEP 160-$\mu$m sources (green, {\tt spiral}; cyan, {\tt starburst}; red, {\tt SF-AGN}; magenta, {\tt AGN2}; blue, {\tt AGN1}), in three redshift bins (from $left$ to $right$): 0.8$<$$z$$<$1.25; 1.25$<$$z$$<$1.8; and 1.8$<$$z$$<$2.2. The relation known as main sequence is plotted as a solid line (from \citealt{elb07} in the lower redshift bin, rescaled as (1$+$$z$)$^{2.8}$ in the central bin and from \citealt{rod11} at $z$$\sim$2), while the dashed-line shows the same relation 0.6 dex higher, indicating the separation between MS and above MS objects adopted by \citet{rod11}.  
The horizontal dotted lines show the nominal SFR limits of the GOODS-S (lower) and COSMOS (upper) samples in the different redshift intervals. The orange open squares and the dark-green open diamonds mark the two sub-classes of {\tt SF-AGN} galaxies: the {\tt SF-AGN(SB)} and {\tt SF-AGN(Spiral)} respectively.
}
\label{fig:ssfrclass}
\end{figure}

Having computed stellar masses and SFRs for each source, we can check
how the PACS selected sources
populate the SFR--stellar mass plane and the
so called main sequence (MS) of star-forming galaxies, as
a function of redshift.
This relation (i.e. SFR versus stellar mass)  has been shown to be
quite tight in the local Universe (\citealt{peng10}, 2011) and well
established at  redshift $z$$\sim$1  (\citealt{elb07}) and up to
$z$$\sim$2 (\citealt{dad07}, \citealt{rod11}) and $z$$\sim$3
(\citealt{magd10}), with 
normalisation scaling as $\sim$(1$+$$z$)$^{2.8}$ out to $z$$\sim$2, as shown by \citet{sarg12} (see also \citealt{elb07}, \citealt{rod10b}, \citealt{pann09}, \citealt{kar11}).
At $z$$\sim$2 and $\sim$1.5 we assume a slope of 0.79 for the MS in the SFR versus stellar-mass plane (according to \citealt{rod11} and \citealt{sarg12}), while
at $z$$\sim$1 we assume a slope of 0.9, as found by \citet{elb07}, and we limit our investigation to the redshift range 0.8$<$$z$$<$2.2.

By combining UV and far-IR data, \citet{rod11}
re-evaluated the locus of the MS at $z$$\sim$2,
showing that objects lying a factor of 4 above the MS (in SFR) can be
considered as outliers with respect to the average locus
where smoothly star-forming galaxies spend most of their lives in a
secular and steady regime.
In that work, off-sequence sources (characterized by very high specific-SFRs) are
assumed to be in a starburst mode, and are found to contribute
only 2 per cent of mass-selected star-forming galaxies and to account for only
10 per cent of the cosmic SFR density at $z$$\sim$2 (\citealt{rod11}).

In order to check what kind of objects we could classify as on- and off-MS sources 
in our IR sample compared to previous findings, based either on IR or on optical surveys
(e.g. \citealt{rod11}; \citealt{sarg12}), we have splitted our sample into off-MS and on-MS.
For consistency with previous studies we have applied the same criterion as \citet{rod11} (0.6 dex above the MS) over the 
whole 0.8$<$$z$$<$2.2 redshift range, by using as a reference MS the relation found by \citet{rod11} at
$z$$\sim$2, scaled as described above at $z$$\sim$1.5, and the relation found by \citet{elb07} at $z$$\sim$1.
In Fig.~\ref{fig:ssfrclass}, we show 
the SFR versus stellar mass distributions 
in three redshift bins (0.8$<$$z$$<$1.25, 1.25$<$$z$$<$1.8 and 1.8$<$$z$$<$2.2),
for the PACS sources included in the computation of the luminosity
functions presented in this work.
The colour code marks the different SED-classes to which each source
belongs. 
We also report the typical loci of the MS at the various redshifts
(scaling as (1$+$$z$)$^{2.8}$, as mentioned above).
Details are given in the caption of the Figure. The typical far-IR selection bias 
(PACS-{\em Herschel} in this case) appears as an approximate horizontal 
SFR cut (\citealt{rod11}, \citealt{wuyt11}), shown as thin dotted line in Fig.~\ref{fig:ssfrclass}.
We note that the trends of mid/far-IR SEDs with offset from the main sequence observed in Fig.\ref{fig:ssfrclass} 
(and widely discussed in Section~\ref{sec_discuss}) are in good agreement with the results of \citet{elb11} and \citet{nor12}.

\begin{figure*}
\includegraphics[width=18cm]{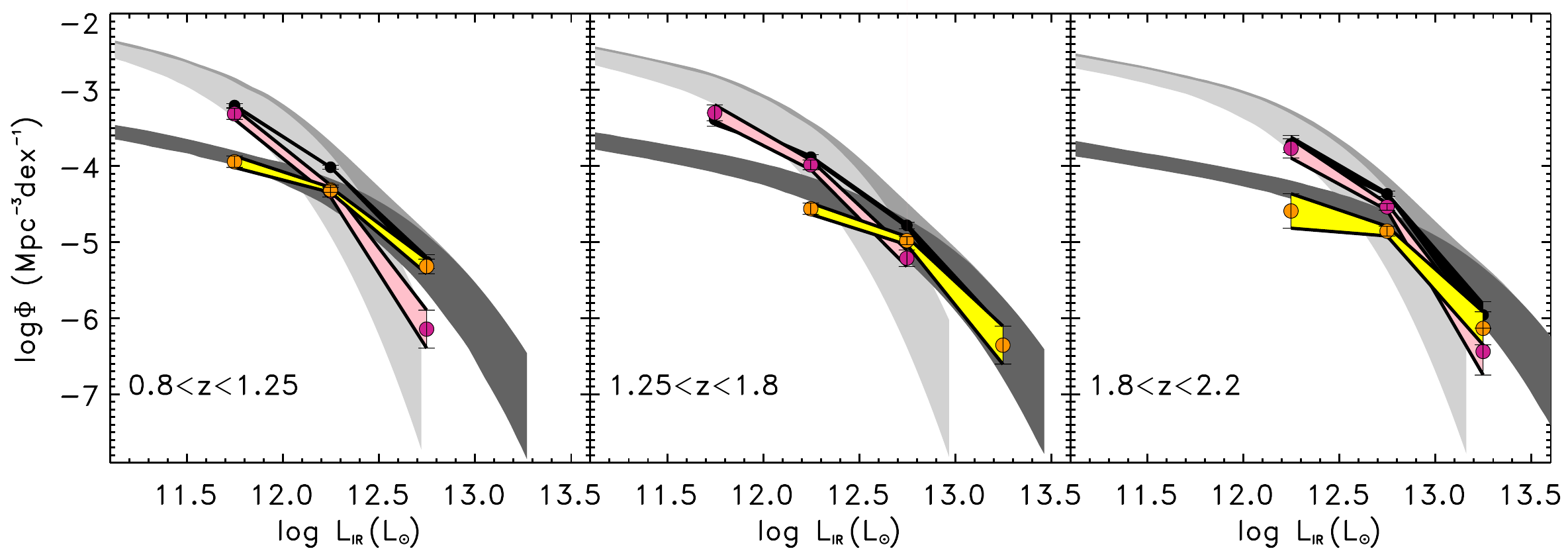} 
\caption{Contribution of MS ($\pm$1$\sigma$ uncertainty region: pink shaded area) and off-MS ($\pm$1$\sigma$ uncertainty region: yellow shaded area) galaxies to the total IR LF (black filled dots and area) in three different redshift bins. For comparison, the recent esimates of \citet{sarg12} in similar $z$-bins are shown as grey (total LF, in background), light grey (MS LF) and dark grey (off-MS LF) filled regions.}
\label{fig:lfms}
\end{figure*}

With the selection based on \citet{rod11} and overplotted in 
Fig.~\ref{fig:ssfrclass} (sources qualify as ``off-MS" if they lie more than 0.6 dex above the observed SFR-stellar mass relation)
, we can
compute the contribution of off-MS (also called ``starburst'' in the literature)
and on-MS (``steady star-formers'' in the literature) sources to the total IR LFs.
This is presented in Fig.~\ref{fig:lfms}, where the total IR LFs of on- and off-MS sources have been computed
independently for the three redshift bins.
The pink and yellow filled areas correspond to the $\pm$1$\sigma$
uncertainty regions of the total IR LFs estimated for the on-MS
and off-MS populations, respectively, while the black filled area
marks the global population.
Our results are compared with the recent estimates by \citet{sarg12}, a non-parametric approach
that is based on three basic observables: the redshift
evolution of the stellar mass function
for star-forming galaxies; the evolution of the sSFR of MS galaxies; and
a double-Gaussian decomposition of the sSFR distribution at fixed
stellar mass into a contribution (assumed
redshift- and mass-invariant) from MS and off-MS (i.e. starburst) activity.
The evolution of the two populations found both in our data and the \citet{sarg12} model are very similar. Both data and estimates indicate that the bright-end of the total IR LF is dominated by off-MS sources. However, although consistent within the uncertainties, the relative contribution of off-MS sources seems to be stronger in the \citet{sarg12} model than observed in the present computation, where we find that the bright-ends of the PEP IR LFs are more similarly populated by MS and off-MS sources (especially at $z$$\sim$2). 
This difference can be at least partly ascribed to the sharp cut we apply to separate MS from off-MS sources, while \citet{sarg12} model the off- and on-MS populations with two continuous log-normal distributions centred at 0.6 dex above the MS and exactly on the MS respectively, of which the one describing the ``starburst'' population has wings that extend into our on-MS selection region (hence attributing more sources to the starburst category than are selected in our off-MS class).
A better agreement between our data and the estimates of  \citet{sarg12} 
is found at the faint-end of the LFs that appear to be completely dominated by the ``normal'' MS galaxies at all redshifts (although the total and relative contributions at log($L_{\rm IR}$/L$_{\odot}$)$<$12 are slightly lower in the data than predicted by \citealt{sarg12}). Good agreement is also found with respect to the evolution of the cross-over luminosity (i.e. where the contributions from on- and off-MS sources are equal); in the model of \citet{sarg12} the cross-over luminosity 
simply shifts to higher luminosities and lower densities (according to the luminosity and density evolution considered).
Note that the model assumptions rely on results from different surveys, selected at different wavelengths, complete in mass and with good sampling of the MS. On the other hand, our selection is in SFR and our sources do not follow any clear sequence in stellar mass--SFR plane, because, except at the highest masses, the data are not deep enough to reach
well into the main sequence.
These different selection effects are likely to lead to some differences between our LFs and the estimates of \citet{sarg12}.

To quantify the relative contribution of the two populations,
our observed data have been fitted with a modified Schechter function,
in order to integrate them and compute their comoving number and luminosity densities
as functions of redshift (see next Section).

\section{Number Density and IR Luminosity Density}
\label{IRlumdens}
We derive the evolution of the comoving number and luminosity density (total, see Fig~\ref{figldensz}) of the PEP sources, either belonging to the different SED classes (Fig~\ref{figldensz}, $left$), to the on- and off-MS categories ($middle$) and to the different mass intervals ($right$), by integrating the total IR LF in the different redshift bins from $z$$\sim$0 to $z$$\sim$4. 
To compute the number (and IR luminosity) density, we integrate the Schechter functions that best reproduce the different populations/mass/sSFR-classes, down to log($L$/L$_{\odot}$)$=$8.  
We note that here we consider lower limits the number and luminosity densities at 3.0$<$$z$$<$4.2, since our LF estimate in that redshift bin is likely to be incomplete, as discussed in Section~\ref{sec_totLF}.
We find that the number density of the whole IR population is nearly constant in the $z$$=$0--1.2 redshift range (slightly increasing from $z$$\sim$0 to $z$$\sim$0.5), decreasing at $z$$>$1.2 (see $top$ panels of Fig.~\ref{figldensz}). 
When decomposing the number density according to the different SED classes, we observe that normal {\tt spiral} galaxies dominate the local density, with a smaller contribution also from the {\tt SF-AGN} population
and a negligible one from {\tt starburst}, {\tt AGN1} and {\tt AGN2}. The space density of {\tt spiral} galaxies decreases rapidly at $z$$\geq$0.5, while that of {\tt SF-AGN} stays nearly
constant at 0.5$\lsimeq$$z$$\lsimeq$2.5, largely dominating in that redshift range. {\tt Starburst} galaxies never dominate, while the number density of the bright AGN (both {\tt AGN1} and {\tt AGN2}) increases with redshift, from $\sim$10$^{-4}$ Mpc$^{-3}$ at $z$$\sim$0 to $\sim$1--2$\times$10$^{-3}$ Mpc$^{-3}$ at $z$$\sim$3. At higher redshifts the AGN population largely dominates the number density.\\
If the overall contribution to the IR luminosity density ($\rho_{\rm IR}$) from the AGN components of galaxies is small, $\rho_{\rm IR}$ can be considered as a proxy of the SFR density ($\rho_{\rm SFR}$). 
As a further check, we have therefore studied the evolution of the {\tt SF-AGN} population (which dominates the distribution of sources) by
dividing this class into {\tt SF-AGN(SB)} and {\tt SF-AGN(Spiral)} sub-classes and studying their evolution separately.
Indeed, we have found different evolutionary paths for the two populations, the former dominating at higher redshifts and showing a behaviour similar to that of AGN-dominated sources (e.g. {\tt AGN1} and {\tt AGN2}), the latter 
dominating at intermediate redshifts (between $z$$\sim$1 and 2), rising sharply from $z$$\sim$2 toward the lower redshifts and decreasing, while the {\tt spiral} population rises at $z$$\lsimeq$1.
These evolutionary trends, in terms of number and luminosity density, have been reported in Fig.~\ref{figldensz} as orange dot-dot-dot-dashed ({\tt SF-AGN(SB)})
and dark-green dashed ({\tt SF-AGN(Spiral)}) curves. \\
Galaxies following the SFR--mass relation are always dominant over the off-MS population, at all redshifts (although their space density decreases with increasing $z$, as well as the ``global'' number density), while the number density of the latter population remains nearly constant between $z$$\sim$0.8 and $z$$\sim$2.2. \\
In all the mass bins, the trends with redshift of the galaxy number densities are similar to the ``global'' one, decreasing at higher redshifts, although with slightly different slopes for the different mass intervals.
The number densities of
low mass galaxies (8.5$<$log($M$/M$_{\odot}$)$<$10), reported in the top~right panel of Fig.~\ref{figldensz}, have been computed by integrating the best-fitting modified Schechter function only to $z$$\sim$2, since data were not enough to derive reliable fits at higher redshifts. 
To this redshift, these sources outnumber the higher mass ones, although they fall steeply above $z$$\sim$1, when they reach about the same volume density of higher mass galaxies (10$<$log($M$/M$_{\odot}$)$<$11). Massive objects (log($M$/M$_{\odot}$)$>$11) never dominate (always below 5 per cent) the total number density.
\begin{figure}
\includegraphics[width=8.5cm]{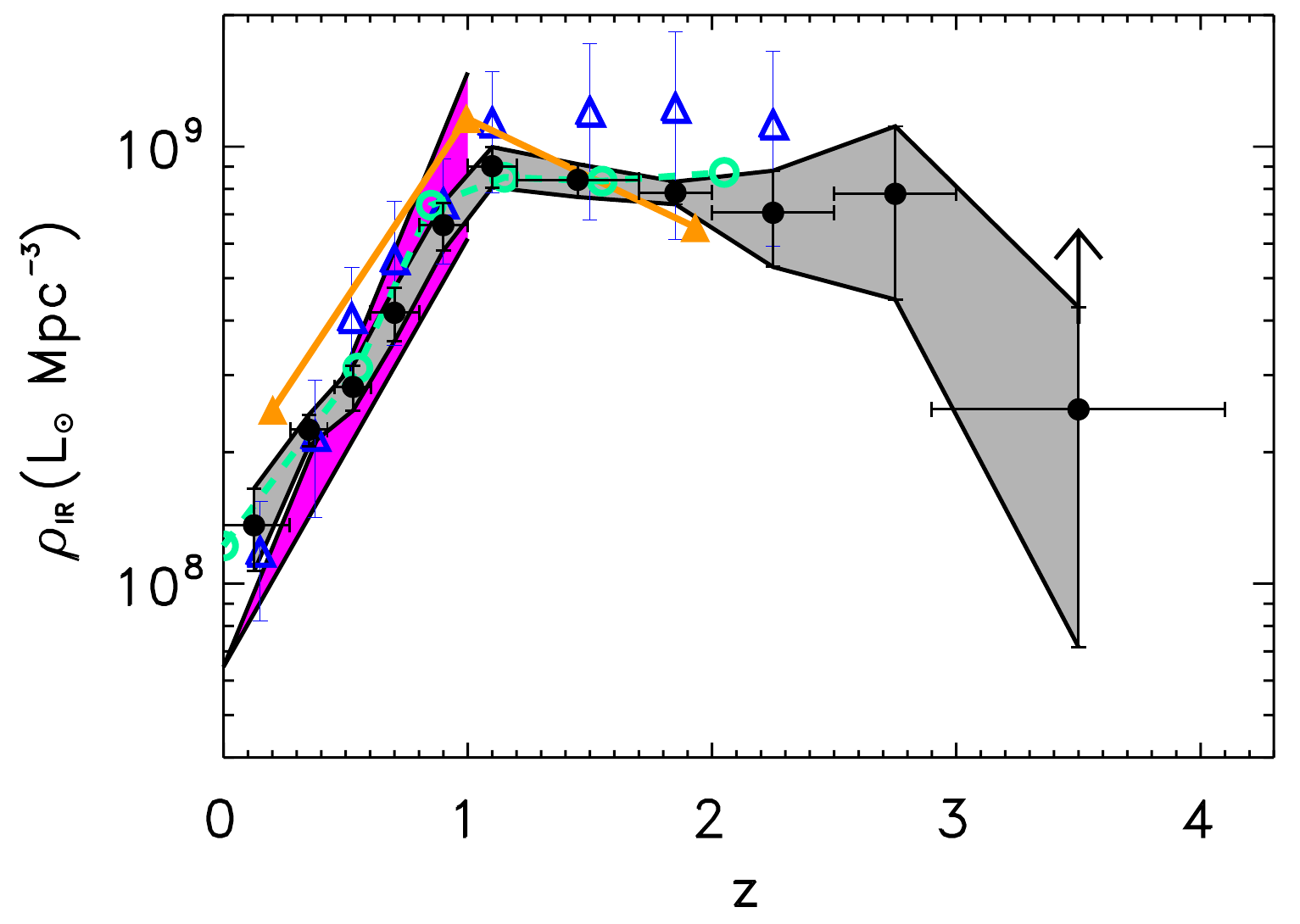}
\caption{Redshift evolution of the total IR luminosity density ($\rho_{\rm IR}$, obtained by integrating the Schechter functions that best reproduce the total IR LF down to log($L$/L$_{\odot}$)$=$8) to $z$$=$4. The results of integrating the best-fitting curve for our observed total IR LF in each $z$-bin are shown as black filled circles (the grey filled area is the $\pm$1$\sigma$ uncertainty locus) and compared with estimates from previous mid-IR surveys (magenta filled area, \citealt{lef05}; orange filled triangles, \citealt{cap07}; blue open triangles, \citealt{rod10a}; and green open circles, \citealt{mag11}). The upward pointing arrow in the highest-$z$ bin means that, due to the large fraction of photometric redshifts and the fact that the PEP selection might miss high-$z$ sources, our 3.0$<$$z$$<$4.2 $\rho_{IR}$ estimate is likely to be a lower limit.}
\label{figldz}
\end{figure}

The total IR LF allows a direct estimate of the total comoving IR luminosity density ($\rho_{\rm IR}$) as a function of $z$, which is a crucial tool for understanding galaxy formation and evolution. Although $\rho_{\rm IR}$ can be converted to a SFR density ($\rho_{\rm SFR}$) under the assumption that the SFR and $L_{\rm IR}$ quantities are connected by the \citet{kenn98} relation, before doing that we must be sure that the total IR luminosity is produced uniquely by star-formation, without contamination from an AGN. The SED decomposition and separation into AGN and SF contributions show a negligible contribution to $L_{\rm IR}$ ($<$10 per cent) from the AGN in most of the {\tt SF-AGN}, and a SF component dominating the far-IR even in the majority of more powerful AGN ({\tt AGN1} and {\tt AGN2}).
Here we prefer to speak in terms of $\rho_{\rm IR}$ rather than of $\rho_{\rm SFR}$, since, especially at high redshift -- where the AGN-dominated sources are more numerous -- the conversion of $\rho_{\rm IR}$ could represent only an upper limit to $\rho_{\rm SFR}$. Note, however, that since this population is never dominant
in our IR survey, we do not expect that contamination related to accretion activity occurring in these objects (mainly at
high-$z$) can significantly affect the results in terms of $\rho_{\rm SFR}$.

\begin{figure*}
\includegraphics[width=18cm]{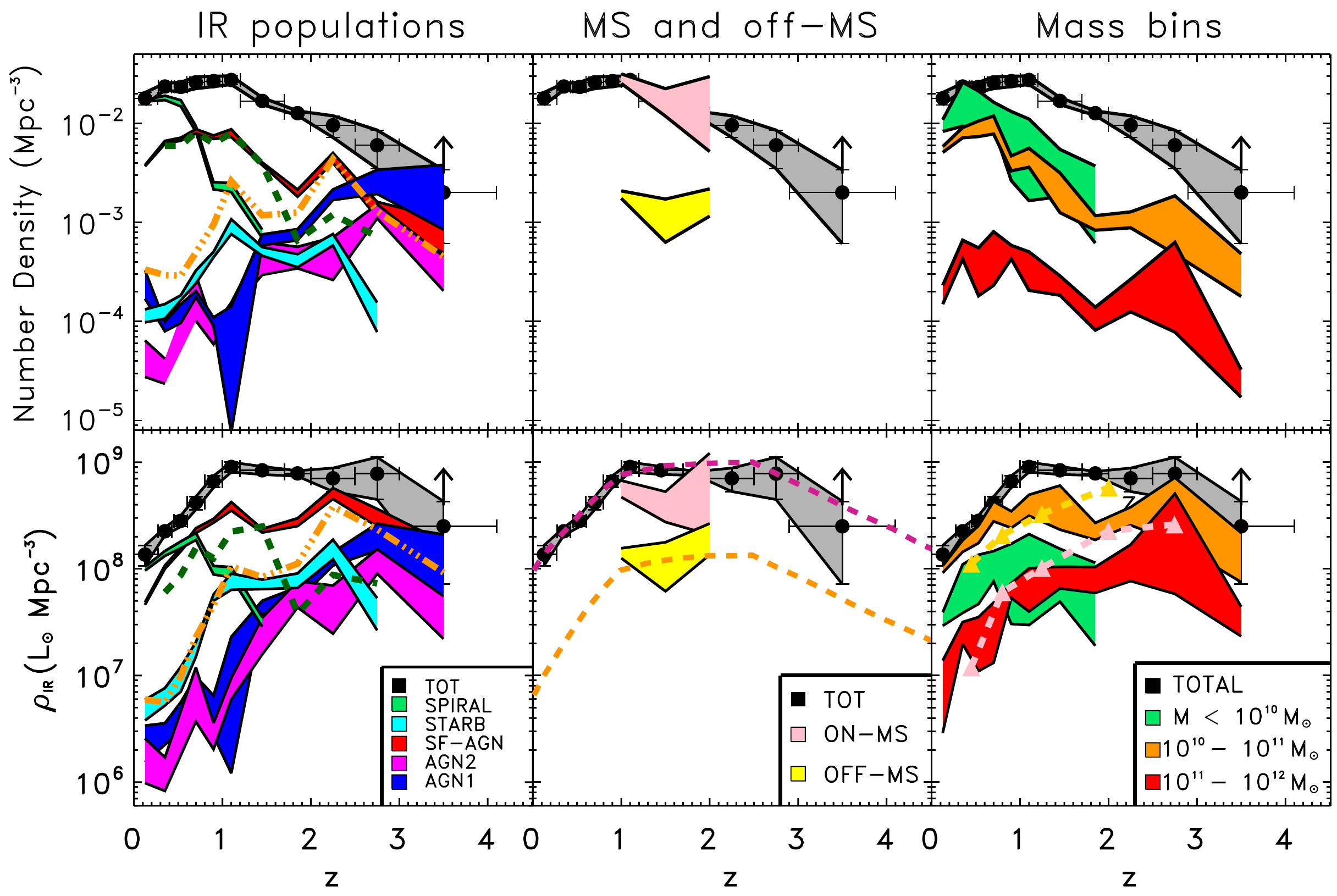}
\caption{$Top$: Evolution of the comoving number density of PEP sources up to $z$$\sim$4 (black filled circles with error-bars within the $\pm$1$\sigma$ uncertainty region, represented by the grey filled area). The upward pointing arrow in the highest-$z$ bin means that our 3.0$<$$z$$<$4.2 estimates are likely to be lower limits. $Bottom$: Redshift evolution of the total IR luminosity density to $z$$=$4. To compute the number and IR luminosity density, we integrate the Schechter functions that best reproduce the different populations/mass/sSFR-classes, down to log($L$/L$_{\odot}$)$=$8.  
The black filled circles and the grey dashed area in all the three panels represent our PEP derived $\rho_{\rm IR}$ and its $\pm$1$\sigma$ uncertainty region, as shown in Fig.~\ref{figldz}. In the {\em left} panels we show the number ($top$) and luminosity density ($bottom$) of the different IR populations (green filled area, {\tt spiral}; cyan, {\tt starburst}; red, {\tt SF-AGN}; magenta, {\tt AGN2}; and blue, {\tt AGN1}). The contribution of {\tt SF-AGN} sources, sub-divided on the basis of their SED resemblance to {\tt spiral} or {\tt starburst} templates, 
are shown by the dark-green dashed ({\tt SF-AGN(Spiral)}) and orange dot-dot-dot-dashed ({\tt SF-AGN(SB)}) lines, respectively.
In the {\em middle} panels we show the uncertainty regions of the relative contribution in number and luminosity density of the sources on- and off- the SFR--stellar mass MS (\citealt{elb07}; \citealt{dad07}), as pink and yellow filled areas, respectively. Our derivations have been compared to those of \citet{beth12} for on- and off-MS sources (converted from $\rho_{\rm SFR}$ to $\rho_{\rm IR}$ using the \citealt{kenn98} relation), which are represented by the purple and orange dashed lines respectively. In the {\em right} panels we show the relative contribution to the number and luminosity density of sources with different masses (green, 8.5$<$log($M$/M$_{\odot}$)$<$10; orange, 10$<$log($M$/M$_{\odot}$)$<$11; and red, 11$<$log($M$/M$_{\odot}$)$<$12). 
For comparison, in the $bottom~right$ panel we plot also the results of \citet{sant09} in the GOODS-S field in similar mass intervals (light-orange triangles and dashed line: 9.77$<$log($M$/M$_{\odot}$)$<$10.77; pink triangles and dashed line: log($M$/M$_{\odot}$)$>$10.77). }
\label{figldensz}
\end{figure*}

In Fig.~\ref{figldz} we show $\rho_{\rm IR}$ estimated from our total IR LF and compare it with results obtained from previous IR surveys ({\citealt{lef05}; \citealt{cap07}; \citealt{rod10a};  
\citealt{mag11}). In the common redshift intervals (0$\lsimeq$$z$$\lsimeq$2--2.5), we find very close agreement with
previous results based on IR data, especially with the \citet{mag11} derivation. 
As well as previous findings, $\rho_{\rm IR}$ from PEP shows the rapid rise from $z$$\sim$0 to $z$$\sim$1, followed by a
flattening at higher redshifts. The indications from our survey are that the intermediate redshift flattening is followed by a 
high redshift decline, which starts around $z$$\sim$3. From our data, $\rho_{\rm IR}$ evolves as (1$+$$z$)$^{3.0\pm0.2}$ up to $z$$\sim$1.1, as (1$+$$z$)$^{-0.3\pm0.1}$ from $z$$\sim$1.1 to $z$$\sim$2.8, then as (1$+$$z$)$^{-6.0\pm0.9}$ up to $z$$\sim$4.

In the bottom panels of Fig.~\ref{figldensz} we plot the different contributions to $\rho_{\rm IR}$ from the different SED populations (left), from the on- and off-MS sources (middle) and from the different mass intervals. We notice a predominance of {\tt spiral}--SED galaxies only at low redshifts ($z$$<$0.5--0.6), when {\tt SF-AGN} begin to dominate $\rho_{\rm IR}$ up to
$z$$\sim$2.5. The {\tt starburst} SED galaxies are never the prevalent population, although their contribution to $\rho_{\rm IR}$ increases rapidly from the local Universe to $z$$\sim$1, then keeps
nearly constant to $z$$\sim$2.5, to decrease at higher redshifts.
The {\tt SF-AGN(SB)} and {\tt SF-AGN(Spiral)} contributions to $\rho_{IR}$ show opposite trends, with the former sharply increasing towards the higher redshifts (dominating at $z$$>$2),
and the latter prevailing between $z$$\sim$1 and $\sim$2, then dropping at higher redshifts.
{\tt AGN1} and {\tt AGN2} start dominating
the IR luminosity density at $z$$\gsimeq$2.5, with their $\rho_{\rm IR}$ always rising from $z$$\sim$0, then remaining almost constant (or slightly decreasing) towards the higher redshifts. Note how these two populations of AGN evolve
similarly, as an indication of their same intrinsic nature and of the absence of any significant bias in the far-IR selection. 

The contributions to $\rho_{\rm IR}$ of MS and off-MS sources (pink and yellow filled regions, respectively, in the $bottom~middle$ panel of Fig.~\ref{figldensz})
stay nearly constant between $z$$\sim$0.8 and $z$$\sim$2.2. 
In particular, the off-MS sources contribution stays around 20 per cent of the total $\rho_{\rm IR}$ over the whole 0.8$<$$z$$<$2.2 redshift interval, showing no 
significant signs of increase (or decrease). 
Our results strengthen the role of MS sources (pink shaded regions)
in the build up of the stellar mass
in galaxies, at all cosmic epochs, with evidence that their role
is even increasing from $z$$\sim$2 to $z$$\sim$1: their number density changes by a factor of $\sim$2, 
while their luminosity/SFR density remains nearly constant.
The importance of the off-MS sources does not show any significant signs of decreasing at lower $z$, their relative (with respect to the total) number (luminosity) density passing from $\sim$9 per cent (22 per cent) at 
$z$$\sim$2 to $\sim$6 per cent (19 per cent) at $z$$\sim$1. 
These fractions are relatively different from those found by \citet{rod11} (off-MS galaxies represent only 2 per cent of mass-selected star-forming galaxies and account for only 10 per cent of the cosmic SFR density at $z$$\sim$2).
However, we must note that \citet{rod11} for their analysis combined far-IR-selected (i.e., SFR-selected) and near-IR-selected (i.e., $M_{\star}$-selected) star-forming samples, well defining the main sequence, while with our data (SFR-selected only), we are not able to observe any correlation between SFR and stellar mass (see Fig.~\ref{fig:ssfrclass}) and barely detect MS objects at $z$$\sim$2.\\    
Our results have been compared to the SFR densities (converted to $\rho_{\rm IR}$ using the \citealt{kenn98} relation) for on- and off-MS sources based on the \citet{sarg12} model  recently derived by \citet{beth12}. These are shown in Fig.~\ref{figldensz} as purple and orange dashed lines respectively. The predicted off-MS $\rho_{\rm SFR}$ agrees well with our
estimate in the common redshift range, while the predicted one for MS sources is higher than that derived from our data (especially at $z$$\sim$1.5). As already discussed regarding the comparison with the
\citet{sarg12} model, this discrepancy is likely to be ascribed to our difficulty to extrapolate the MS to low SFR values and to the different selection criteria used to separate MS from off-MS sources.

In the bottom~right panels of Fig.~\ref{figldensz}, we show the contribution of the different mass populations to the 
luminosity density as a function of redshift. 
Although we detect a similar steep increase of $\rho_{\rm IR}$ versus redshift at $z$$\lsimeq$1 in both low and high mass galaxies, 
the evolution in $\rho_{\rm IR}$ of galaxies with different masses is very different, 
reflecting the downsizing scenario, with $\rho_{\rm IR}$ peaking at higher redshift with increasing mass. 
Indeed, the IR luminosity density of intermediate-mass objects (log($M$/M$_{\odot}$)$=$10--11) always dominates, increasing up to $z$$\sim$1,
then remaining nearly constant at higher redshifts (at least up to $z$$\sim$2.8). 
The IR luminosity density of most massive objects increases even more rapidly with redshift (at $z$$=$2 it was $\sim$5 times higher than today)
and continues to grow up to $z$$=$3, where their contribution to $\rho_{\rm IR}$ is $\sim$30 per cent of the total and close to that of intermediate mass objects
(which contribute $\sim$60 per cent at $z$$>$3). 
We compare our results with those of \citet{sant09}, plotted in the figure as thick dashed lines (light orange: 9.77$<$log($M$/M$_{\odot}$)$<$10.77; pink: log($M$/M$_{\odot}$)$>$10.77; with a Chabrier
IMF used to determine our masses), showing very similar trends and values for both intermediate- and higher-mass galaxies. 
Our analysis of high mass galaxies extends up to $z$$\sim$4, finding that for the most massive galaxies $\rho_{\rm IR}$ continues to rise
even at $z$$\geq$2, with an apparent peak at $z=3$. This result confirms that the formation epoch of galaxies proceeded 
from high- to low-mass systems.

The values of $\rho_{\rm IR}$ in the different redshift intervals, either the total ones or the contributions from the different classes (SED, mass, sSFR), are reported in Table~9. 
\begin{figure*}
\includegraphics[width=12cm]{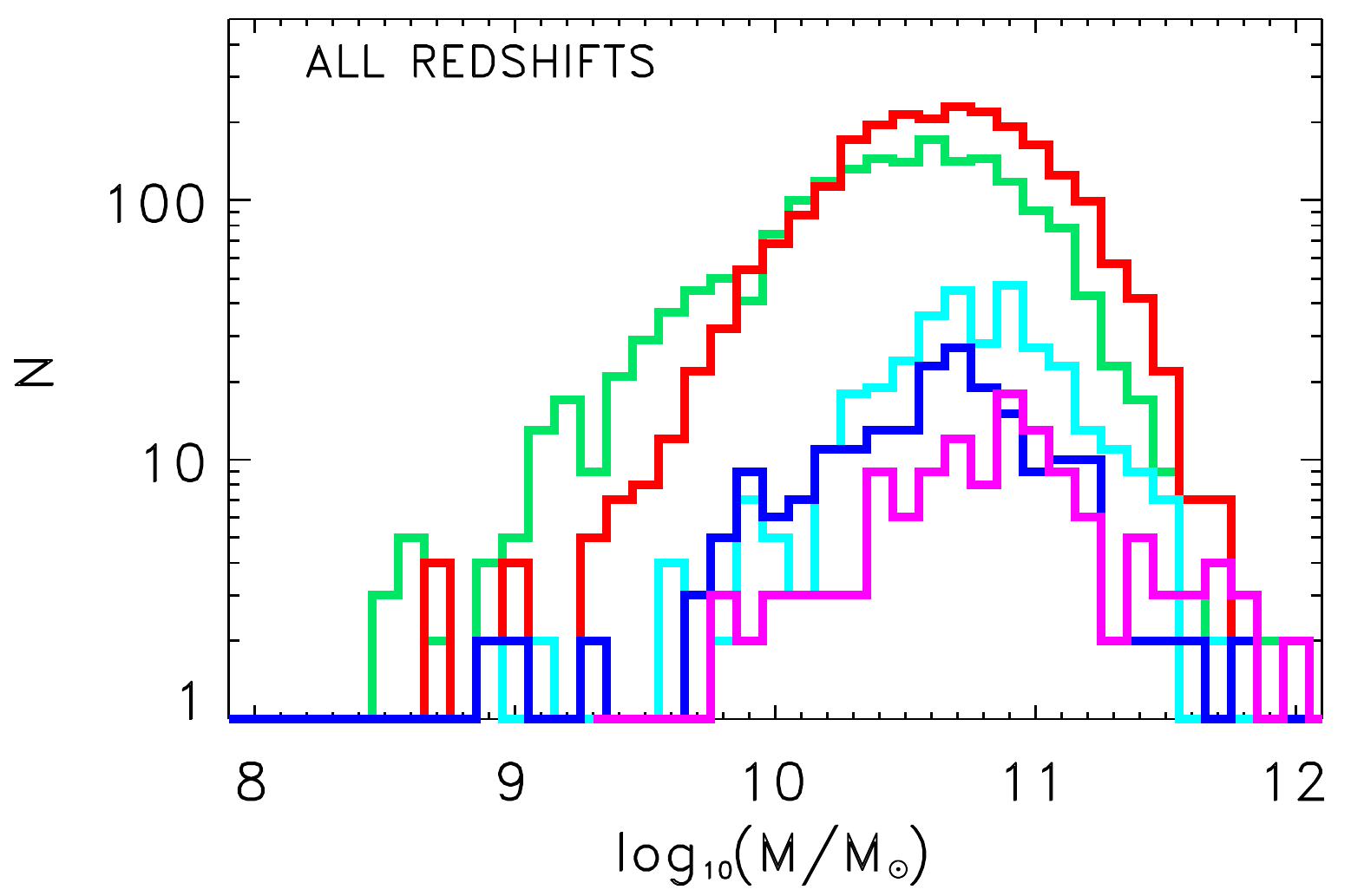}
\includegraphics[width=16cm]{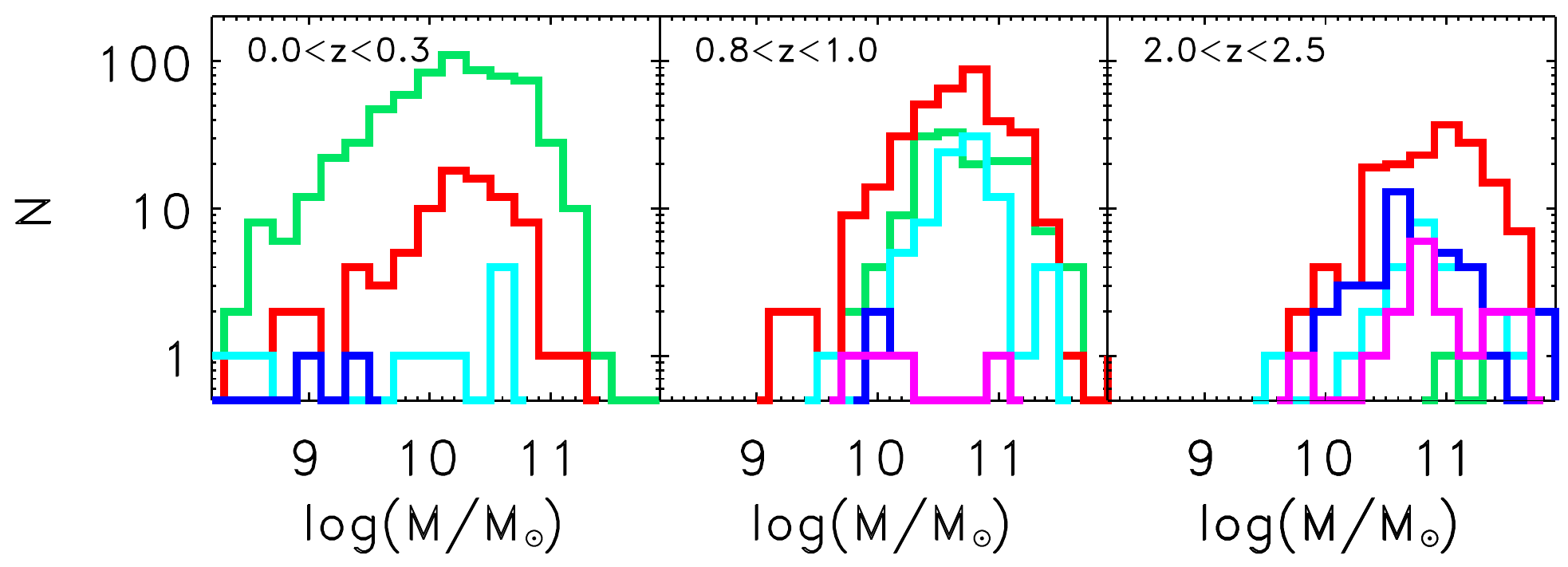}
\caption{Mass distribution of the PEP 160-$\mu$m sources in all the four PEP fields, with different colours corresponding to different SED-classes (green, {\tt spiral}; cyan, {\tt starburst}; red, {\tt SF-AGN}; magenta, {\tt AGN2}; and blue, {\tt AGN1}). The $top$ panel shows the mass distribution of the different IR populations at all redshifts, while the $bottom$ panels report the mass distribution in three representative redshift bins (0$<$$z$$<$0.3, 0.8$<$$z$$<$1.0 and 2.0$<$$z$$<$2.5).}
\label{fig:mass}
\end{figure*}

\section{Discussion}
\label{sec_discuss}
In the previous sections we have discussed the different evolutionary behaviour of different classes of sources, either divided by SED-type, mass or sSFR. In this section we will try to 
understand ``who is who'', discussing which populations are mainly on- or off-MS, which have the larger (smaller) masses, and how and if the relative contributions of these populations
vary with redshift.  

\begin{figure*}
\includegraphics[width=17cm]{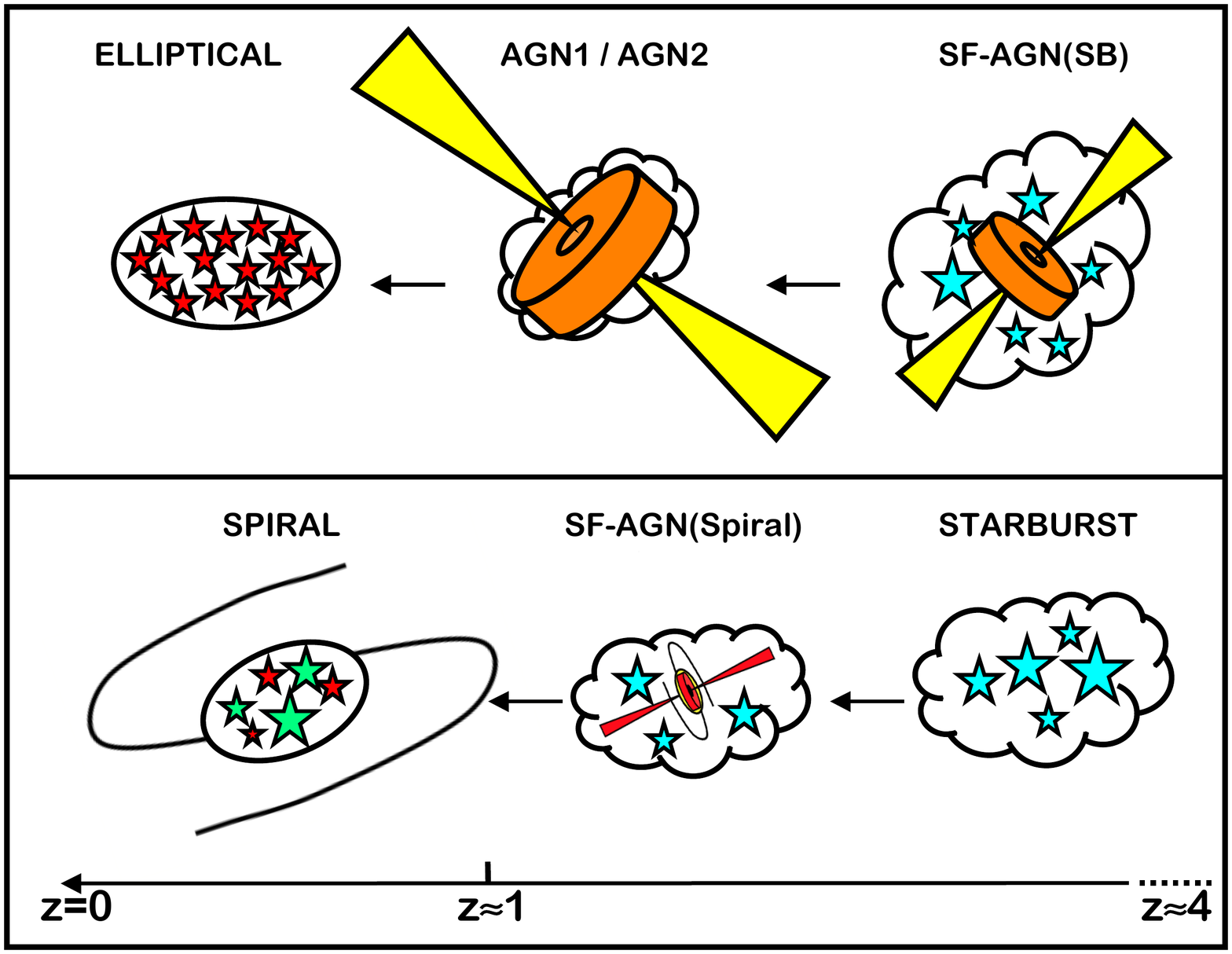}
\caption{A cartoon showing a possible evolutionary scenario involving AGN and star-forming galaxies and leading to the formation, on the one hand, of local elliptical galaxies, on the other hand, of local spiral galaxies. 
{\em Top:} a strong starburst with a growing BH inside ({\tt SF-AGN(SB)}) transforms into an AGN (either {\tt AGN1} or {\tt AGN2}, depending on orientation) when the BH mass and AGN luminosity have grown enough and the star formation is likely to be quenched by feedback processes. Thereafter the galaxy evolves passively toward a local elliptical. {\em Bottom:} the initial moderate starburst ({\tt starburst}, lasting a typical time of $\sim$few 10$^8$ yrs) transforms in a less intense starburst also fueling a low-luminosity AGN ({\tt SF-AGN(Spiral)}), which starts to reveal itself when the starburst activity is fading. Once the AGN is triggered, it heats the dust resulting in an increase in the mid-infrared luminosity (flattening of the SED between 3 and 8\,$\mu$m). The {\tt SF-AGN(Spiral)} system could last about 7$\times$ longer than the pure starburst phase before becoming a steady spiral galaxy ({\tt spiral}).}
\label{figcartoon}
\end{figure*}
From Fig.~\ref{fig:ssfrclass} we note that the off-MS sources are dominated by galaxies with AGN-type SEDs. In the lower redshift bin considered, MS sources are mainly {\tt spirals} and {\tt SF-AGN(Spiral)}, with a small tail of these populations contributing also to the off-MS. {\tt AGN1} and {\tt AGN2} are prevalently 0.6~dex above the MS (off-MS). 
Although the mass estimate of {\tt AGN1} suffers of the largest uncertainties, our result suggests that the off-MS population is largely constituted by sources containing an AGN, with a
higher fraction of AGN-dominated objects at higher $z$. This should indicate that the major merger episode likely associated with what is considered a different mode of star-formation (e.g. \citealt{wuyt11} finds that off-MS galaxies are very compact from their S\'ersic index, and quite likely mergers), could also trigger an intense AGN activity, whose presence strongly influences the SFR within the host galaxy.
While AGN show preferentially high sSFRs, in our sample they do not seem to prefer higher-mass systems (see Fig.~\ref{fig:mass}). However, this may again be due to our 
SFR selection, so we miss normal main sequence galaxies especially at low mass (cf. \citealt{rod11})
Therefore, our AGN are still in an intense SF phase, where they are still increasing their stellar mass by actively forming stars. 
Due to the far-IR selection, in fact, we miss the further phase of quiescent, more massive population, with the SF totally quenched and all the stellar mass already formed and in place. 
Moreover, our result is not in conflict with the consensus in the literature (i.e. that X-ray AGN above a certain X-ray luminosity do prefer massive galaxies), since our classification is not in AGN luminosity cut, but more a 
cut in $L_{\rm AGN}$/ $L_{\rm SF}$ ratio. The recent result of \citet{mul12}, that $L_{\rm X}$/$L_{\rm IR}$ (which is a proxy of $L_{\rm AGN}$/$L_{\rm SF}$) is on average almost independent of stellar mass, implies that our selected AGN are not necessarily 
hosted by very massive galaxies, as indeed we find.  \\
Our results seem to confirm those of \citet{sant12}, finding evidence of a higher average SF activity in AGN hosts with respect to inactive galaxies and a more pronounced level of SF enhancement in the hosts of luminous AGN (i.e. our {\tt AGN1} and {\tt AGN2}).
Most of the {\tt starburst} galaxies (i.e. with SEDs fitted by local ``starburst'' templates) are classified on-MS at any redshifts, though they are always in the region between the MS and the threshold. The {\tt starburst} population seems to 
enhance its sSFR from $z$$\sim$2 to $z$$\sim$1, occupying the region around the MS at higher $z$ and shifting to higher SFRs (peaking at $\sim$2$\times$ the MS) at lower $z$. The bulk of the {\tt spiral} population remains around the MS at all redshifts, although they almost disappear 
from our sample at $z$$\sim$2. 
From Fig.~\ref{fig:ssfrclass} we also note that, while the {\tt SF-AGN} population as a whole occupies both the loci of MS and off-MS sources, when divided into {\tt SF-AGN(SB)} and {\tt SF-AGN(Spiral)},
it shows a very clear segregation (suggesting a different mode of star-formation for the 
two sub-classes). In fact, {\tt SF-AGN(SB)} are mainly concentrated in the off-MS region, while the bulk of the {\tt SF-AGN(Spiral)} is on-MS. 

If we mix together all the ``ingredients'' presented above, we can try to give a global interpretation to our results in terms of galaxy evolution. 
In agreement with other {\em Herschel} findings (i.e. \citealt{shao10}; \citealt{lapi11}; \citealt{sant12}; \citealt{ros12}; \citealt{page12}; \citealt{mul12}), we propose the following twofold evolutionary scenario for galaxies and AGN (sketched in a cartoon in Fig.~\ref{figcartoon}): 
\begin{itemize}
\item On the one hand, we observe that the sources with AGN-dominated SEDs (either {\tt AGN1} or {\tt AGN2}) and those with a starburst-like SED, but containing a non-dominant AGN ({\tt SF-AGN(SB)}) have a peak in number and luminosity density at $z$$\sim$2--2.5, dominating at higher redshifts and rapidly decreasing at lower z.
The evolution of {\tt AGN1} and {\tt AGN2}, both in luminosity and in density, is very similar, suggesting the same nature for type 1 and 2 AGN, and the unbiased power of observations in the 
far-IR band towards orientation/obscuration (affecting both optical and X-ray observations). While dominating the mid-IR part of the SED of these objects, the AGN is not able to explain the 
high observed far-IR emission, which is mostly powered by star-formation (as for the {\tt SF-AGN(SB)} population, where the AGN never dominates the energetic output, probably due to dust-obscuration). 
The hosts of these AGN appear to form stars in a very efficient way, placing a large fraction of them above the known
stellar mass--SFR MS (see Fig.~\ref{fig:ssfrclass}). {\tt AGN1}, {\tt AGN2} and {\tt SF-AGN(SB)} are likely the progenitors of the elliptical galaxies observed nowadays in optical and near-IR surveys, forming through an intense burst of star-formation occurring during major mergers or in dense nuclear star-forming regions (\citealt{gran01}; \citealt{dad10}; \citealt{wuyt11}), then followed by a phase of nuclear activity during which their SMBHs grow (i.e. \citealt{hopk08a}, 2008b; \citealt{lapi11}). It is generally agreed that the SMBHs and their host galaxies are tightly related, with major-mergers being considered the likely process responsible for transporting large amount of gas towards the centre of the merging system, feeding the SMBH and triggering the SF activity. After the intense starburst phase, the AGN are believed to suppress the SF (i.e. \citealt{dimat05}), so that the remnant quickly evolves to a red massive spheroid. This picture is strongly supported by the recent {\em Herschel} results  from PEP (\citealt{ros12}), HerMES (\citealt{page12}; \citealt{har12}) and H-ATLAS (\citealt{lapi11}) Surveys. 
The latter work, in particular, has shown that the bright-end of the IR LFs and counts at high redshift ($>$1.5) are consistent with the picture (e.g. \citealt{gran01}) predicting the presence of a population of strongly-obscured, star-forming galaxies with SED appreciably different from those of the local starbursts. X-ray and mid-IR spectroscopic observations (e.g. \citealt{alex05}, \citealt{val07}) of sub-mm selected sources have revealed in many of them the presence of a growing SMBH, powering an obscured AGNs. In this framework, the most likely evolutionary path envisages first a {\tt SF-AGN(SB)} phase (star-forming galaxy with a growing SMBH inside), then an AGN-dominated phase, and finally the formation of an elliptical galaxy in passive evolution.
\item On the other hand, we observe that a significant fraction of our IR selected sources is constituted by moderately star-forming galaxies characterised by an SED similar to that of spiral galaxies, but also suggesting the presence of a low-luminosity AGN ({\tt SF-AGN(Spiral)}), best-fitted by local Seyfert 1.8/2 templates. The bulk of these objects and their principal contribution to $\rho_{IR}$ are
at intermediate redshifts (1$\lsimeq$$z$$\lsimeq$2), while they decrease between $z$$\sim$1 and $z$$=$0, as the {\tt spiral} population rises. Most of the {\tt SF-AGN(Spiral)-}, {\tt spiral-} and {\tt starburst-}SED
galaxies occupy the region around the SFR-stellar mass MS at any redshifts, suggesting a steady mode of star-formation  rather than a ``starburst'' one for these three populations, whose evolution is likely connected.
Indeed, \citet{mul12} have recently shown that at least up to $z$$\sim$2, SMBHs have grown together with their host galaxies in star-forming galaxies, irrespective of host galaxy mass and triggering mechanism.
Given the number densities (Fig.~\ref{figldensz}), evolutionary trends (Fig.~\ref{fig:lphievol}), SFRs and masses (Figs.~\ref{fig:ssfrclass} and \ref{fig:mass}) of the {\tt SF-AGN(Spiral)}, {\tt starburst} and {\tt spiral} populations, we suggest that these three classes of objects might constitute different phases in the life of a galaxy undergoing secular evolution. 
The gas in moderate {\tt starburst} galaxies, undergoing a burst of enhanced SF due either to gravitational interactions or disk instabilities (typical burst duration time of the order of a few 10$^{8}$ yr), might also fuel a low-luminosity AGN, which starts to reveal itself when the starburst activity is fading. 
Given the relatively high stellar masses found for the bulk of the {\tt SF-AGN(Spiral)} (log($M$/M$_{\odot}$)$\sim$10--11, see Fig.~\ref{fig:ssfrclass}), and the almost constant $M_{\rm BH}$/M$_{\star}$ ratio (of $\sim$1$-$2$\times$10$^{-3}$) recently suggested by \citet{mul12} for all the 0.5$<$$z$$<$2.5 star-forming galaxies, they are likely to contain relatively massive BHs ($M_{\rm BH}$$\sim$10$^7$--10$^8$ M$_{\odot}$). They can therefore have either low values of their radiative efficiency or low values of their accretion mass rate $\dot{m}$ ($\dot{m}$$=$$\dot{M}/\dot{M}_{\rm Edd}$$<$0.01).
AGN with low radiative efficiency or low accretion mass rates are generally called radiatively inefficient accretion flows (RIAFs -- \citealt{nar95}--, which include also the advection-dominated accretion flow, ADAF -- \citealt{nar94}). Low $\dot{m}$ AGN are more difficult to detect, since they often are less luminous than their hosts. Because of this, the nuclear emission is diluted by the host galaxy's emission and many of them are likely to be classified as ``normal galaxies'' in most surveys if the AGN luminosity is less than that of the host (e.g. \citealt{hopk09}). 
Given the relative number densities of the {\tt starburst} and {\tt SF-AGN(Spiral)} populations at 1$<$$z$$<$2, we hypothesize a typical duration 
of the {\tt SF-AGN(Spiral)} phase about 7 times longer than the {\tt starburst} one (typical burst duration $\sim$10$^8$ yr). Then, after a typical time of $\sim$7$\times$10$^8$ yr, the AGN activity stops and these 
objects, whose number density decreases at $z$$\lsimeq$1, are likely to become steady {\tt spiral} galaxies (rapidly increasing between $z$$\sim$1 and $z$$=$0) at lower redshifts.

\end{itemize}

\section{Conclusions}
\label{sec_concl}
We have used the 70-, 100-, 160-, 250-, 350- and 500-$\mu$m data from the cosmological guaranteed time {\em Herschel} surveys, PEP and HerMES, in the GOODS-S and -N, ECDFS and COSMOS, to characterise the 
evolution of the IR luminosity function and luminosity density of PACS selected sources across the redshift range 0$\lsimeq$$z$$\lsimeq$4. Evolution is well constrained by our data up to $z$$\sim$3, strong hints
of evolution are derived at 3$<$$z$$<$4.
In the present work we have:
   \begin{enumerate}
       \item completely characterised the multi-wavelength SEDs of the PEP sources by performing a detailed SED-fitting analysis and comparison with known template library of IR populations. Sources have been classified, based on their broad-band SEDs, in five main classes: {\tt spiral}, {\tt starburst}, {\tt SF-AGN}, {\tt AGN1} and {\tt AGN2}.
      \item computed the rest-frame LFs at 35, 60 and 90\,$\mu$m up to $z$$\sim$4 from the 70-, 100- and 160-$\mu$m selected samples respectively.
      \item integrated the SEDs over $\lambda_{\rm rest}$$=$8--1000\,$\mu$m and computed the total IR LF up to $z$$\sim$4 and studied its evolution with redshift,
      finding strong luminosity evolution $\propto$(1$+$$z$)$^{3.55\pm0.10}$ up to $z$$\sim$1.85, and $\propto$(1$+$$z$)$^{1.62\pm0.51}$ between $z$$\sim$1.85 and $z$$\sim$4, combined 
      with a negative density evolution $\propto$(1$+$$z$)$^{-0.57\pm0.22}$ up to $z$$\sim$1.1 and $\propto$(1$+$$z$)$^{-3.92\pm0.34}$ at $z$$>$1.1 and up to $z$$\sim$4.
      \item derived the evolution of the comoving total IR luminosity density, which is found to increase as (1$+$$z$)$^{3.0\pm0.2}$ up to $z$$\sim$1.1, then to remain nearly constant (decrease as (1$+$$z$)$^{-0.3\pm0.1}$) from $z$$\sim$1.1 to $z$$\sim$2.8, and to decrease as (1$+$$z$)$^{-6.0\pm0.9}$ up to $z$$\sim$4.
      \item found that the evolution derived for the global IR LF is indeed a combination of different evolutionary paths: the IR population does not evolve all together ``as a whole'', as is often assumed in the literature, but is composed of different galaxy classes evolving differently: the {\tt spiral}--SED galaxies dominate $\rho_{\rm IR}$ only at low redshifts ($z$$\lsimeq$0.5--0.6), then {\tt SF-AGN} dominate up to $z$$\sim$2.5, while {\tt AGN1} and {\tt AGN2} start dominating the IR luminosity density only at $z$$\gsimeq$2.5.
      \item derived the relative contribution to $\rho_{\rm IR}$ of MS and off-MS sources, which keep nearly constant between $z$$\sim$0.8 and $z$$\sim$2.2, with the MS population always dominating. The contribution to $\rho_{\rm IR}$ of the off-MS sources shows no significant signs of increase with $z$ (from $\sim$19 per cent at $z$$\sim$0.8--1.25 to $\gsimeq$22 per cent at $z$$\sim$1.8--2.2).
      \item derived very different evolutionary behaviour in terms of different contributions to $\rho_{\rm IR}$, for galaxies with different masses, reflecting the downsizing scenario ($\rho_{\rm IR}$ peaks at higher redshift with increasing mass). Intermediate-mass objects (log($M$/M$_{\odot}$)$=$10--11) always dominate the IR luminosity density, increasing with redshift up to $z$$\sim$1,
then remaining nearly constant at higher redshifts (at least up to $z$$\sim$2.8), while the contribution of most massive objects increases even more rapidly with $z$ (at $z$$\sim$2 it was $\sim$5 times higher than today) and continues to grow up to $z$$\sim$3.
    \item described a possible twofold evolutionary scenario for IR sources: on the one hand, {\tt AGN1} and {\tt AGN2}, representing the same population, after an intense starburst phase (due to a major merging event,  appearing as {\tt SF-AGN(SB)} SED galaxies), suppress the SF (and shine as AGN) and evolve to red massive spheroids; on the other hand, the {\tt SF-AGN(Spiral)} galaxies represent a phase in the life of a star-forming galaxy, following a moderate burst of SF ({\tt starburst}, with SEDs like those of local starburst galaxies) and preceding the formation of a steady {\tt spiral} galaxy as we observed in the local Universe. 

            \end{enumerate}

\section*{Acknowledgments}
PACS has been developed by a consortium of institutes led by MPE (Germany) and including: UVIE
(Austria); KU Leuven, CSL, IMEC (Belgium); CEA, LAM (France); MPIA (Germany); INAF-IFSI/
OAA/OAP/OAT, LENS, SISSA (Italy); and IAC (Spain). This development has been supported by the
funding agencies BMVIT (Austria), ESA-PRODEX (Belgium), CEA/CNES (France), DLR (Germany),
ASI/INAF (Italy), and CICYT/MCYT (Spain). 
SPIRE has been developed by a consortium of institutes led
by Cardiff Univ. (UK) and including: Univ. Lethbridge (Canada);
NAOC (China); CEA, LAM (France); IFSI, Univ. Padua (Italy);
IAC (Spain); Stockholm Observatory (Sweden); Imperial College
London, RAL, UCL-MSSL, UKATC, Univ. Sussex (UK); and Caltech,
JPL, NHSC, Univ. Colorado (USA). This development has been
supported by national funding agencies: CSA (Canada); NAOC
(China); CEA, CNES, CNRS (France); ASI (Italy); MCINN (Spain);
SNSB (Sweden); STFC, UKSA (UK); and NASA (USA).
CG and FP acknowledge financial contribution from the contracts 
PRIN-INAF 1.06.09.05 and ASI-INAF I\/005\/07/1 and I\/005\/11\/0.
PM thanks the University of Trieste for the grant FRA2009.

\bsp

\begin{figure*}
\begin{minipage}{180mm}
\includegraphics[width=17cm]{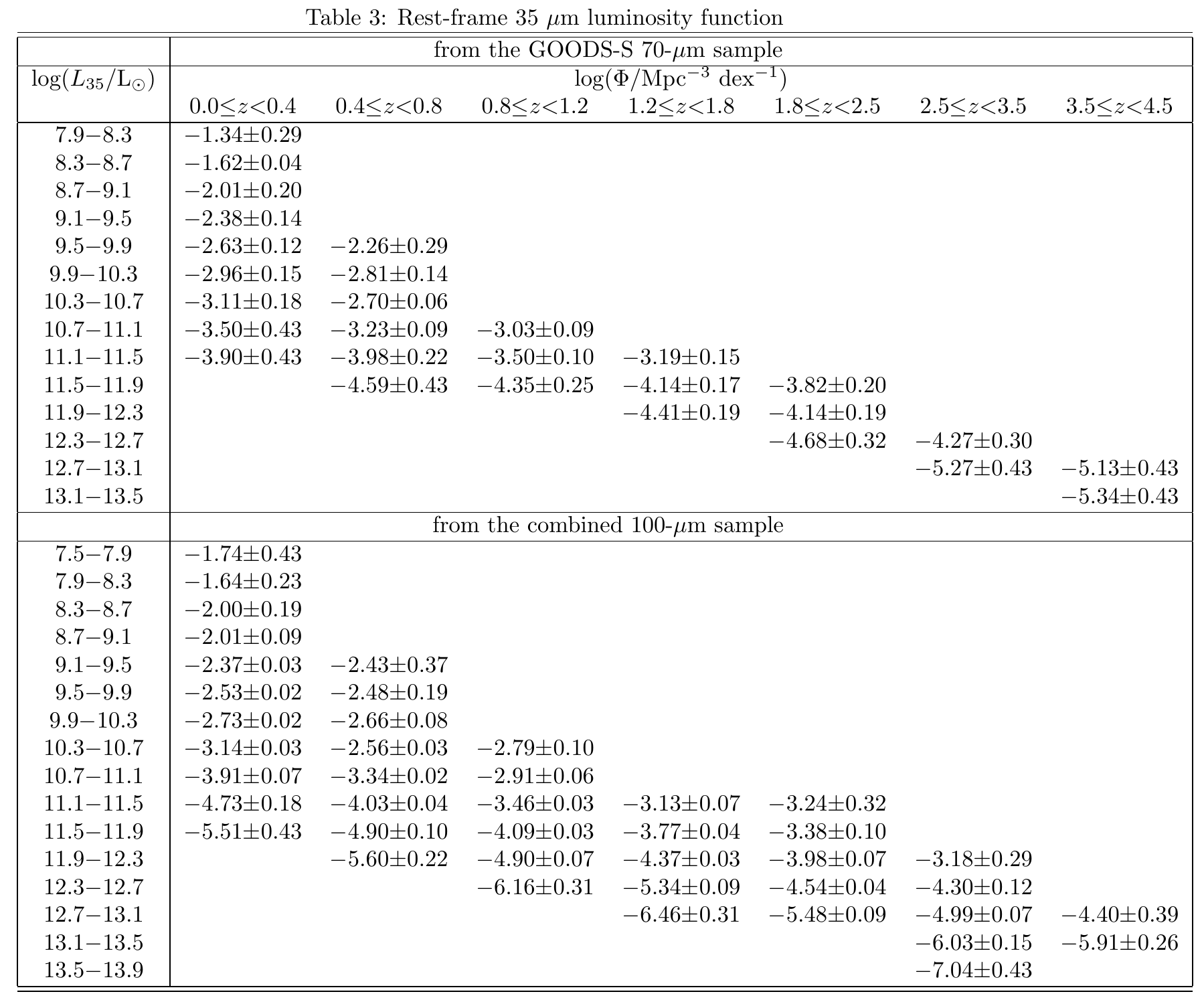}
\label{TabLF35um}
\end{minipage}
\end{figure*}
\begin{figure*}
\begin{minipage}{180mm}
\includegraphics[width=17cm]{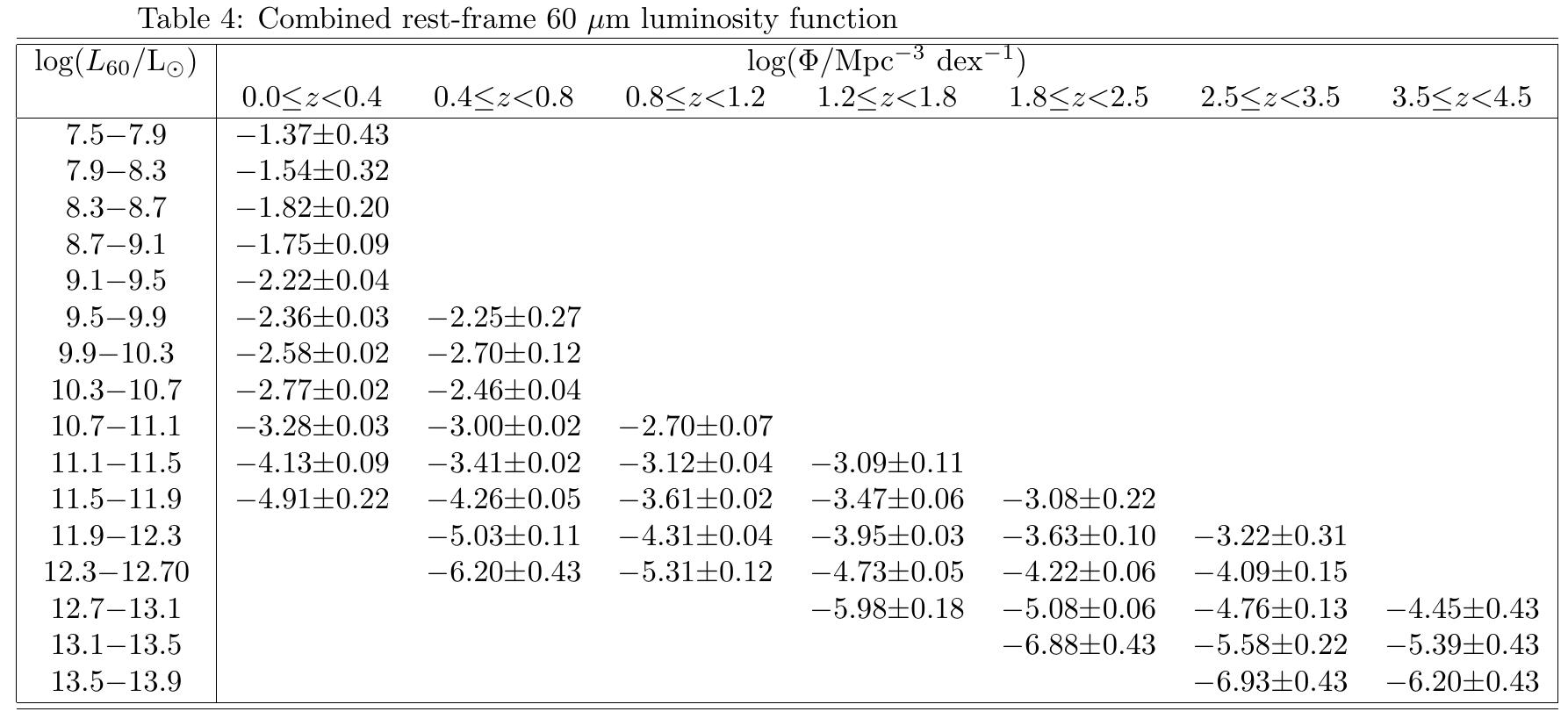}
\label{TabLF60um}
\end{minipage}
\end{figure*}
\begin{figure*}
\begin{minipage}{180mm}
\includegraphics[width=17cm]{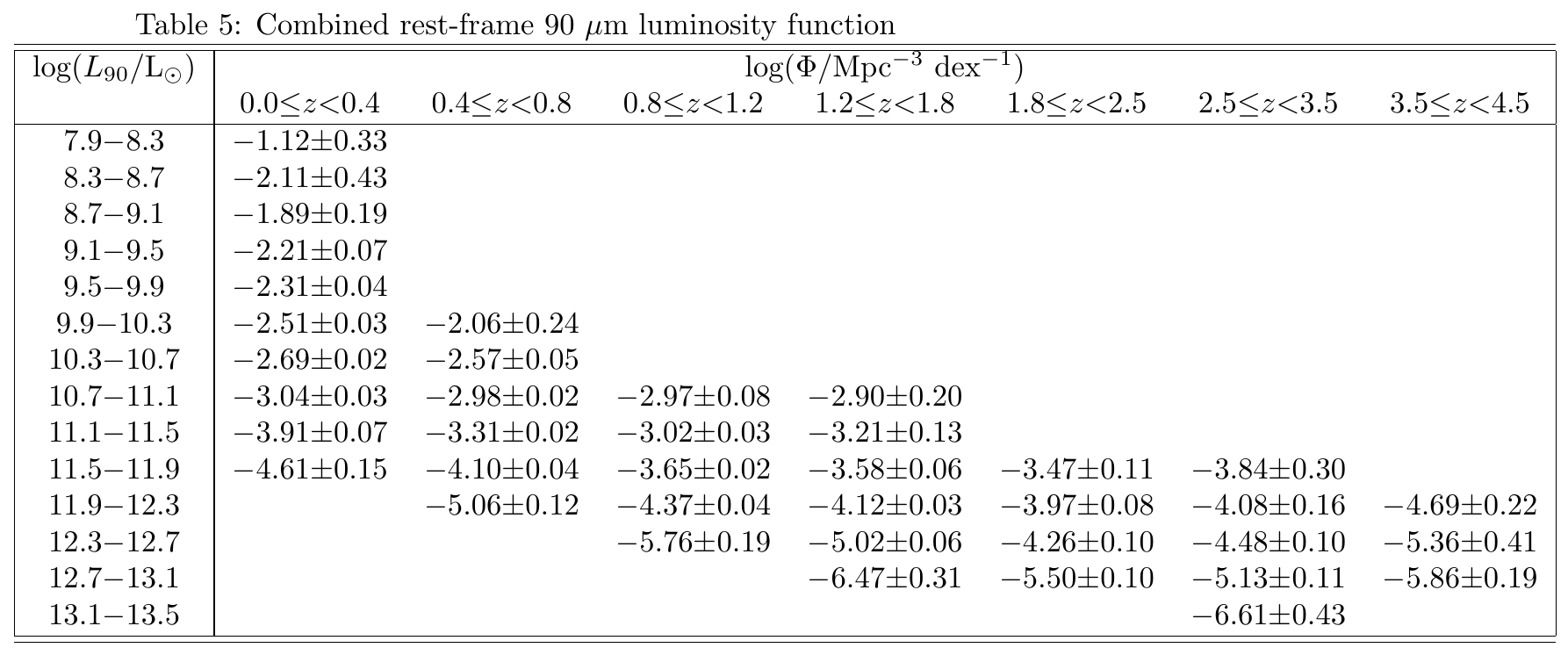}
\label{TabLF90um}
\end{minipage}
\end{figure*}

\begin{figure*}
\begin{minipage}{185mm}
\includegraphics[width=18cm]{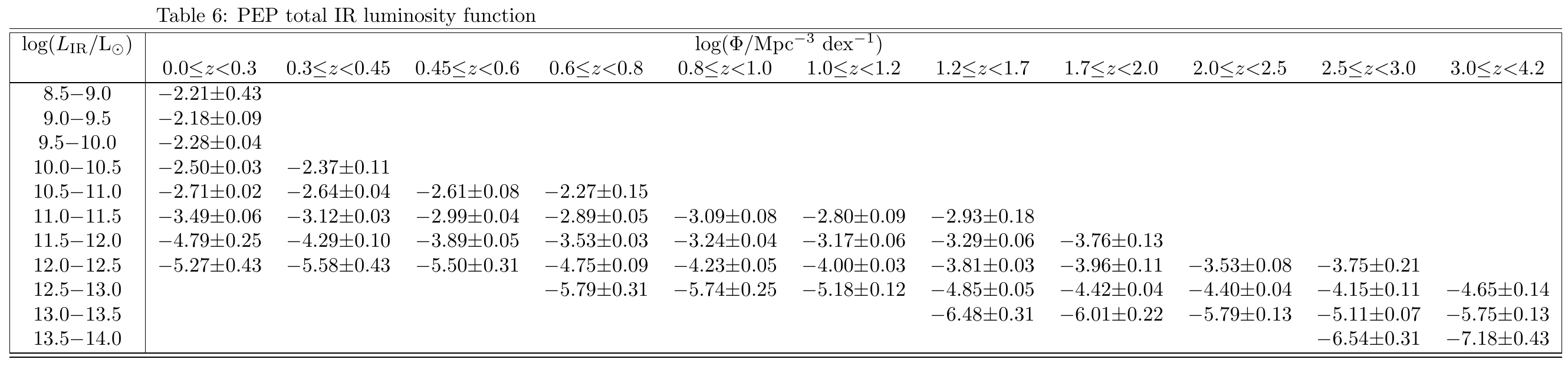}
\label{TabLFtot}
\end{minipage}
\end{figure*}

\begin{figure*}
\begin{minipage}{185mm}
\includegraphics[width=18cm]{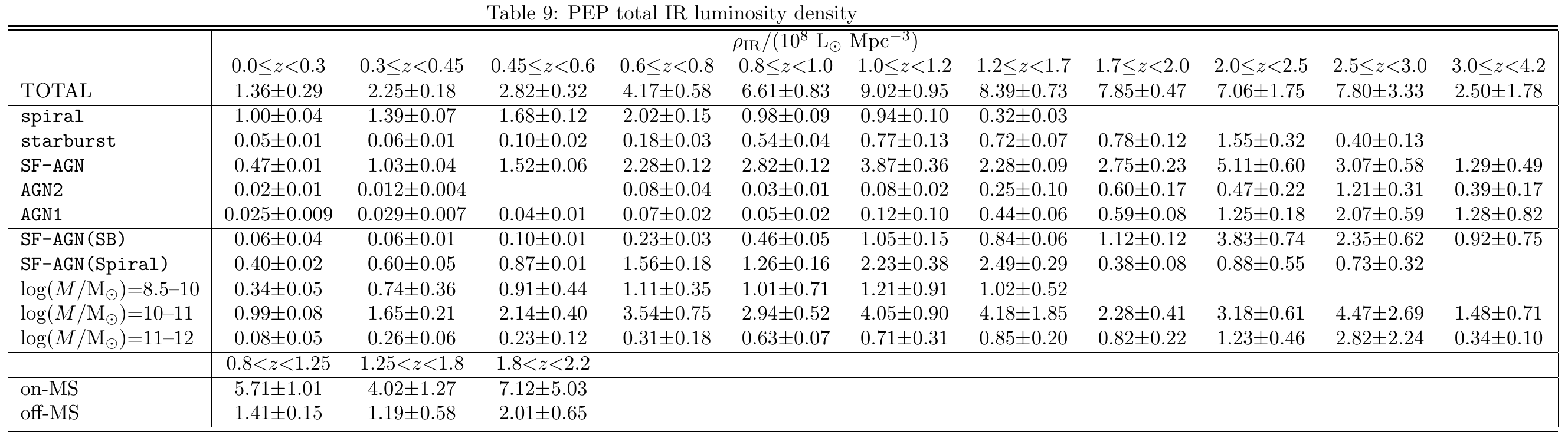}
\label{tab:ldz}
\end{minipage}
\end{figure*}

\label{lastpage}

\end{document}